\documentclass[english,twocolumn,superscriptaddress,showpacs,aps,prb,reprint,floatfix,nobibnotes,longbibliography]{revtex4-2}

\usepackage[utf8]{inputenc}
\usepackage{graphicx,subfigure}     
\usepackage{bm}             
\usepackage{verbatim}
\usepackage{epsf}
\usepackage{amssymb,stmaryrd}   
\usepackage{amsmath}
\usepackage{dcolumn}        
\usepackage{siunitx}
\usepackage{xcolor}
\usepackage{epstopdf}
\usepackage{mathrsfs}
\usepackage{xspace}
\usepackage{placeins}
\usepackage{babel}
\graphicspath{{images/}}  
\usepackage{wrapfig}
\usepackage{tikz}  
\usetikzlibrary{arrows,shapes,trees,positioning}  
\usepackage{comment}
\usepackage{textcomp}
\usepackage[normalem]{ulem}   

\usepackage[colorlinks=true,citecolor=blue]{hyperref}

\usepackage{soul}
\setstcolor{red}
\usepackage{ulem}
\hypersetup{
	colorlinks=true,
	linkcolor=blue,
	citecolor=blue,
	urlcolor=blue,
}
\DeclareUnicodeCharacter{0096}{_}

\newcommand{\FthGT}{Fe\texorpdfstring{$_3$}{3}GeTe\texorpdfstring{$_2$}{2}\xspace}
\newcommand{\FnGT}{Fe\texorpdfstring{$_n$}{n}GeTe\texorpdfstring{$_2$}{2}\xspace}
\newcommand{\HIIab}{$\textbf{H} \perp \textbf{c}$\xspace}
\newcommand{\HIIc}{$\textbf{H} \parallel \textbf{c}$\xspace}

\newcommand{\Hres}{H_\mathrm{res}}
\newcommand{\HD}{$H_\mathrm{D}$\xspace}
\newcommand{\Hr}{$H_\mathrm{res}$\xspace}
\newcommand{\Kint}{$K_\mathrm{int}$\xspace}
\newcommand{\muB}{\mu_\mathrm{B}}
\newcommand{\TC}{$T_\mathrm{C}$\xspace}

\newcommand{\delH}{$\Delta H$\xspace}
\newcommand{\Ha}{$H_\mathrm{a}$\xspace}

\definecolor{darkgreen}{rgb}{0, 0.5, 0}

\DeclareUnicodeCharacter{2212}{\textminus}
\DeclareUnicodeCharacter{1515}{\textminus}

\begin{document}


\title{Magnetic anisotropy in the near-stoichiometric van der Waals ferromagnet \FthGT}

\author{Riju Pal}
\email{rijupal07@gmail.com}
\affiliation{Leibniz Institute for Solid State and Materials Research, Helmholtzstr. 20, Dresden, D-01069, Germany}
\affiliation{Institute for Solid State and Materials Physics, TU Dresden, Dresden, D-01062, Germany}
\affiliation{Department of Condensed Matter and Materials Physics, S. N. Bose National Centre for Basic Sciences, Block JD, Sector III, Salt Lake, Kolkata, 700106, India}

\author{Joyal John Abraham}
\affiliation{Leibniz Institute for Solid State and Materials Research, Helmholtzstr. 20, Dresden, D-01069, Germany}
\affiliation{Institute for Solid State and Materials Physics, TU Dresden, Dresden, D-01062, Germany}

\author{Satyabrata Bera}
\affiliation{School of Physical Sciences, Indian Association for the Cultivation of Science, Jadavpur, Kolkata, 700032, India}

\author{Mintu Mondal}
\affiliation{School of Physical Sciences, Indian Association for the Cultivation of Science, Jadavpur, Kolkata, 700032, India}

\author{Atindra Nath Pal}
\affiliation{Department of Condensed Matter and Materials Physics, S. N. Bose National Centre for Basic Sciences, Block JD, Sector III, Salt Lake, Kolkata, 700106, India}

\author{Bernd Büchner}
\affiliation{Leibniz Institute for Solid State and Materials Research, Helmholtzstr. 20, Dresden, D-01069, Germany}
\affiliation{Institute for Solid State and Materials Physics, TU Dresden, Dresden, D-01062, Germany}
\affiliation{Institute for Solid State and Materials Physics and Würzburg-Dresden Cluster of Excellence ctd.qmat, TU Dresden, Dresden, D-01069, Germany}

\author{Vladislav Kataev}
\affiliation{Leibniz Institute for Solid State and Materials Research, Helmholtzstr. 20, Dresden, D-01069, Germany}

\author{Alexey Alfonsov}
\email{a.alfonsov@ifw-dresden.de}
\affiliation{Leibniz Institute for Solid State and Materials Research, Helmholtzstr. 20, Dresden, D-01069, Germany}




\begin{abstract}
 Quasi-two-dimensional (2D) van der Waals (vdW) ferromagnets such as the series Fe$_{3-x}$GeTe$_2$, with a relatively high Curie temperature and robust metallicity, offer an ideal platform for investigating itinerant magnetism in reduced dimensions. Here, we present a comprehensive electron spin resonance (ESR) investigation of single-crystalline almost stoichiometric Fe$_{3.03 \pm 0.03}$GeTe$_{2}$ across wide ranges of frequencies, temperatures, and magnetic fields to gain quantitative insights into its magnetic anisotropy and spin dynamics. Frequency-dependent ESR measurements establish \FthGT\ as an easy-axis ferromagnet. Temperature-dependent high-field ESR reveals a large internal field that gradually decreases at higher temperatures. Remarkably, this internal field persists even above \TC, evidencing short-range spin correlations in \FthGT. Analysis of spin-wave modes yields a strong uniaxial magnetocrystalline anisotropy $K_{\text{int}} \approx - 5 \times 10^6$~erg\,cm$^{-3}$ at 3~K and a large magnon gap $\Delta(3~\mathrm{K})$ $\approx$ 87.8 $\pm$ 13.7 GHz ($\approx$ 0.363 $\pm$ 0.057 meV). Our study highlights that small variations in Fe content in \FthGT lead to substantial changes in the magnon gap at low temperatures, indicating the extreme sensitivity of spin dynamics to the chemical composition. These results establish \FthGT as a model vdW ferromagnet for exploring tunable anisotropies and magnon excitations in metallic 2D magnets.
\end{abstract}


\maketitle


\section{Introduction}

Layered vdW crystals comprise a versatile class of quantum materials that continue to attract significant attention due to their novel fundamental properties and technological potential \cite{Novoselov2016, Ahn2020a}. Of particular interest are 2D magnetic vdW materials, which can retain long-range magnetic order even in the atomically thin limit, offering promising routes toward miniaturized magnetic memory and spintronic devices \cite{Gong2019, Hao2022, burch2018}. In such systems, the inherently weak interlayer vdW bonding renders many compounds effectively quasi-2D even in their bulk form, since strong intralayer exchange dominates over much weaker interlayer couplings. Additionally, magnetic anisotropy plays an important role in stabilizing the magnetic ground state and controlling the spin dynamics, even in the monolayer limit \cite{MerminWagner, Morrish, Coey}. A quantitative understanding of such anisotropy is therefore essential for describing the microscopic origin and robustness of 2D magnetism and realizing novel spintronic functionalities. In this context, magnetic resonance methods, in particular high-frequency / high-field electron-spin-resonance (HF-ESR) technique, is a sensitive and direct probe of anisotropic fields, magnon excitation gaps, and the temperature dependence of spin relaxation, as well as overall spin dynamics in low-dimensional vdW magnets \cite{Kataev2024, Cho2023fmr, Pal2024_ESR, Moro2022, Lee2020fmr, Pal_MMS_2025, Wang2023fmr, Beier2025, Alfonsov2021b, Alahmed2021, Zeisner2020}.

\begin{figure*}[ht]
    \centering
    \includegraphics[width=0.93\textwidth]{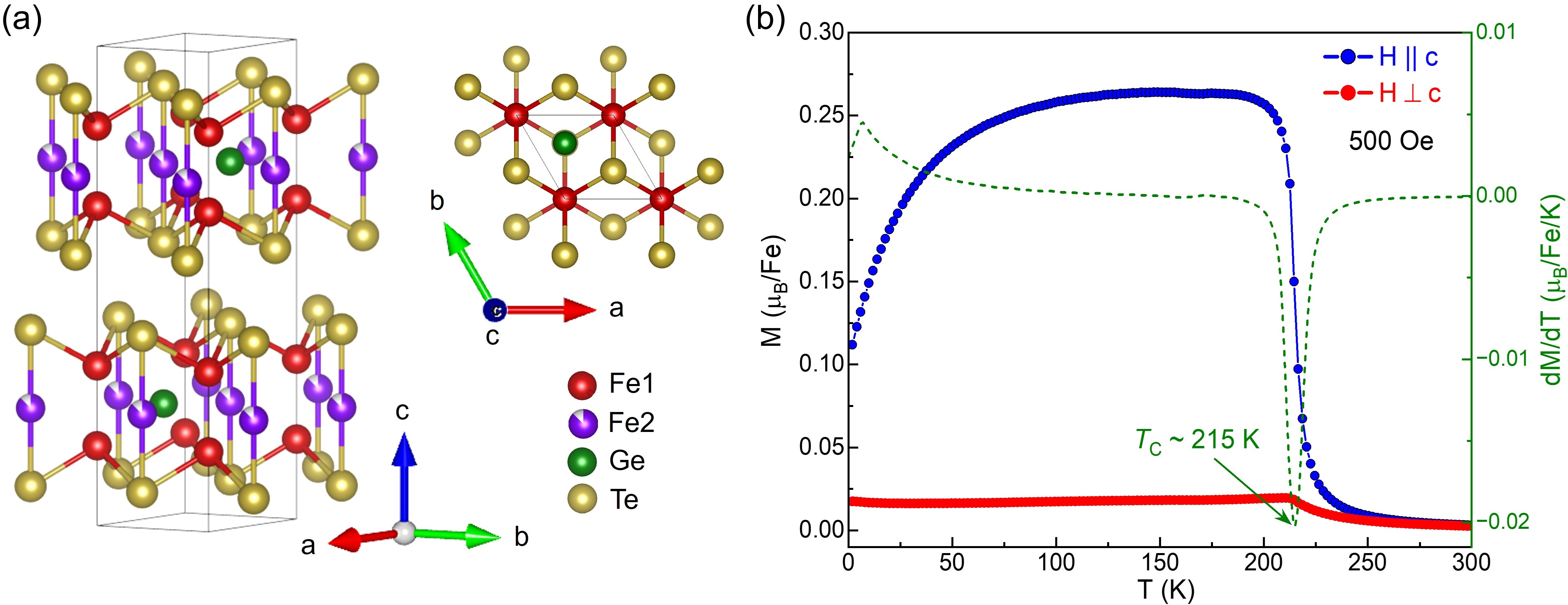}
    \caption{(a) Crystal structure of \FthGT. The unit cell (left) and the top view (right) of a single vdW layer of \FthGT are shown here. The unit cell is enclosed using the dotted lines. The top view of the structure shows the hexagonal symmetry of the ab plane. Two inequivalent Fe sites are termed Fe1 and Fe2. (b) The temperature dependence of magnetization and its temperature derivative, measured under a magnetic field of 500 Oe in the \HIIc and \HIIab configurations. The minimum in the temperature derivative of the magnetization (dM/dT) reveals the Curie temperature \TC $\sim$ 215 K of \FthGT.}
    \label{Fig1}
\end{figure*}

Following the experimental discovery of intrinsic 2D ferromagnetism in insulating compounds such as CrI$_3$ and Cr$_2$Ge$_2$Te$_6$ \cite{Huang2017, Gong2017}, numerous vdW magnets, including CrBr$_3$, CrSiTe$_3$, FePS$_3$, and NiPS$_3$, have been investigated \cite{Jiang2021, Shen2021, Williams2015, Lee2016b, Wang2022c}, revealing unconventional electronic and magnetic properties. However, most of these materials are insulating or semiconducting, with their magnetic ordering temperatures being well below room temperature. In contrast, the Fe$_n$GeTe$_2$ family ($n \approx 3$–$5$) exhibits itinerant ferromagnetism with Curie temperatures (\TC) approaching or even exceeding room temperature \cite{Deng2018b, Pal2024, Seo2020, Pal2024_ESR, Bera2023, Tan2021, Pal_Raman_2025, May2019, Fei2018, Pal_2024_aip}, providing a promising platform for exploring 2D magnetism in metallic systems. \FthGT\ is particularly notable due to its robust metallic ferromagnetism, strong uniaxial magnetic anisotropy, relative environmental stability, and thickness-dependent magnetic properties \cite{Albarakati2019a, Zhou2022b}.

\FthGT\ crystallizes in a hexagonal structure (space group $P$6$_3$/$mmc$), consisting of Fe$_3$Ge layers sandwiched between two Te layers, where adjacent Te planes are held together by weak vdW forces (see Fig. \ref{Fig1}(a)) \cite{Chen2013, Wang2017, Deng2018b, Fei2018}. The Fe sublattice contains two inequivalent crystallographic sites, commonly associated with Fe$^{3+}$ (Fe1) and Fe$^{2+}$ (Fe2). 
While both possess partially filled $d$ orbitals, theoretical studies indicate that they contribute differently to the magnetism: Fe1 generally hosts more localized moments, whereas Fe2 exhibits a more itinerant character, reflecting the mixed itinerant and localized nature of magnetism in this compound \cite{Zhang2018, Xu2024, Kim2022a}. Bulk crystals exhibit a relatively high ferromagnetic ordering temperature ($T_\mathrm{C} \approx 215$~K), with the magnetic easy axis aligned along the crystallographic $c$-direction (Fig.~\ref{Fig1}(b)) \cite{Fei2018, Deng2018b}. Recent experiments have revealed a rich spectrum of correlated electronic and magnetic phenomena in \FthGT, including a giant anomalous Hall effect \cite{Kim2018b, Wang2017}, signatures of Kondo-lattice and heavy-fermion behavior \cite{Zhang2018}, and topological spin textures such as skyrmions \cite{Ding2020a}. Moreover, its magnetic properties can be tuned by thickness \cite{Deng2018b, Tan2018, Fei2018}, electrostatic gating \cite{Deng2018b}, or strain \cite{Hu2020d, ThicknessF3GT}, offering versatile control over its magnetic ground state. 
A particularly intriguing aspect of \FthGT is its intrinsic non-stoichiometric nature. The Fe sublattice can accommodate vacancies, leading to compositions commonly written as Fe$_{3-x}$GeTe$_2$, where even small variations in Fe occupancy significantly modify the magnetic moments, coercive fields, exchange interactions, and \TC \cite{Liu2017c, May2016, Mayoh2021, Backes2024}. While the impact of Fe content on static magnetic properties is well established \cite{May2016, Mayoh2021, Backes2024}, its influence on magnon excitation gaps and magnetocrystalline anisotropy remains insufficiently understood. A recent study of a Fe-deficient crystal ($x = 0.08$) revealed markedly different spin-dynamic behavior \cite{Beier2025}, underscoring the exceptional sensitivity of this material to stoichiometry. A detailed investigation of the magnetic anisotropy and spin excitations in a near-stoichiometric \FthGT is therefore crucial to elucidate the role of Fe stoichiometry in spin dynamics, particularly near the stoichiometric limit ($n \approx 3$), and to establish it as a key tuning parameter for controlling magnetic anisotropy and spin excitations in \FthGT.

In this work, we present a comprehensive ESR investigation of single-crystalline \FthGT with a near-stoichiometric, slightly Fe-rich composition (Fe content $\sim$ 3.03 $\pm$ 0.03) over wide ranges of frequency, temperature, and magnetic field orientation. HF-ESR measurements confirm \FthGT as an easy-axis ferromagnet with a large anisotropy field of $\sim$3.2 T at 3 K, which remains seemingly temperature independent up to $\sim$100~K before gradually decreasing at higher temperatures. A finite shift of the ferromagnetic resonance lines persists even above \TC, evidencing the presence of short-range spin correlations in \FthGT. The intrinsic uniaxial magnetocrystalline anisotropy energy reaches $\sim$ $-5 \times 10^{6}$~erg cm$^{-3}$ at 3~K, yielding a sizable magnon excitation gap of $\Delta(3~\mathrm{K})$ $\approx$ 87.8 $\pm$ 13.7 GHz ($\approx$ 0.363 $\pm$ 0.057 meV), comparatively large among vdW ferromagnets. Comparison with previously reported Fe-deficient compositions reveals pronounced variations in both anisotropy and magnon gap, highlighting the extreme sensitivity of the anisotropy and spin dynamics in \FthGT to its Fe content. Overall, these results demonstrate Fe stoichiometry as an effective microscopic control parameter for tuning magnetic anisotropy and magnon excitation gaps in \FthGT, thereby positioning it as a model metallic vdW ferromagnet for engineering tunable spin excitations in low-dimensional magnetic systems.

\section{Experimental Details}

High-quality single crystals of \FthGT were synthesized \cite{Bera2023a} following the conventional chemical vapor transport (CVT) method \cite{Chen2013, Fei2018, Yi2017b}, using iodine (I$_{2}$) as the transport agent. A stoichiometric mixture of high-purity elemental powders of Fe, Ge, and Te (purity 99.99 \%) in a molar ratio of 3:1:2 was thoroughly ground using an agate mortar. Subsequently, I$_{2}$ powder was added at a concentration of 5 mg cm$^{-3}$, and the entire mixture was sealed in an evacuated quartz tube under a pressure of 2 × 10$^{-4}$ mbar. The sealed quartz tube was then placed horizontally in a two-zone tube furnace, where the temperature was maintained at 750$^{0}$ C on the source side and 700$^{0}$ C on the growth side for a period of seven days. After the growth process, multiple high-quality, shiny, thin single crystals were obtained at the colder end, with typical dimensions around 2 $\times$ 1.5 $\times$ 0.1 mm$^3$. Structural phase analysis was carried out on some of these crystals using X-ray diffraction (XRD) on a Rigaku diffractometer with Cu-K$_\alpha$ radiation ($\lambda$ = 1.54056 \AA), while micro-morphology was examined via field emission scanning electron microscopy (FESEM, JEOL LSM-6500). Energy-dispersive X-ray spectroscopy (EDX) on the freshly cleaved surface of the measured bulk crystal confirms it is slightly Fe-rich, with an average elemental composition Fe$_{3.03(3)}$Ge$_{0.89(4)}$Te$_{2.00(1)}$ (see Appendix \ref{A}). The synthesis procedure, along with detailed structural and physical characterizations confirming the high quality of the grown crystals, has been reported in the earlier work \cite{Bera2023a}.

Temperature and magnetic field-dependent magnetization measurements ($M-T$ and $M-H$) of the \FthGT crystal were performed using a magnetic properties measurement system (MPMS-SQUID by Quantum Design, USA). The $M-T$ measurements were carried out in the temperature range of 1.8 – 300 K under an applied magnetic field of 500 Oe oriented both in the \textit{ab}-plane and along the \textit{c}-axis (see Fig. \ref{Fig1}(b)). To minimize the remanent or residual magnetization present in the sample, an external magnetic field of 5 T was applied at 300 K and then gradually reduced to zero while periodically changing the field's polarity. All measurements were conducted under zero-field-cooled (ZFC) conditions: the sample was cooled from 300 K down to 1.8 K in the absence of an external field. Magnetization data were collected during the warming cycle under a constant applied field. 

A well-characterized bulk single crystal of \FthGT, with dimensions approximately 1.74 mm $\times$ 1.47 mm $\times$ 0.24 mm, was chosen for electron paramagnetic resonance (EPR) and ferromagnetic resonance (FMR) measurements. Conceptually, the primary distinction between EPR and ferromagnetic resonance (FMR) lies in the nature of their excitations: EPR probes the resonant excitation of individual paramagnetic spins within a magnetic system, whereas FMR involves the collective excitations of the net magnetization ($\mathbf{M}$) in a ferromagnetically ordered state \cite{Zeisner2020, Farle1998}. Since the instruments used for the detection of both resonances are the same, in the text below, we will use a more general term, electron spin resonance (ESR). High-frequency and high-field (HF-) ESR studies were conducted using a custom-built spectrometer. Microwaves (MW) in the frequency range of 75 GHz to 330 GHz were generated and detected using the PNA-X vector network analyzer (Keysight Technologies) with frequency extensions from Virginia Diodes, Inc. (VDI), as well as with a set of amplifier multiplier chains (AMC) and zero-bias detectors (ZBD) also from VDI. Continuous magnetic field sweeps were performed using a superconducting magnet (Oxford Instruments). A home-made probehead equipped with oversized waveguides was inserted into a variable temperature insert (VTI) of a He$^4$ cryostat (Oxford Instruments), allowing the stabilization of sample temperature between 3\, K and 300\, K. All the measurements were carried out by continuously sweeping the magnetic field from 0 to 16 T and back at fixed frequencies and temperatures. In the HF-ESR system, utilizing a vector network analyzer, some instrumental mixing of the absorption and dispersion signals is unavoidable due to the complex impedance of the broadband probehead. Since the spectral lines deviate from a Lorentzian shape, and therefore it was impossible to account for such a mixing during the fitting procedure, the \Hr and \delH were determined by manual picking of the position of the minimum of the resonance line and by calculating the full width at half of this minimum, respectively. In order to extract these values \Hr and \delH from such a mixed signal at the detector, we employed the approach reported in \cite{Pal2024_ESR}. 

\begin{figure*}[ht]
    \centering
    \includegraphics[width=\textwidth]{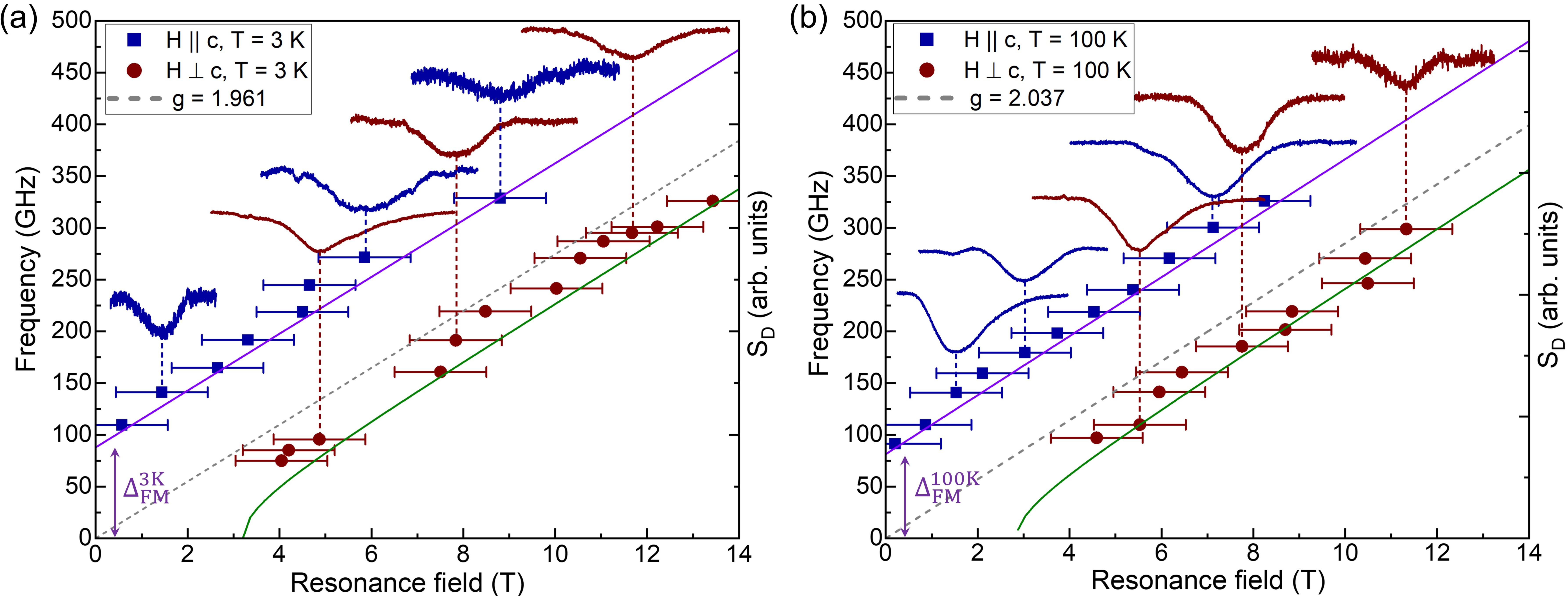}
    \caption{Frequency dependence of the resonance fields \Hr at $T = $ 3\,K (a) and 100\,K (b) measured in \HIIc (blue squares) and in \HIIab (red circles) configurations, respectively. Dashed lines represent the paramagnetic response estimated with Eq.~\ref{eq:Hnu} with the average $g$-factors given in the legends. Solid lines indicate the simulated best fit results using Eq. \ref{Eq:EAHIIc} and \ref{Eq:EAHIIab}. Representative normalized and vertically shifted spectra are shown for some of the frequencies (right vertical axis) with the colors following respective magnetic field orientations.}
    \label{Fig2}
\end{figure*}

\section{Results and Discussion}

\subsection{Frequency dependence of HF-ESR response}
\label{sec:freq_dep}

To investigate the magnetic anisotropies and $g$-factors of \FthGT in detail, we performed frequency-dependent ESR measurements at some selected temperatures -- in the magnetically ordered state (at 3 K and 100 K) and well above the magnetic phase transition (at 300 K) in both configurations of the magnetic field \HIIc and \HIIab. The corresponding frequency vs. magnetic field ($\nu$–$H$) diagrams are presented in Fig. \ref{Fig2}(a), (b), and Appendix \ref{B}. Exemplary spectra are displayed on the right vertical axis of Fig. \ref{Fig2}. At all temperatures, the ESR spectra exhibit a single, relatively broad resonance line, with an average full width at half maximum (FWHM) of approximately 1 T. To account for potential uncertainties in the extracted parameters due to the spectral quality, we have assigned maximum possible error bars equal to the average spectral linewidth ($\approx$1 T) for all temperatures  (3 K, 100 K, and 300 K) in the $\nu$–$H$ diagrams. 

The expected paramagnetic resonance position at a resonance field $H_{res}^{para}$ can be estimated from the following relation \cite{Abragam2012}:
\begin{equation}
		h\nu = g\muB \mu_{0} H_{res}^{para}
		\label{eq:Hnu}
\end{equation}
where $\nu$ is the microwave frequency, $h$ is Planck's constant, $g$ is the $g$-factor of the resonating spins, $\mu_{0}$ is the vacuum permeability, $\muB$ is the Bohr magneton.

The analytical expressions of the spin wave energies for a uniaxial easy-axis ferromagnet are \cite{Turov, Holstein1940, Alfonsov2021b}:
\begin{subequations}
	\label{Eq:EA}
	\renewcommand{\theequation}{\theparentequation.\arabic{equation}}
	\begin{align}
	&\text{\HIIc: \ }  & h \nu &= \text{g} \muB \mu_{0} (H+|H_a|)  \label{Eq:EAHIIc}\\
	&\text{\HIIab: \ } & h \nu &= \text{g} \muB \mu_{0} \sqrt{H(H-|H_a|)} \ \text{for } H \geq |H_a| \label{Eq:EAHIIab}\\
    &  & h \nu &= \text{g}\mu_B\mu_0 \sqrt{|H_a|^2 - H^2} \ \text{for } H \leq |H_a| 
\end{align}
\end{subequations}
where \Ha is the total magnetic anisotropy field.

At 300 K ($\sim$85 K above \TC ), a linear dependence of frequency on the resonance field is observed in the $\nu$-$\Hres$ diagram for both field orientations (see Appendix \ref{B}). This behavior is characteristic of the paramagnetic state of \FthGT and can be described by the conventional paramagnetic resonance condition (see Eq. \ref{eq:Hnu}). Fits to the data according to Eq.~\ref{eq:Hnu} are shown in Fig. \ref{FigS1} in Appendix \ref{B} as dashed lines, yielding an average expected paramagnetic $g$ = 2.07 $\pm$ 0.04. 

\begin{figure*}[ht]
	\centering
	\includegraphics[width=0.95\textwidth]{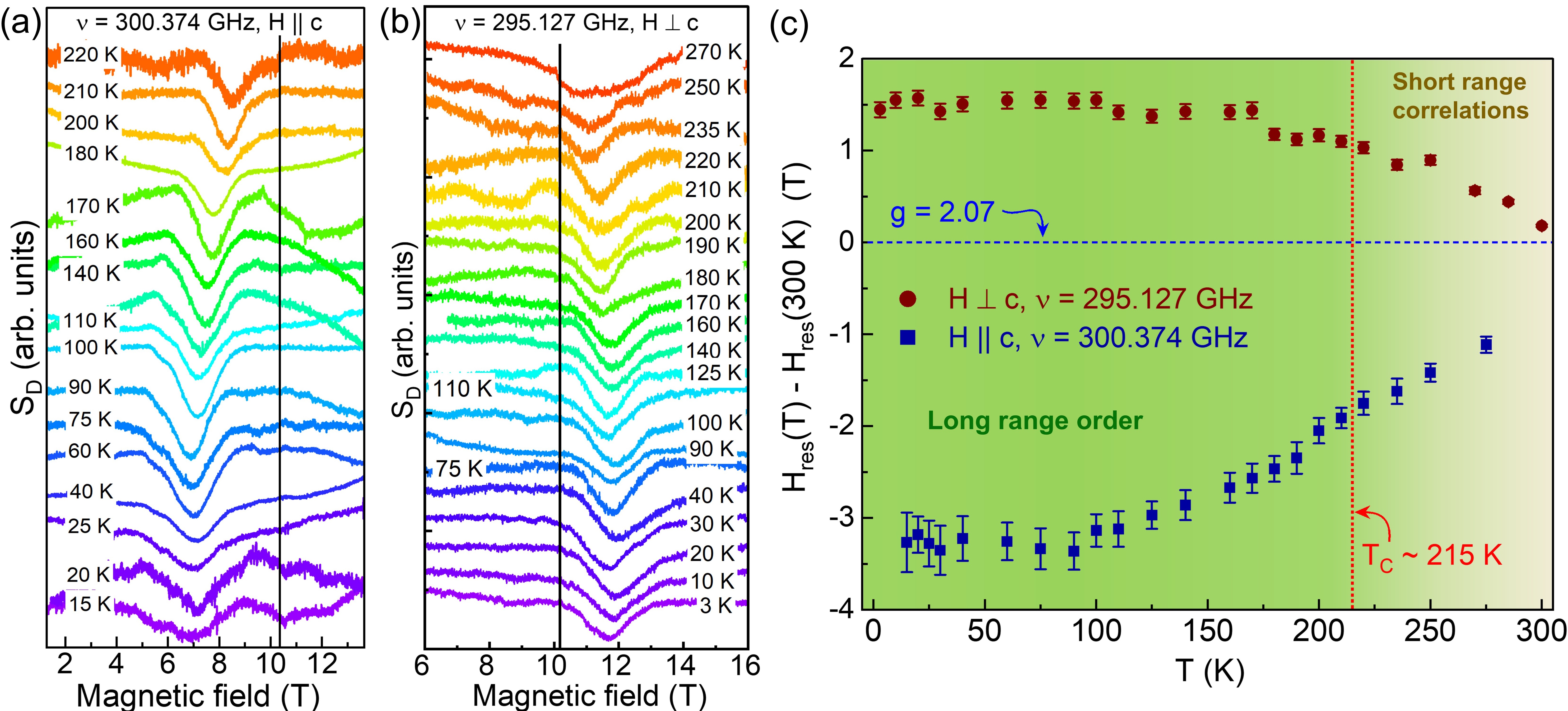}
	\caption{Temperature evolution of the HF-ESR spectra in the field direction \HIIc (a) and \HIIab (b), measured at fixed excitation frequencies of approximately 300 GHz. For clarity, all spectra are normalized to unity and vertically offset. Vertical black solid lines indicate the expected paramagnetic resonance position as given by Eq. \ref{eq:Hnu}. (c) Temperature dependence of the resonance shift, $\delta H(T)$, at a microwave frequency $\nu$ of $\approx$300 GHz with an external magnetic field applied parallel (blue squares) and perpendicular (red circles) to the c axis. A blue dashed line indicates the zero shift in the paramagnet state with the $g$-factor, $g$ = 2.07. The red dotted line shows the ordering temperature (\TC $\approx $ 215 K) of \FthGT, determined at an applied magnetic field of 500 Oe. The color gradient illustrates the regions of long-range ordering and short-range correlations.}
    \label{Fig3}
\end{figure*}
In the magnetically ordered state at $T = 3$~K (see Fig.~\ref{Fig2}(a)), the resonance field data for \HIIc show a positive intercept on the frequency axis, suggesting a ferromagnetic behavior with a uniaxial easy-axis anisotropy. This magnetic anisotropy can also be associated with the presence of static internal fields on the ESR timescale, which shift the position of the ESR line. To determine the g-factors, unconstrained fitting of Eq.~\ref{Eq:EA} to the measured data was used in this case. For \HIIc, the fit using Eq.~\ref{Eq:EAHIIc} yields $g_\parallel$ = 1.95 $\pm$ 0.11, whereas for \HIIab, $g_{\perp}$ = 1.97 $\pm$ 0.06. The average $g$-factor is thus $g_{av}^{3 K}$ $\approx$ 1.96 $\pm$ 0.08. Now, using this averaged g-factor, the best-fit lines are calculated for \HIIab and \HIIc simultaneously, based on a uniaxial easy-axis ferromagnet model (Eq. \ref{Eq:EA}). These lines show good agreement with the experimental $\nu$-$\Hres$ data within the error bars (see Fig.~\ref{Fig2}(a)). Therefore, at 3 K, the overall $\nu$-$\Hres$ dependence for both magnetic field orientations is found to be consistent with the expected behavior of a conventional uniaxial ferromagnet with easy-axis anisotropy and can be well described by Eqs.~\ref{Eq:EA}. The ferromagnetic zero-field magnon excitation gap ($\Delta$) for $\mathbf{H} \parallel c$, estimated from the intercept with the frequency axis, yields the value $\Delta_{\mathrm{FM}}|_{H=0} \equiv \Delta_{\mathrm{FM}}^{3 K}$ $\approx$ 87.8 $\pm$ 13.7 GHz ($\approx$ 0.363 $\pm$ 0.057 meV).

Similarly, at $T = 100$ K (Fig. \ref{Fig2}(b)), the $\nu$-$\Hres$ dependence for both magnetic field orientations shows an easy-axis anisotropy. The best fit using Eqs.~\ref{Eq:EA} provides $g_\parallel$ = 2.07 $\pm$ 0.04 for \HIIc, and $g_\perp$ = 2.01 $\pm$ 0.09 for \HIIab. This results in an average $g$ factor $g_{av}^{100 K}$ $\approx$ 2.04 $\pm$ 0.07. The estimated magnitude of the zero-field magnon excitation gap at $T$ = 100 K is $\Delta_{\mathrm{FM}}^{\mathrm{100 K}}$ $\approx$ 81.3 $\pm$ 7.1 GHz ($\approx$ 0.336 $\pm$ 0.029 meV), indicating a similar gap, within the error bar, to the value observed at $\sim3$ K. Notably, this remarkably large and persistent magnon excitation gap in \FthGT, exceeding those reported for other vdW magnetic materials \cite{Alahmed2021, Pal2024_ESR, Yang2025, Shen2021, Jonak2022, Dillon1965, Li2022anomalous, khan2019, Zeisner2018, wang2023esr}, underscores the strong magnetic anisotropy present in this system. 

To examine the possible in-plane (\(ab\)-plane) anisotropy, the \FthGT crystal was rotated by 90$^\circ$ from its initial in-plane orientation (IP1). In this orthogonal in-plane configuration (IP2), the frequency dependence of the resonance field ($\nu$ vs $H_\mathrm{res}$) was measured at 3 K and 100 K. A comparison between the results from the IP1 and IP2 configurations is presented in Appendix \ref{C}. Within experimental uncertainty, both orientations yield nearly identical behavior at both temperatures, indicating the absence of measurable in-plane anisotropy in this \FthGT\ sample. This observation is consistent with earlier findings from magnetization studies \cite{Wang2017}.

\subsection{Temperature dependence of HF-ESR response}

The temperature evolution of the HF-ESR response in \FthGT was investigated over the range $T = 3 - 300$\,K at excitation frequencies close to 300\,GHz for both magnetic field orientations, \HIIc and \HIIab. The corresponding ESR spectra are presented in Fig.~\ref{Fig3}(a) and (b), respectively. The vertical solid line indicates the expected paramagnetic resonance position ($H_{\mathrm{res}}^{\mathrm{para}}$), as determined from Eq.~\ref{eq:Hnu} using the g-factor measured at 300 K.

At the lowest measured temperatures, the ESR responses are observed at magnetic fields either lower (\HIIc) or higher (\HIIab) than the expected paramagnetic resonance position, respectively. This is in line with frequency-dependent measurements (see Sec.~\ref{sec:freq_dep}). As the temperature increases, for both magnetic field configurations, the resonance lines are gradually shifting toward the expected paramagnetic resonance line (see Fig. \ref{Fig3}). To obtain a deeper understanding of how the magnetic anisotropy evolves with temperature, the resonance fields $H_{\mathrm{res}}$, and their temperature-dependent shifts, defined as $\delta H(T) = H_{\mathrm{res}}(T) - H_{\mathrm{res}}(300\,\mathrm{K})$, are plotted for both \HIIc and \HIIab configurations as shown in Appendix \ref{E} and Fig.~\ref{Fig3}(c), respectively. The reference resonance field, $H_{\mathrm{res}}(300~\mathrm{K})$, was determined using Eq.~\ref{eq:Hnu}, with the $g$-factor value extracted from the $\nu$-$\Hres$ data measured at $T = 300$~K. This shift $\delta H(T)$ serves as a measure of the deviation of the resonance field from the expected paramagnetic resonance position at each temperature. As depicted in Fig.~\ref{Fig3}(c), the $\delta H(T)$ dependence for two magnetic field orientations displays an asymmetric behavior relative to the zero-shift ($g$ = 2.07) line. In general, the resonance shifts are more pronounced when the magnetic field is applied along the direction of the magnetic anisotropy axis -- here along the $c$-axis of the \FthGT crystal, as also reflected in Fig. \ref{Fig2}. This observation highlights the $c$-axis as the dominant magnetic easy-axis at all temperatures, aligning with the magnetization data (see Appendix \ref{D}) ~\cite{Fei2018, Wang2017, Tan2018}.

\begin{figure*}[!ht]
	\centering
	\includegraphics[width=0.95\textwidth]{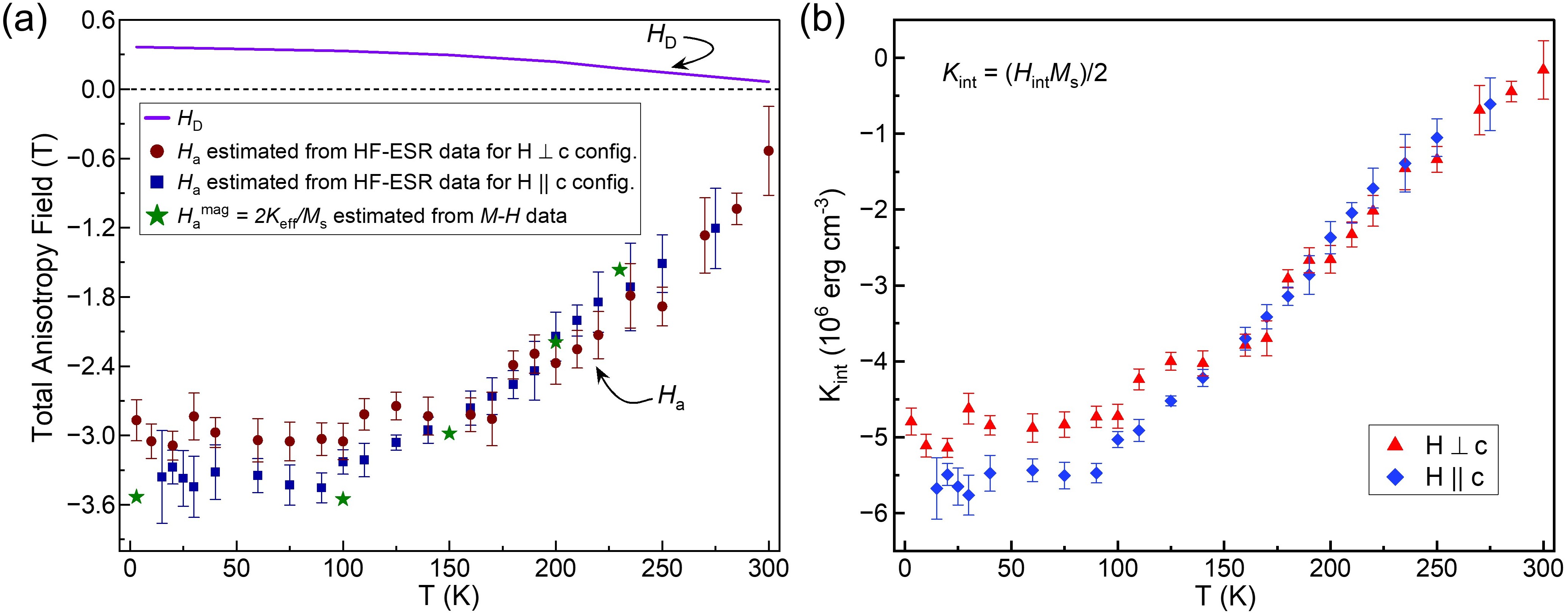}
	\caption{(a) Total anisotropy field \Ha as a function of temperature for both \HIIab and \HIIc configurations, extracted from the temperature dependence of the resonance fields. The purple solid line is the calculated shape anisotropy field \HD given by the shape of the crystal. Green stars are the total anisotropy field calculated from the $M-H$ data. (b) Temperature dependence of the intrinsic MA constant (\Kint) estimated from the measurements in the \HIIab and \HIIc configurations, respectively.}
	\label{Fig5}
\end{figure*}

Notably, the resonance shift ($\delta H(T)$) remains finite for both field orientations even at temperatures up to $\sim$ 300 K (Fig. \ref{Fig3}(c)), approximately 85\,K above the $T_{\mathrm{C}}$ $\approx $ 215 K of \FthGT, determined at an applied magnetic field of 500 Oe. We observed a pronounced enhancement of ferromagnetism with increasing external magnetic field \cite{Liu2017c} (see Fig.~\ref{FigS10} in Appendix.~\ref{J}): For fields above 0.5~T, \TC increases monotonically, reaching $\sim$245~K at 7~T and can be extrapolated to \TC $\approx$ 255~K at 10~T. As shown in Fig. \ref{FigS10} in Appendix.~\ref{J}, the ferromagnetic phase transition at this field is broadened over a region of $\pm$ 25 K. The HF-ESR data shown in Fig. \ref{Fig3}(c) were obtained at magnetic fields of $\sim$ 10 T, suggesting that the observed resonance shift up to $\sim$ 255 K arises from magnetic field-assisted enhancement of ferromagnetic correlations. However, a finite resonance shift persists up to room temperature ($\sim$300~K). This behavior indicates the possible presence of short-range spin correlations in \FthGT that remain static on the HF-ESR timescale ($\sim$10~ps) and develop in the close vicinity of the paramagnetic regime above \TC, which is characteristic of low-dimensional spin systems \cite{Alfonsov2021b, Zeisner2018}.

\subsection{Estimation of the temperature dependence of magnetic anisotropy}

To determine the temperature evolution of the total magnetic anisotropy field $H_{\mathrm{a}}$, the temperature dependence of $H_{\mathrm{res}}$ measured at $\approx 300$ GHz for both \HIIc and \HIIab configurations was analyzed using Eqs. \ref{Eq:EAHIIc} and \ref{Eq:EAHIIab}, respectively (see Appendix \ref{E}). In this process, the $g$-factor was fixed at the paramagnetic value of $g \approx 2.07$. Consequently, $H_{\mathrm{a}}$ parameter was extracted, and its temperature evolution is presented in Fig. \ref{Fig5}(a) for both magnetic field orientations. Here, the negative sign of \Ha defines the easy-axis type of anisotropy in \FthGT. As expected for uniaxial ferromagnets (Eq. \ref{Eq:EA}),  for both \HIIc and \HIIab, the values of the total anisotropy field are found to be practically the same at all temperatures. As can be seen in Fig. \ref{Fig5}(a), \Ha is gradually increasing with decreasing temperature, and it becomes almost constant below $\sim$100 K, within the error bar. For comparison, the estimates of \Ha from the static magnetization ($M-H$) curves, measured at several selected temperatures, are plotted in Fig.~\ref{Fig5}(a) by the star symbols, showing a reasonable agreement with the HF-ESR results (see Appendix \ref{F} and \ref{G} for more details).

To gain a deeper understanding of the magnetic anisotropy in \FthGT, it is essential to disentangle and quantify the intrinsic contribution from the total anisotropy by eliminating the extrinsic effect arising from the sample shape. The total anisotropy field at each temperature, obtained from the HF-ESR measurements, can then be expressed as the sum of two components \cite{Pal2024_ESR}:
\begin{equation}
    H_a = H_{\text{int}} + H_{\text{D}}
    \label{eq:Ha}
\end{equation}
where \( H_{\text{int}} \) is the intrinsic magnetocrystalline anisotropy field. The term \( H_{\text{D}} \) corresponds to the shape anisotropy field, which arises due to the sample geometry. This contribution can be quantitatively described by:
\begin{equation}
H_{\text{D}} = 4 \pi N M_\text{s},
\label{eq:HD}
\end{equation}
with $N$ representing the demagnetization factor for the plate-like geometry of the sample, and \( M_\text{s} \) denoting the saturation magnetization.

To investigate the origin of the total anisotropy field, the shape anisotropy field was first estimated at various temperatures. Considering the plate-like shape of the sample with dimensions 1.74 mm $\times$ 1.47 mm $\times$ 0.24 mm, the demagnetization factor was approximated as $N$ $\approx 0.76 $ for the measured crystal \cite{Osborn1945, Cronemeyer1991}. The shape anisotropy field (\HD) was then calculated using Eq. \ref{eq:HD} with the saturation magnetization values \( M_\text{s}(T) \) measured at different temperatures (see Appendix \ref{H}). The result is plotted as a solid line in Fig. \ref{Fig5}(a) and Fig. \ref{FigS6}(a) in Appendix \ref{I}. This estimate of \( H_D \) enables determination of the intrinsic magnetocrystalline anisotropy field \( H_{\text{int}} \) using Eq. \ref{eq:Ha}.

The estimated \( H_{\text{int}} \) is negative over the entire measurement temperature range (see Fig. \ref{FigS6}(b) in Appendix \ref{I}), confirming that the intrinsic magnetocrystalline anisotropy of \FthGT is of the easy-axis type. Furthermore, the magnitude of \( H_{\text{int}} \) is significantly larger than that of \( H_D \) at all temperatures. Finally, the intrinsic magnetocrystalline anisotropy constant \( K_{\text{int}} = H_{\text{int}} M_\text{s}/2 \) was evaluated for both magnetic field orientations and is presented in Fig.~\ref{Fig5}(b). At 3 K, the value of $K_{\text{int}}$ is $\approx$ -5 $\times$ 10$^6$ erg cm$^{-3}$, which is nearly a factor of two smaller than that reported for the closely related compound with lower Fe content (= 2.92) \cite{Beier2025}, and yet remains significantly larger than the values reported for other vdW ferromagnetic materials (see Fig. \ref{FigS9}(a)) \cite{Alahmed2021, Pal2024_ESR, Yang2025, Shen2021, Jonak2022, Dillon1965, Li2022anomalous, khan2019, Zeisner2018, wang2023esr}.

To visualize the magnetic anisotropy of a single sublattice in \FthGT, we can express the anisotropy energy density of a uniaxial ferromagnet whose easy axis lies along the crystallographic $c$-axis (taken as the $z$-axis) in polar coordinates as \cite{Coey, Morrish, Cullity}
\begin{equation}
E_{\mathrm{ani}} = K_u \cos^2\theta,
\label{eq:anisotropy}
\end{equation}
where $K_u < 0$ corresponds to \textit{easy-axis} anisotropy and $\theta$ is the polar angle between the magnetization and the $z$-axis. For $K_u < 0$, the energy is minimized at $\theta = 0$ or $\pi$ ($E_{\mathrm{ani}} = -K_u$), corresponding to magnetization along $\pm z$ direction, and maximized at $\theta = \pi/2$ ($E_{\mathrm{ani}} = 0$), i.e., when the magnetization lies in the $xy$-plane. Thus, the $\pm z$ direction defines the easy axis, while the $xy$-plane represents the hard-axis orientations.

\begin{figure}[ht]
	\centering
	\includegraphics[width=0.95\linewidth]{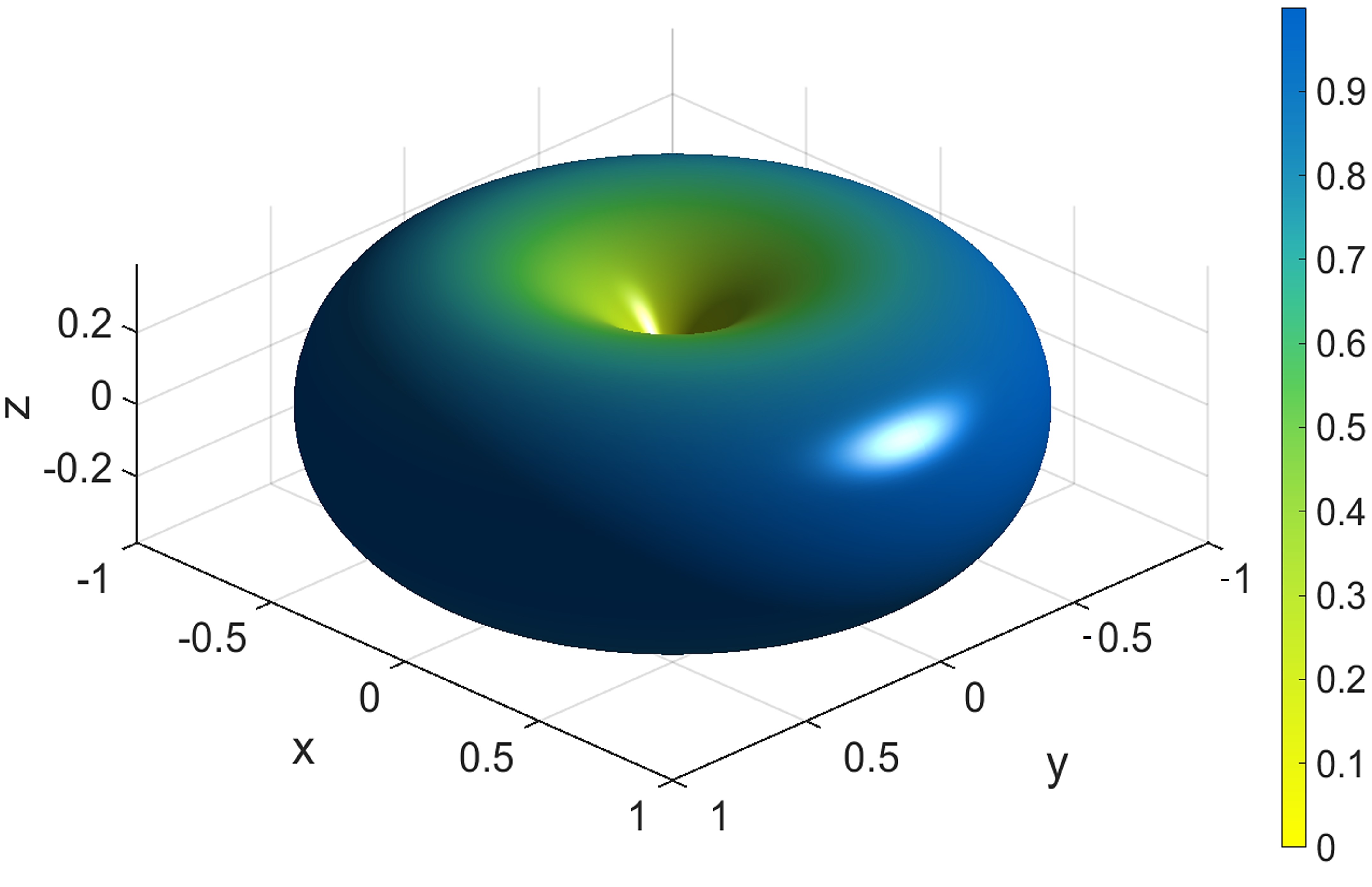}
	\caption{Anisotropy energy density demonstration in Cartesian coordinates (following Eq.~\ref{eq:anisotropy}), illustrating magnetic anisotropy of a single sublattice of \FthGT, as follows from the analysis of the HF-ESR results. Here, the yellow color corresponds to the energetically favorable direction (easy axis along the $c$-axis), while blue indicates higher-energy regions (hard plane). The surface is rotationally symmetric about the $z$-axis, reflecting the uniaxial magnetic anisotropy of the crystal.}
	\label{Fig8a}
\end{figure}

Figure~\ref{Fig8a} illustrates the three-dimensional (3D) anisotropy energy density following Eq.~(\ref{eq:anisotropy}). The color scale represents the magnitude of the anisotropy energy, with the yellow regions indicating the lowest-energy configurations (the easy-axis along the $z$ direction) and blue regions corresponding to higher-energy orientations (along the $xy$-plane). The surface is rotationally symmetric about the $z$-axis, reflecting the uniaxial symmetry of the crystal: all azimuthal directions in the basal plane are energetically equivalent. This visualization highlights the easy-axis anisotropy and orientation dependence of the magnetization in \FthGT, consistent with its layered structure and spin-orbit-induced anisotropy \cite{Zhu2016, Liu2022}.

\begin{figure*}[ht]
	\centering
	\includegraphics[width=\textwidth]{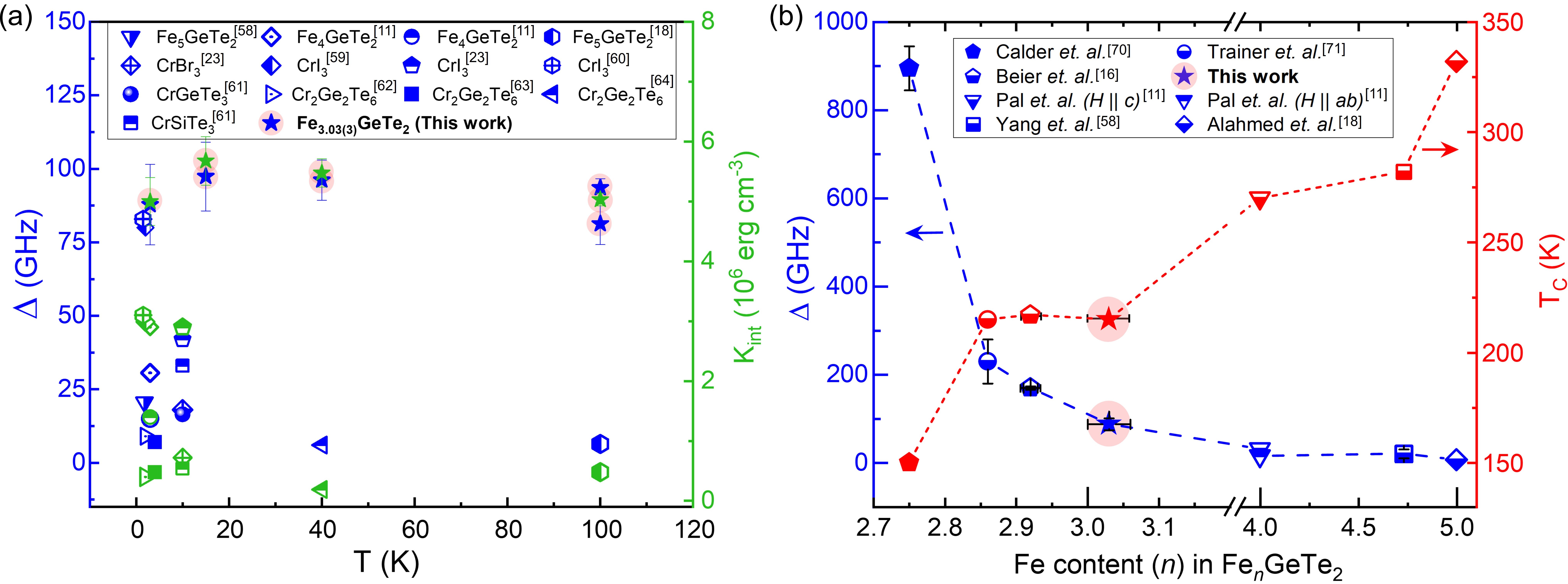}
	\caption{(a) Comparison of zero-field magnon excitation gap (left axis and blue symbols) and intrinsic magnetocrystalline anisotropy energy (right axis and green symbols) of the measured Fe$_{3.03(3)}$GeTe$_2$ crystal (close to stoichiometric \FthGT) with those of other vdW ferromagnetic materials. The star symbol represents the values obtained from frequency- and temperature-dependent resonance field data in the present work. Data for the other materials are compiled from Refs. \cite{Alahmed2021, Pal2024_ESR, Yang2025, Shen2021, Jonak2022, Dillon1965, Li2022anomalous, khan2019, Zeisner2018, wang2023esr}. (b) Zero-field magnon excitation gap (left axis with blue symbols) and the ferromagnetic ordering temperature (right axis with red symbols) as a function of the Fe content in the \FnGT family. The magnon gap exhibits a strong sensitivity to the Fe content, particularly near $n \approx 3$, where it decreases rapidly with increasing Fe occupancy. In contrast, \TC increases nearly linearly with $n$. In addition to the present work (indicated by the star symbol), data of different Fe contents are taken from inelastic neutron scattering, magnetization, and FMR measurements reported in Refs. \cite{Calder2019, Trainer2022, Beier2025, Pal2024_ESR, Yang2025, Alahmed2021}.}
	\label{FigS9}
\end{figure*}

As seen from Fig.~\ref{Fig5}(b), with increasing temperature, \( K_{\text{int}} \) remains nearly constant up to $\approx$ 100~K within the error bars, and then decreases rapidly at higher temperatures. Several studies on \FthGT\ have reported a temperature-driven reconstruction of its non-trivial electronic band structure below \TC, giving rise to significant electronic correlation effects, electron mass-renormalization, Fermi surface enlargement, and even a gradual formation of Kondo-lattice below T$^{*}$ $\sim$ 100~K \cite{Zhang2018, Bao2023, Deng2018b, Bai2022, Bao2022neutron, cora2020, Liu2024}, possibly related to the coexistence and complex interplay between localized moments and itinerant electrons occupying different 3$d$ orbitals of Fe atoms in this metallic ferromagnet \cite{Bao2022neutron, Bai2022, Zhang2018}. Such weakening of the temperature dependence of magnetocrystalline anisotropy energy density \( K_{\text{int}} \) below $\sim$ 100~K (see Fig.~\ref{Fig5}(b)) might be related to these electronic phenomena. Therefore, this finding calls for density functional theory (DFT) calculations, including spin-orbit coupling and Hubbard potential $U$, to obtain deeper insights into a possible influence of the observed electronic correlation effects on magnetocrystalline anisotropy.

To enable a direct comparison of the magnon excitation gap and intrinsic magnetocrystalline anisotropy energy of \FthGT with those of other vdW ferromagnets measured at different temperatures, we evaluate the ferromagnetic zero-field magnon gap from our HF-ESR data. As the magnon gap is a manifestation of the out-of-plane magnetic anisotropy, it can be approximately estimated from the temperature dependence of total anisotropy field \( H_a \) for the \HIIc\ configuration (Fig.~\ref{Fig5}(a)) using Eq.~\ref{Eq:EAHIIc} as:
\begin{equation}
\Delta (T) = \nu \big|_{H=0} = \frac{g \mu_B \mu_0}{h} \left| H_a (T) \right|.
\end{equation}
At $T$ = 15 K, the zero-field magnon gap is estimated as $\Delta\big|_{15 K}$ = 97.4 $\pm$ 11.7 GHz ($\approx$ 0.403 $\pm$ 0.045 meV). This large gap remains almost constant up to temperatures of approximately 100~K and agrees, within error bars, with the value obtained from the \(\nu\)--\(H\) data at the same temperatures (Fig. \ref{Fig2}). A comparison of the estimated magnon excitation gap and intrinsic magnetocrystalline anisotropy energy of different vdW ferromagnetic materials is plotted in Fig. \ref{FigS9} (a).

Importantly, in \FnGT\ family, a close examination reveals that the zero-field magnon excitation gap is highly sensitive to the Fe content (\(n\)), particularly in the vicinity of \(n \approx 3\). A comparison across different Fe compositions shows that the gap decreases sharply with increasing Fe occupancy, following an approximately exponential trend, as plotted in Fig. \ref{FigS9} (b). For instance, the gap reduces from \(\sim 895\)~GHz for \(n = 2.75\) \cite{Calder2019} to \(\sim 230\)~GHz (\(n = 2.86\)) \cite{Trainer2022}, \(\sim 170\)~GHz (\(n = 2.92\)) \cite{Beier2025}, and further to \(\sim 88\)~GHz for the near-stoichiometric, slightly Fe-rich composition (\(n = 3.03 \pm 0.03\)) investigated in the present work. This decreasing trend continues for higher members of the series, reaching \(\sim 30.6\) and 15~GHz for \(n = 4\) \cite{Pal2024_ESR}, 20.6 GHz for \(n = 4.73\) \cite{Yang2025}, and \(\sim 6.4\)~GHz for \(n = 5\) \cite{Alahmed2021}. In contrast, the Curie temperature \(T_\mathrm{C}\) exhibits a markedly different behavior, increasing nearly linearly with Fe content \cite{May2016, Mayoh2021}, from \(\sim 150\)~K at \(n = 2.75\) to \(\sim 215\)--217~K near \(n \approx 3\), and further to \(\sim 270\)~K (\(n = 4\)), \(\sim 282\)~K (\(n = 4.73\)) and \(\sim 332\)~K (\(n = 5\)) \cite{Calder2019, Trainer2022, Beier2025, Pal2024_ESR, Yang2025, Alahmed2021, Shu2024}. These results underscore the nontrivial and inverse relationship between different magnetic properties in \FnGT, establishing Fe stoichiometry as a crucial microscopic control parameter governing spin excitations.

\section{Conclusion}

In summary, we have carried out a comprehensive ESR investigation of near-stoichiometric single-crystalline quasi-2D vdW ferromagnet \FthGT across a wide range of frequencies, temperatures, and magnetic field orientations, enabling us to disentangle and quantify the intrinsic magnetocrystalline anisotropy. Frequency-dependent ESR data establish Fe$_{3.03(3)}$GeTe$_2$ as an easy-axis ferromagnet with the easy axis aligned along the crystallographic $c$-direction. Temperature-dependent HF-ESR measurements reveal the evolution of magnetic anisotropy, with the internal field remaining almost constant up to 100 K ($\sim$ 3.2 T at 3 K). The intrinsic magnetocrystalline anisotropy shows an easy-axis character throughout the measured temperature range, reaching $\sim -5 \times 10^6$ erg cm$^{-3}$ at 3 K -- comparatively larger than typical values reported for other vdW ferromagnets outside the Fe$_{3-x}$GeTe$_2$ family -- confirming strong uniaxial anisotropy of \FthGT. The magnon excitation gap, $\Delta|_{3\mathrm{K}}$ = 87.8 $\pm$ 13.7 GHz ($\approx$ 0.363 $\pm$ 0.057 meV) further underscores the large anisotropy in this system. Notably, our study concludes that even small variations in Fe content can lead to substantial changes in the magnon gap at low temperatures, highlighting the extreme sensitivity of spin dynamics to chemical composition in this \FthGT. Furthermore, the shift of the ESR lines persists at temperatures even above \TC, suggesting the enhancement of ferromagnetic short-range spin correlations in \FthGT. These findings establish the \FthGT\ as a model platform for exploring tunable spin-wave excitations and anisotropies in metallic vdW ferromagnets.\\

\begin{acknowledgments}
The authors acknowledge R. Kluge and S. Wurmehl for the EDX measurements. R. Pal acknowledges the Department of Science and Technology (DST), SNBNCBS, and IFW Dresden for funding. S. Bera thanks CSIR, Govt. of India, for the Research Fellowship with Grant No. 09/080(1110)/2019-EMR-I. M. Mondal acknowledges funding from the DST, Government of India (CRG/2023/001100), and CSIR-Human Resource Development Group (HRDG) (03/1511/23/EMR-II). A. N. Pal acknowledges funding from the DST Nano Mission with grant No. DST/NM/TUE/QM-10/2019. This work was supported by the Deutsche Forschungsgemeinschaft (DFG) through grants No. AL 1771/8-1 (Project No. 499461434), and within the Collaborative Research Center SFB 1143 ``Correlated Magnetism – From Frustration to Topology'' (project-id 247310070), and the Dresden-Würzburg Cluster of Excellence (EXC 2147) ``ctd.qmat - Complexity, Topology and Dynamics in Quantum Matter'' (project-id 390858490).

\end{acknowledgments}


\onecolumngrid
\FloatBarrier

\section*{Appendix}

\appendix

\renewcommand{\thefigure}{A\arabic{figure}}     
\setcounter{figure}{0}

\section{Energy-dispersive X-ray spectroscopy (EDX) data} \label{A}
\begin{figure}[ht]
	\centering
	\includegraphics[width=0.55\linewidth]{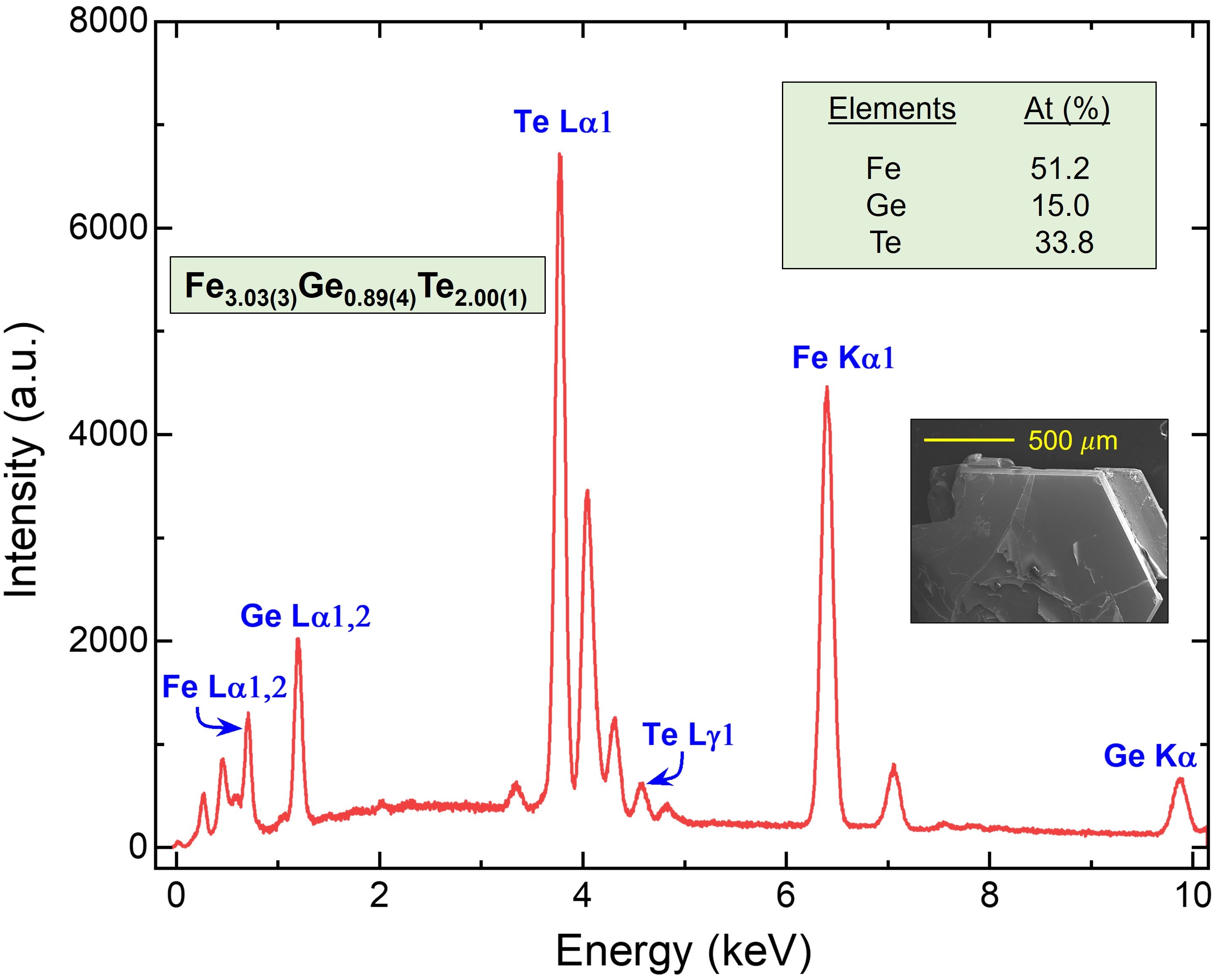}
	\caption{Energy-dispersive X-ray spectroscopy (EDX) spectra of the bulk \FthGT single crystal. Measurements were performed at multiple positions across the crystal, demonstrating good compositional homogeneity with small error bars. Inset: Atomic percentages of each element (top right) and a scanning electron microscopy image of the freshly cleaved crystal (bottom right). The averaged elemental composition indicates a slightly Fe-rich stoichiometry of Fe$_{3.03(3)}$Ge$_{0.89(4)}$Te$_{2.00(1)}$.}
	\label{FigS11}
\end{figure}

Energy-dispersive X-ray spectroscopy (EDX) measurements were carried out on the freshly cleaved bulk single-crystal sample of \FthGT. The corresponding EDX spectra, atomic percentages, and the scanning electron microscopy image are shown in Fig. \ref{FigS11}. Measurements performed at multiple (more than ten) cleaved positions across the crystal indicate good compositional homogeneity with small error bars. The analysis reveals a slightly Fe-rich and near-stoichiometric composition, with average elemental contents of Fe = 3.03 $\pm$ 0.03, Ge = 0.89 $\pm$ 0.04, and Te = 2.00 $\pm$ 0.01, yielding an average stoichiometry of Fe$_{3.03(3)}$Ge$_{0.89(4)}$Te$_{2.00(1)}$.

\clearpage

\section{Frequency dependence of the resonance fields at 300 K}
\label{B}
\begin{figure}[ht]
	\centering
	\includegraphics[width=0.55\linewidth]{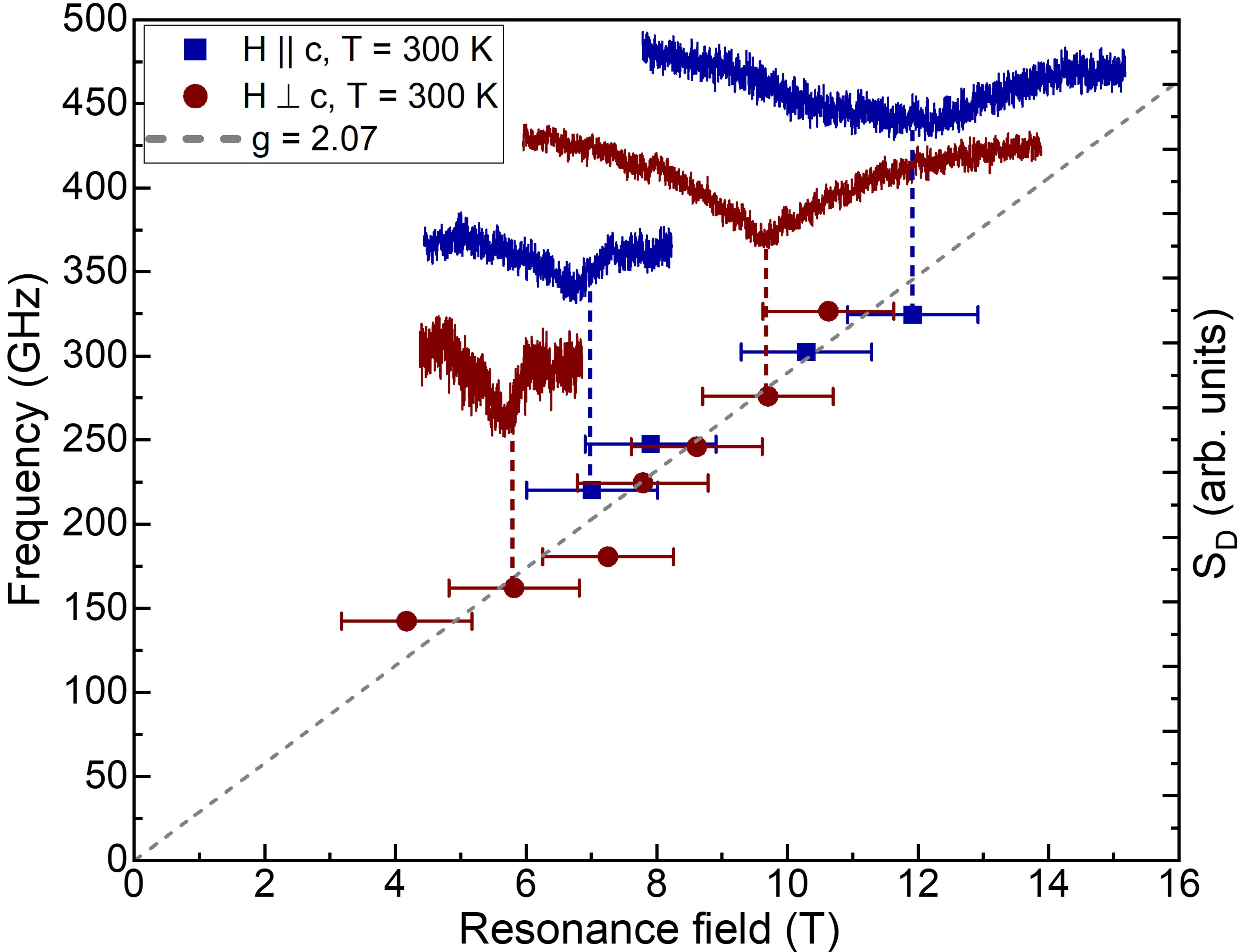}
	\caption{Frequency dependence of the resonance fields \Hr at $T = $ 300 K measured in \HIIc (blue squares) and in \HIIab (red circles) configurations. Dashed lines represent the paramagnetic response estimated with Eq.~1 (main text) with the average $g$-factors given in the legends. Representative normalized and vertically shifted spectra are shown for some of the frequencies (right vertical axis) with respective colors following respective magnetic field orientations.}
	\label{FigS1}
\end{figure}

\section{In-plane anisotropy from HF-ESR Measurements}
\label{C}

\begin{figure}[ht]
	\centering
	\includegraphics[width=\linewidth]{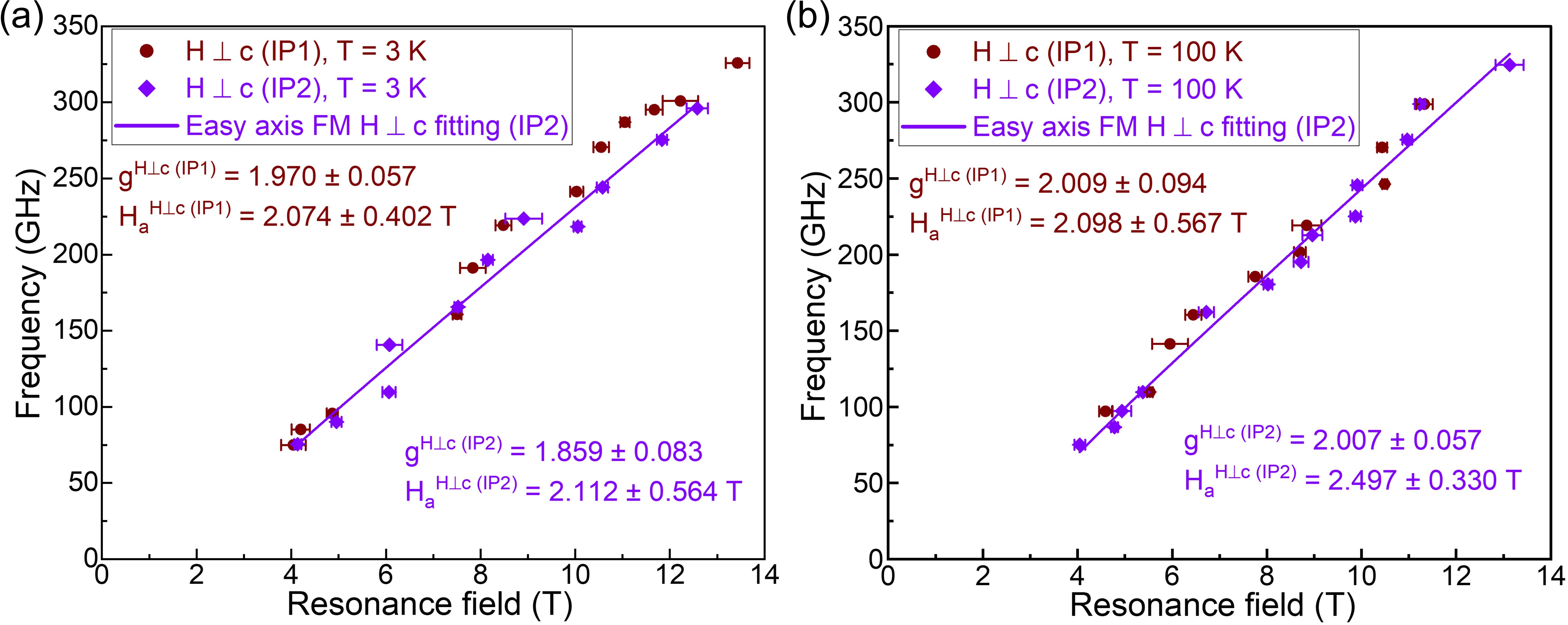}
	\caption{Frequency dependence of the resonance fields \Hr measured at (a) 3 K and (b) 100 K along two mutually orthogonal in-plane directions, labeled IP1 (\HIIab) and IP2 (newly measured, perpendicular to the IP1 in-plane). The violet line represents the fit to the IP2 data. The fitting parameters are displayed in the respective insets.}
	\label{FigS2}
\end{figure}

To examine the anisotropy within the \(ab\)-plane, the \FthGT crystal was rotated by 90° from its initial in-plane orientation (IP1). In this alternative in-plane setup (IP2), the frequency dependence of the resonance fields ($\nu$ vs $H$) was recorded at temperatures of 3 K and 100 K. A comparison between the results from configurations IP1 and IP2 is presented in Fig.~\ref{FigS2}. The data demonstrate that, within the margin of error, both orientations produce comparable outcomes at both temperatures, indicating the absence of any measurable in-plane anisotropy in this \FthGT sample, which is consistent with earlier findings from magnetization studies \cite{Wang2017}.

\section{Temperature dependence of the resonance field}\label{E}
\begin{figure}[ht]
	\centering
	\includegraphics[width=0.5\linewidth]{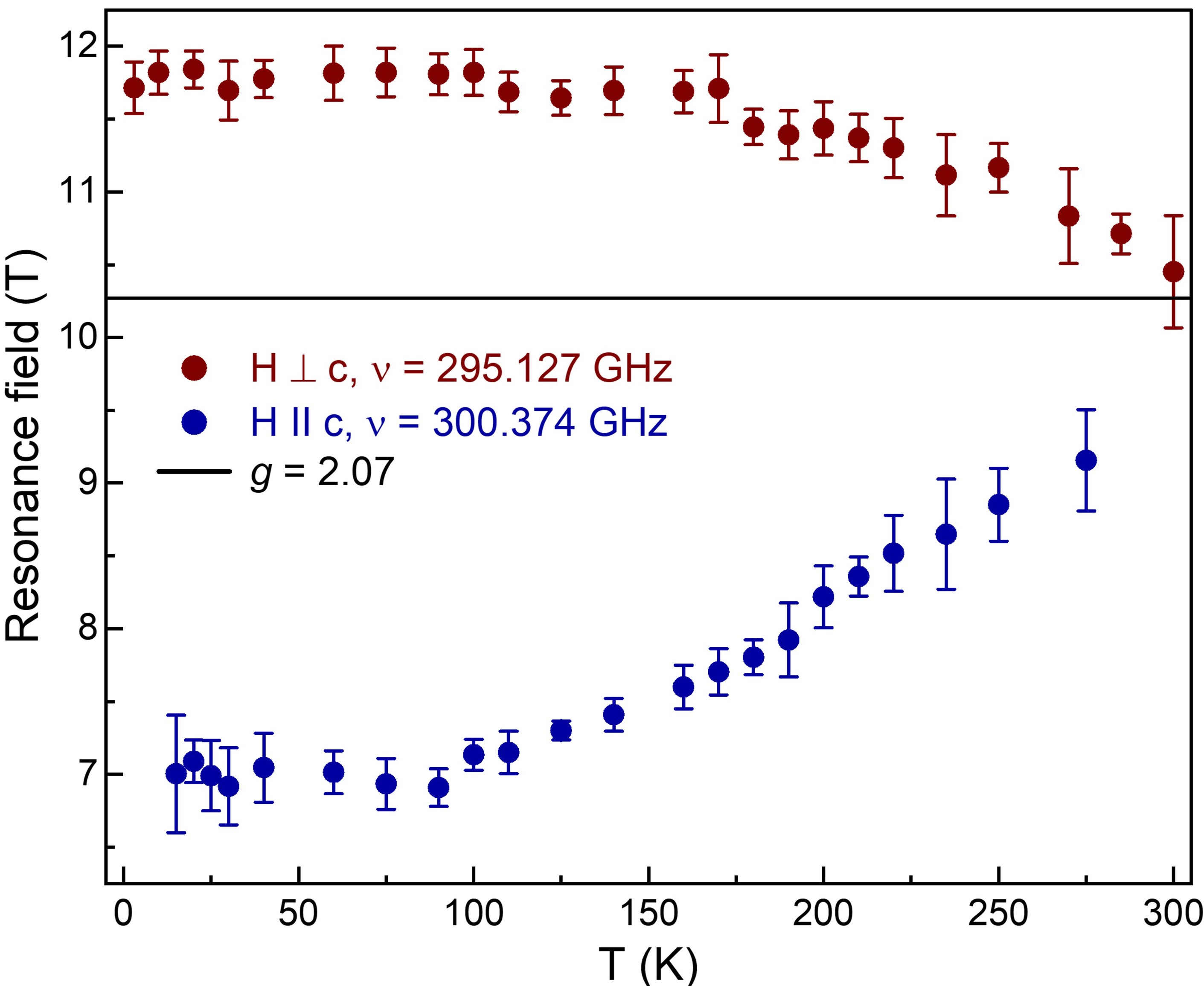}
	\caption{Temperature dependence of the resonance field, $H_{res}$, at a microwave frequency $\nu$ of $\approx$ 300 GHz with an external magnetic field applied parallel (blue circles) and perpendicular (red circles) to the c axis. The black horizontal line indicates the expected paramagnetic line with the $g$-factor, $g$ = 2.07.}
	\label{FigS5}
\end{figure}

\section{Magnetic field dependence of the magnetization}\label{D}
\begin{figure}[ht]
	\centering
	\includegraphics[width=\textwidth]{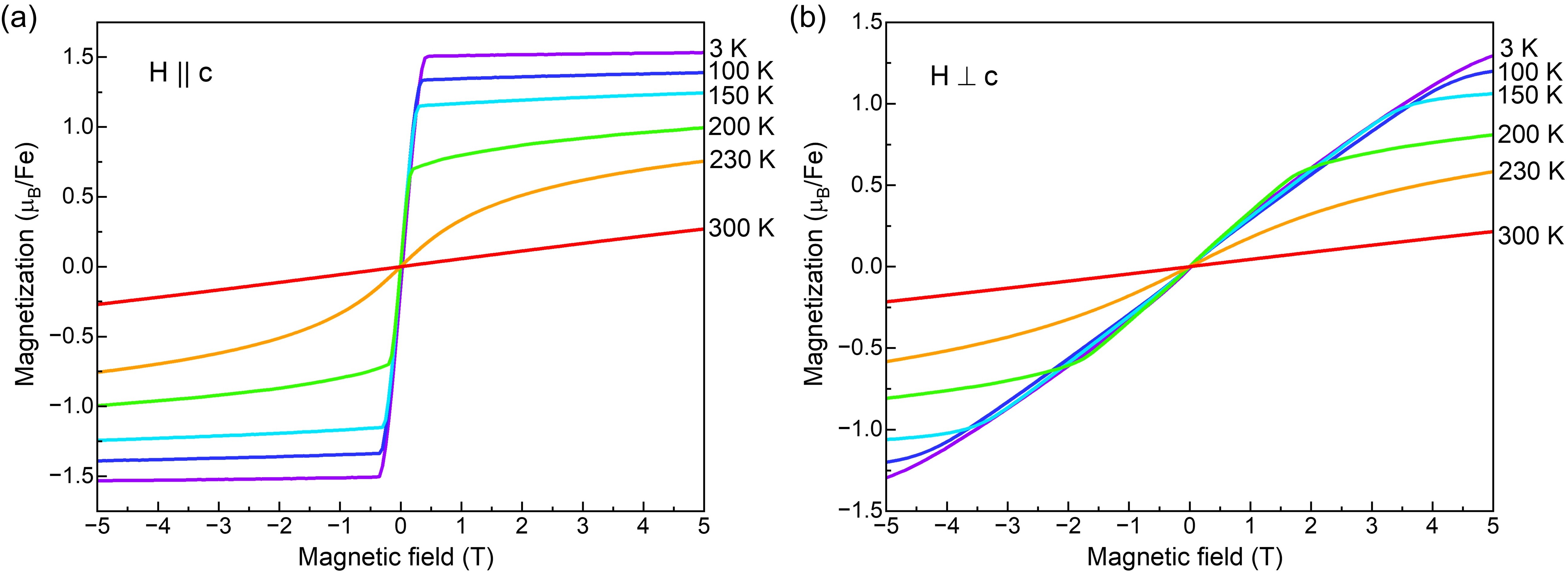}
	\caption{Magnetic field dependence of the magnetization measured for (a) \HIIc and (b) \HIIab configurations at different temperatures, showing that the magnetic easy axis is the c-axis of the \FthGT crystal.}
	\label{FigS4}
\end{figure}

\section{Magnetic-field-induced increase of the ordering temperature}\label{J}

\begin{figure}[ht]
	\centering
	\includegraphics[width=\linewidth]{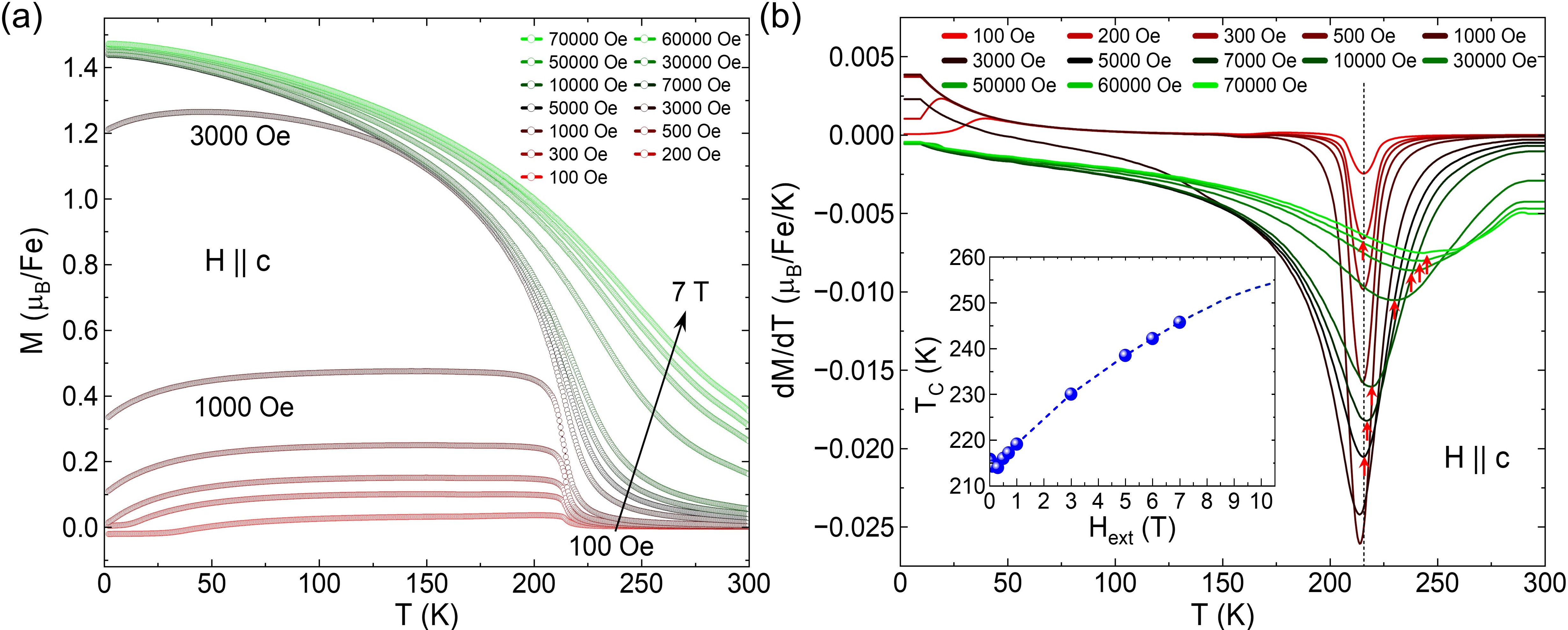}
	\caption{(a) Temperature-dependent magnetization ($M$–$T$) measurements under various external magnetic fields ($H$) applied along the $c$ axis of \FthGT. (b) Temperature dependence of the temperature derivative of magnetization, $dM/dT$. The minimum in $dM/dT$ gives the \TC of the crystal, indicated by the red arrow. Inset: External magnetic field dependence of \TC. The dashed line is the guide for the eyes.}
	\label{FigS10}
\end{figure}

To investigate the magnetic-field dependence of Curie temperature, \TC, we performed temperature-dependent magnetization ($M$–$T$) measurements under various external magnetic fields ($H$) applied along the magnetic easy axis ($c$-axis) of the \FthGT\ crystal. The \TC is determined from the minimum in the temperature derivative of the magnetization, $dM/dT$, as shown in Fig.~\ref{FigS10} (b) and Fig. \ref{Fig1}(b) \cite{Liu2017c, Pal2024}. Here, for magnetic fields above 0.5 T, \TC increases monotonically with increasing $H$ and reaches $\sim245$~K at 7~T, as shown in Fig.~\ref{FigS10} (b) (inset). This field-induced enhancement of \TC is expected for ferromagnets, as an external magnetic field favors parallel spin alignment, thereby stabilizing the ordered state against thermal fluctuations \cite{Liu2017c}.

\section{Calculation of total magnetic anisotropy energy (MAE) from static magnetization data} \label{F}

\begin{figure}[ht]
	\centering
	\includegraphics[width=0.55\linewidth]{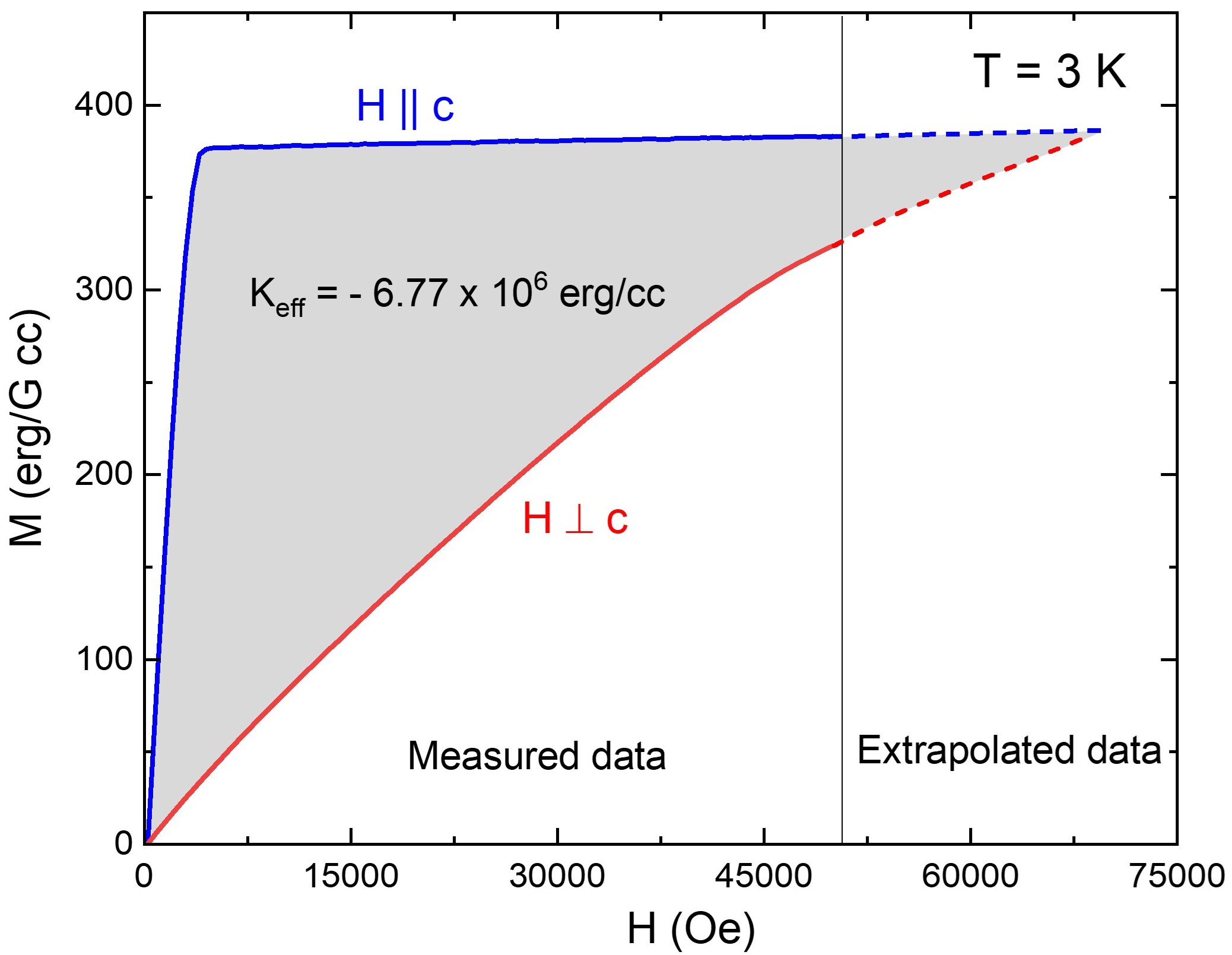}
	\caption{Magnetization isotherms \( M(H) \) measured at 3 K for \HIIc and \HIIab orientations. The gray shaded area indicates the area difference between the two curves, which is the qualitative measure of effective magnetic anisotropy energy (MAE) at 3 K.}
	\label{FigS4a}
\end{figure}

\begin{figure}[ht]
	\centering
	\includegraphics[width=0.5\linewidth]{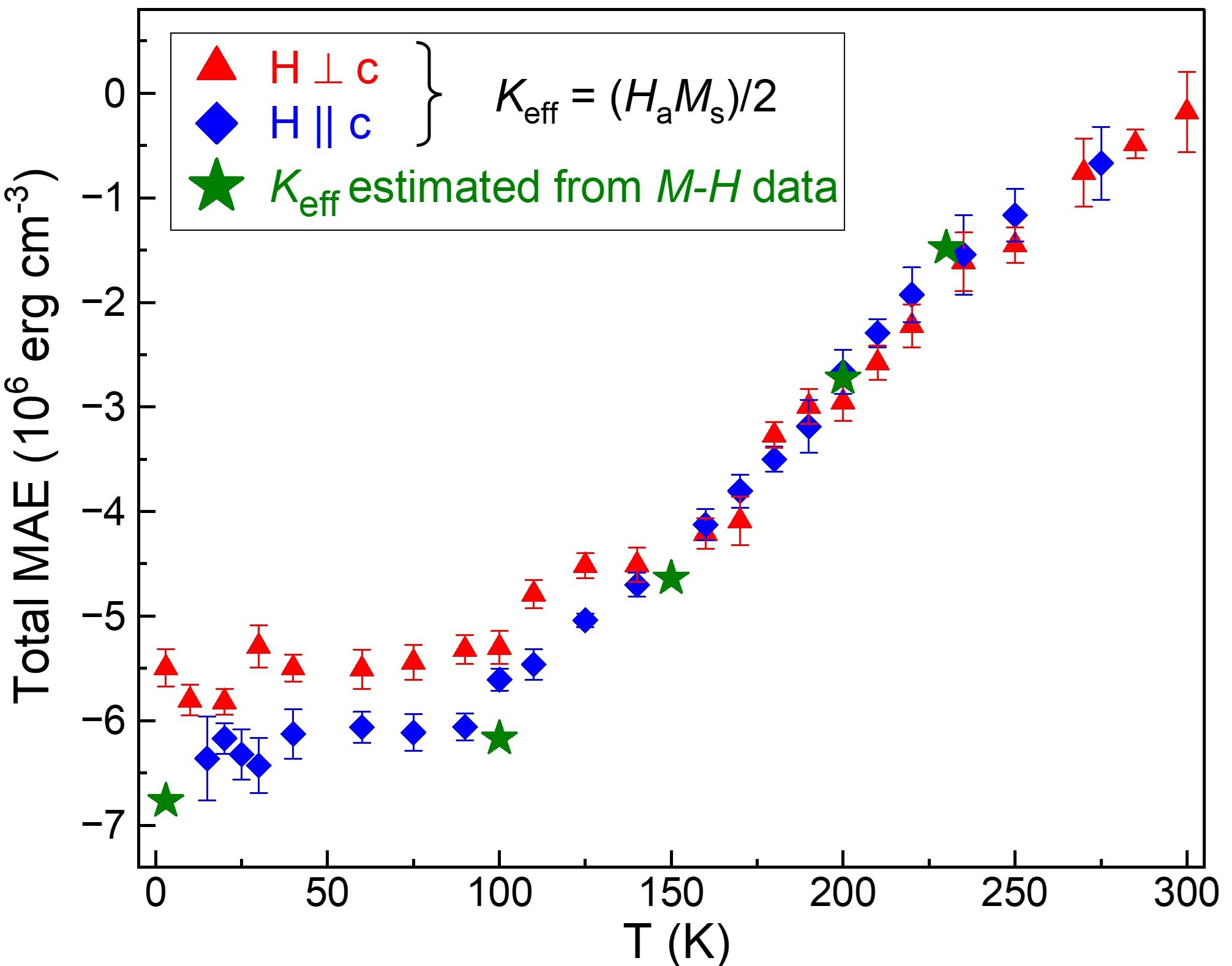}
	\caption{Temperature dependence of total MAE ($K_{\text{eff}}$) is estimated from the HF-ESR data, with an external magnetic field applied parallel (blue diamonds) and perpendicular (red triangles) to the c axis. The total MAE ($K_{\text{eff}}$), estimated from the $M-H$ data followed by Eq. 3 (main text), is plotted by green stars, well matching the results obtained from HF-ESR.}
	\label{FigS6a}
\end{figure}

To qualitatively determine the temperature dependence of the total anisotropy field, another method, known as the \textit{area method}, was employed \cite{Johnson1996, Cullity, Williams1937, Morrish}. This method provides a reasonable estimate of $H_{\mathrm{a}}$ for single-sublattice, uniaxial ferromagnets \cite{Johnson1996, Cullity}. For this, magnetization isotherms \( M(H) \) were recorded at various temperatures for both \HIIc and \HIIab orientations (see Fig. \ref{FigS4}). From these static measurements, the effective magnetic anisotropy energy (MAE) (\( K_{\text{eff}} \)) was qualitatively estimated by integrating the magnetization--field (\( M\text{--}H \)) curves following Refs.~\cite{Johnson1996, Huang2025, Cullity, Williams1937, Morrish, Tran2022}: 
\begin{equation}
K_{\text{eff}} = \mu_0 \int_0^{M_\text{s}} [H_\text{c}(M) - H_{\text{ab}}(M)]\, \text{d}M ,
\label{eq:Keff}
\end{equation}
where \( M_\text{s} \) is the saturation magnetization, and \( H_{\text{ab}} \) and \( H_\text{c} \) denote the in-plane (\HIIab) and out-of-plane (\HIIc) magnetic fields, respectively. The magnetic energy required to change the sample’s magnetization by an infinitesimal amount \(\text{d}M\) under an external magnetic field \(H\), can be represented by \(\mu_0 H\,\text{d}M\) \cite{Johnson1996, Cullity}. Consequently, the integral in Eq.~\ref{eq:Keff} represents the area difference between the two magnetization curves and provides a qualitative measure of the effective MAE (see Fig. \ref{FigS4a}), which includes contributions from both magnetocrystalline and shape anisotropies for easy-axis ferromagnets \cite{Johnson1996}. The temperature-dependent values of \( K_{\text{eff}} \) obtained from this analysis are presented and compared with those obtained from HF-ESR data, as plotted in Fig. \ref{FigS6a}. The corresponding total anisotropy field, calculated using \( H_a^{\mathrm{mag}} = 2K_{\text{eff}}/M_\text{s} \), is plotted in Fig. 4(a) (main text) and exhibits good agreement with those derived from the HF-ESR measurements.\\

\section{Temperature dependence of total magnetic anisotropy energy (MAE)}\label{G}
\clearpage

\section{Temperature dependence of the saturation magnetization}\label{H}
\begin{figure}[ht]
	\centering
	\includegraphics[width=0.55\linewidth]{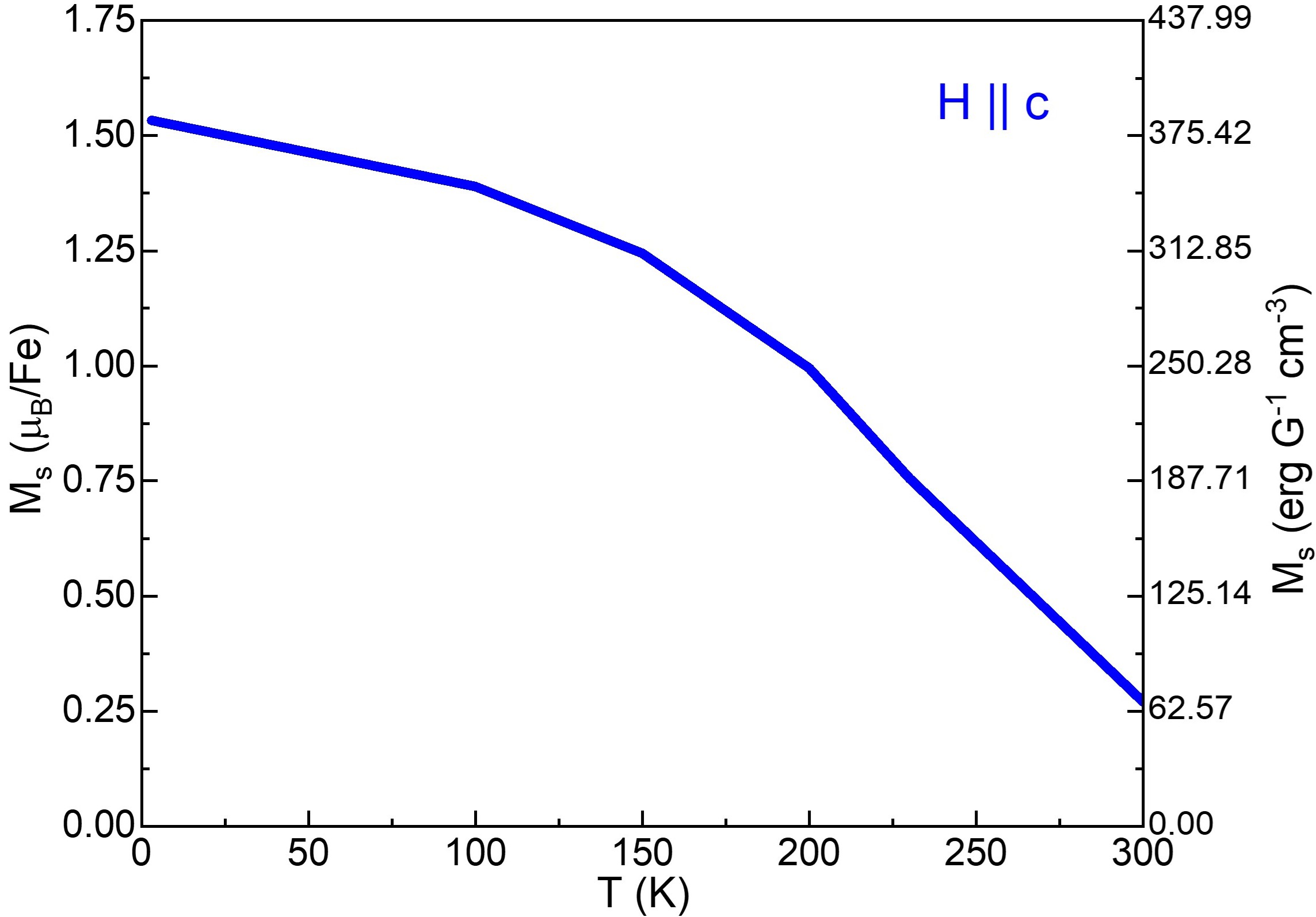}
	\caption{Temperature dependence of the saturation magnetization $M_\mathrm{s}$ for \HIIc configuration.}
	\label{FigS3}
\end{figure}

\section{Temperature dependence of the demagnetization field and estimated intrinsic magnetic anisotropy field} \label{I}
\begin{figure}[ht]
	\centering
	\includegraphics[width=\textwidth]{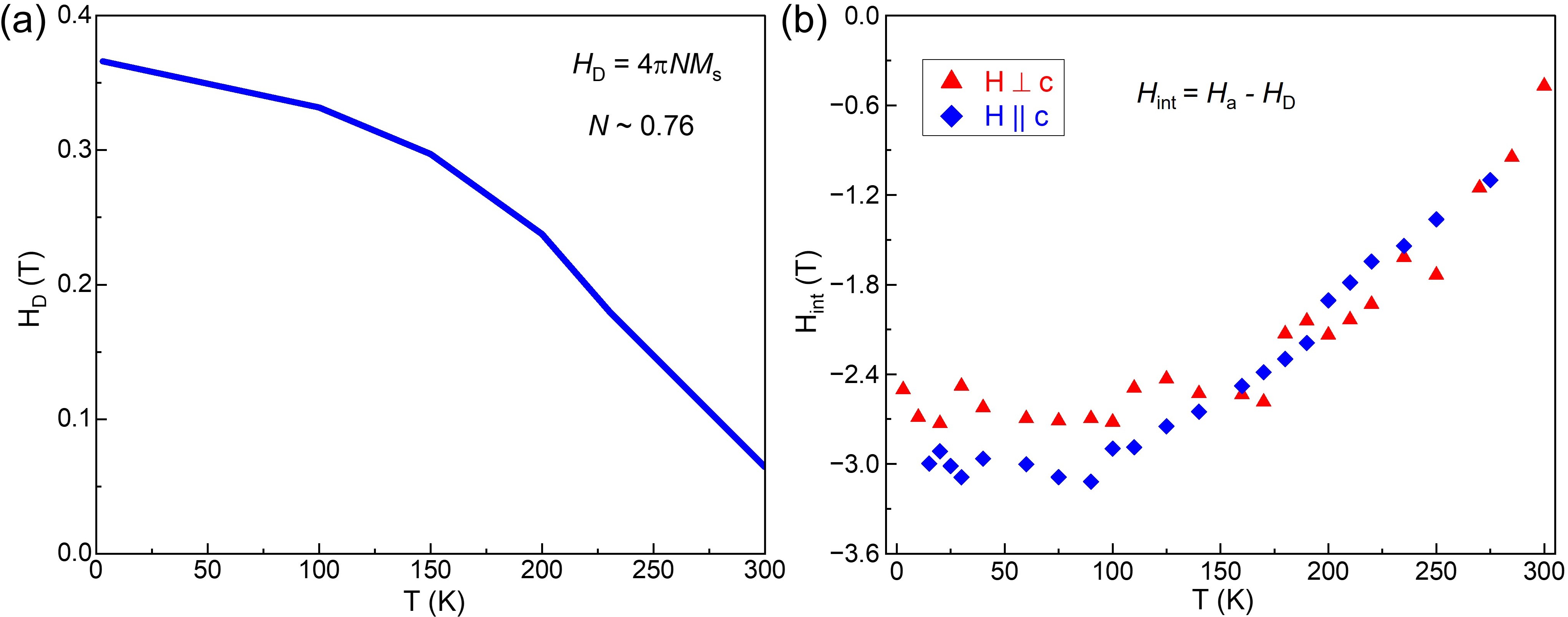}
	\caption{(a) Temperature dependence of the demagnetization field $H_{D}$, estimated from the temperature-dependent saturation magnetization data and the calculated demagnetization factor. (b) Temperature dependence of the intrinsic magnetic anisotropy field ($H_{\text{int}}$), estimated from total magnetic anisotropy field and demagnetization field, in the \HIIab and \HIIc configurations, respectively.}
	\label{FigS6}
\end{figure}

\vspace{-10pt}

\FloatBarrier

\twocolumngrid

\FloatBarrier


\bibliographystyle{apsrev4-2}

\begin{thebibliography}{81}%
\makeatletter
\providecommand \@ifxundefined [1]{%
 \@ifx{#1\undefined}
}%
\providecommand \@ifnum [1]{%
 \ifnum #1\expandafter \@firstoftwo
 \else \expandafter \@secondoftwo
 \fi
}%
\providecommand \@ifx [1]{%
 \ifx #1\expandafter \@firstoftwo
 \else \expandafter \@secondoftwo
 \fi
}%
\providecommand \natexlab [1]{#1}%
\providecommand \enquote  [1]{``#1''}%
\providecommand \bibnamefont  [1]{#1}%
\providecommand \bibfnamefont [1]{#1}%
\providecommand \citenamefont [1]{#1}%
\providecommand \href@noop [0]{\@secondoftwo}%
\providecommand \href [0]{\begingroup \@sanitize@url \@href}%
\providecommand \@href[1]{\@@startlink{#1}\@@href}%
\providecommand \@@href[1]{\endgroup#1\@@endlink}%
\providecommand \@sanitize@url [0]{\catcode `\\12\catcode `\$12\catcode `\&12\catcode `\#12\catcode `\^12\catcode `\_12\catcode `\%12\relax}%
\providecommand \@@startlink[1]{}%
\providecommand \@@endlink[0]{}%
\providecommand \url  [0]{\begingroup\@sanitize@url \@url }%
\providecommand \@url [1]{\endgroup\@href {#1}{\urlprefix }}%
\providecommand \urlprefix  [0]{URL }%
\providecommand \Eprint [0]{\href }%
\providecommand \doibase [0]{https://doi.org/}%
\providecommand \selectlanguage [0]{\@gobble}%
\providecommand \bibinfo  [0]{\@secondoftwo}%
\providecommand \bibfield  [0]{\@secondoftwo}%
\providecommand \translation [1]{[#1]}%
\providecommand \BibitemOpen [0]{}%
\providecommand \bibitemStop [0]{}%
\providecommand \bibitemNoStop [0]{.\EOS\space}%
\providecommand \EOS [0]{\spacefactor3000\relax}%
\providecommand \BibitemShut  [1]{\csname bibitem#1\endcsname}%
\let\auto@bib@innerbib\@empty
\bibitem [{\citenamefont {Novoselov}\ \emph {et~al.}(2016)\citenamefont {Novoselov}, \citenamefont {Mishchenko}, \citenamefont {Carvalho},\ and\ \citenamefont {Castro~Neto}}]{Novoselov2016}%
  \BibitemOpen
  \bibfield  {author} {\bibinfo {author} {\bibfnamefont {K.~S.}\ \bibnamefont {Novoselov}}, \bibinfo {author} {\bibfnamefont {A.}~\bibnamefont {Mishchenko}}, \bibinfo {author} {\bibfnamefont {A.}~\bibnamefont {Carvalho}},\ and\ \bibinfo {author} {\bibfnamefont {A.~H.}\ \bibnamefont {Castro~Neto}},\ }\bibfield  {title} {\bibinfo {title} {{2D materials and van der Waals heterostructures}},\ }\href {https://www.science.org/doi/10.1126/science.aac9439} {\bibfield  {journal} {\bibinfo  {journal} {Science}\ }\textbf {\bibinfo {volume} {353}},\ \bibinfo {pages} {9439} (\bibinfo {year} {2016})}\BibitemShut {NoStop}%
\bibitem [{\citenamefont {Ahn}(2020)}]{Ahn2020a}%
  \BibitemOpen
  \bibfield  {author} {\bibinfo {author} {\bibfnamefont {E.~C.}\ \bibnamefont {Ahn}},\ }\bibfield  {title} {\bibinfo {title} {{2D materials for spintronic devices}},\ }\href {http://dx.doi.org/10.1038/s41699-020-0152-0} {\bibfield  {journal} {\bibinfo  {journal} {npj 2D Materials and Applications}\ }\textbf {\bibinfo {volume} {4}} (\bibinfo {year} {2020})}\BibitemShut {NoStop}%
\bibitem [{\citenamefont {Gong}\ and\ \citenamefont {Zhang}(2019)}]{Gong2019}%
  \BibitemOpen
  \bibfield  {author} {\bibinfo {author} {\bibfnamefont {C.}~\bibnamefont {Gong}}\ and\ \bibinfo {author} {\bibfnamefont {X.}~\bibnamefont {Zhang}},\ }\bibfield  {title} {\bibinfo {title} {{Two-dimensional magnetic crystals and emergent heterostructure devices}},\ }\href {http://dx.doi.org/10.1126/science.aav4450} {\bibfield  {journal} {\bibinfo  {journal} {Science}\ }\textbf {\bibinfo {volume} {363}} (\bibinfo {year} {2019})}\BibitemShut {NoStop}%
\bibitem [{\citenamefont {Hao}\ \emph {et~al.}(2022)\citenamefont {Hao}, \citenamefont {Dai}, \citenamefont {Cai}, \citenamefont {Chen}, \citenamefont {Xing}, \citenamefont {Chen}, \citenamefont {Zhai}, \citenamefont {Wang},\ and\ \citenamefont {Han}}]{Hao2022}%
  \BibitemOpen
  \bibfield  {author} {\bibinfo {author} {\bibfnamefont {Q.}~\bibnamefont {Hao}}, \bibinfo {author} {\bibfnamefont {H.}~\bibnamefont {Dai}}, \bibinfo {author} {\bibfnamefont {M.}~\bibnamefont {Cai}}, \bibinfo {author} {\bibfnamefont {X.}~\bibnamefont {Chen}}, \bibinfo {author} {\bibfnamefont {Y.}~\bibnamefont {Xing}}, \bibinfo {author} {\bibfnamefont {H.}~\bibnamefont {Chen}}, \bibinfo {author} {\bibfnamefont {T.}~\bibnamefont {Zhai}}, \bibinfo {author} {\bibfnamefont {X.}~\bibnamefont {Wang}},\ and\ \bibinfo {author} {\bibfnamefont {J.}~\bibnamefont {Han}},\ }\bibfield  {title} {\bibinfo {title} {{2D Magnetic Heterostructures and Emergent Spintronic Devices}},\ }\href {https://onlinelibrary.wiley.com/doi/10.1002/aelm.202200164} {\bibfield  {journal} {\bibinfo  {journal} {Advanced Electronic Materials}\ }\textbf {\bibinfo {volume} {8}},\ \bibinfo {pages} {1} (\bibinfo {year} {2022})}\BibitemShut {NoStop}%
\bibitem [{\citenamefont {Burch}\ \emph {et~al.}(2018)\citenamefont {Burch}, \citenamefont {Mandrus},\ and\ \citenamefont {Park}}]{burch2018}%
  \BibitemOpen
  \bibfield  {author} {\bibinfo {author} {\bibfnamefont {K.~S.}\ \bibnamefont {Burch}}, \bibinfo {author} {\bibfnamefont {D.}~\bibnamefont {Mandrus}},\ and\ \bibinfo {author} {\bibfnamefont {J.-G.}\ \bibnamefont {Park}},\ }\bibfield  {title} {\bibinfo {title} {{Magnetism in two-dimensional van der Waals materials}},\ }\href {https://doi.org/10.1038/s41586-018-0631-z} {\bibfield  {journal} {\bibinfo  {journal} {Nature}\ }\textbf {\bibinfo {volume} {563}},\ \bibinfo {pages} {47} (\bibinfo {year} {2018})}\BibitemShut {NoStop}%
\bibitem [{\citenamefont {Mermin}\ and\ \citenamefont {Wagner}(1966)}]{MerminWagner}%
  \BibitemOpen
  \bibfield  {author} {\bibinfo {author} {\bibfnamefont {N.~D.}\ \bibnamefont {Mermin}}\ and\ \bibinfo {author} {\bibfnamefont {H.}~\bibnamefont {Wagner}},\ }\bibfield  {title} {\bibinfo {title} {Absence of ferromagnetism or antiferromagnetism in one- or two-dimensional isotropic heisenberg models},\ }\href {https://doi.org/10.1103/PhysRevLett.17.1133} {\bibfield  {journal} {\bibinfo  {journal} {Physical Review Letters}\ }\textbf {\bibinfo {volume} {17}},\ \bibinfo {pages} {1133} (\bibinfo {year} {1966})}\BibitemShut {NoStop}%
\bibitem [{\citenamefont {Morrish}(2001)}]{Morrish}%
  \BibitemOpen
  \bibfield  {author} {\bibinfo {author} {\bibfnamefont {A.~H.}\ \bibnamefont {Morrish}},\ }\href@noop {} {\emph {\bibinfo {title} {The physical principles of magnetism}}}\ (\bibinfo  {publisher} {Wiley-IEEE Press},\ \bibinfo {year} {2001})\BibitemShut {NoStop}%
\bibitem [{\citenamefont {Coey}(2010)}]{Coey}%
  \BibitemOpen
  \bibfield  {author} {\bibinfo {author} {\bibfnamefont {J.~M.}\ \bibnamefont {Coey}},\ }\href@noop {} {\emph {\bibinfo {title} {Magnetism and Magnetic Materials}}}\ (\bibinfo  {publisher} {Cambridge University Press},\ \bibinfo {year} {2010})\BibitemShut {NoStop}%
\bibitem [{\citenamefont {Kataev}\ \emph {et~al.}(2024)\citenamefont {Kataev}, \citenamefont {B{\"{u}}chner},\ and\ \citenamefont {Alfonsov}}]{Kataev2024}%
  \BibitemOpen
  \bibfield  {author} {\bibinfo {author} {\bibfnamefont {V.}~\bibnamefont {Kataev}}, \bibinfo {author} {\bibfnamefont {B.}~\bibnamefont {B{\"{u}}chner}},\ and\ \bibinfo {author} {\bibfnamefont {A.}~\bibnamefont {Alfonsov}},\ }\bibfield  {title} {\bibinfo {title} {{Electron Spin Resonance Spectroscopy on Magnetic van der Waals Compounds}},\ }\href {https://link.springer.com/article/10.1007/s00723-024-01671-x} {\bibfield  {journal} {\bibinfo  {journal} {Applied Magnetic Resonance}\ }\textbf {\bibinfo {volume} {55}},\ \bibinfo {pages} {923} (\bibinfo {year} {2024})}\BibitemShut {NoStop}%
\bibitem [{\citenamefont {Cho}\ \emph {et~al.}(2023)\citenamefont {Cho}, \citenamefont {Pawbake}, \citenamefont {Aubergier}, \citenamefont {Barra}, \citenamefont {Mosina}, \citenamefont {Sofer}, \citenamefont {Zhitomirsky}, \citenamefont {Faugeras},\ and\ \citenamefont {Piot}}]{Cho2023fmr}%
  \BibitemOpen
  \bibfield  {author} {\bibinfo {author} {\bibfnamefont {C.}~\bibnamefont {Cho}}, \bibinfo {author} {\bibfnamefont {A.}~\bibnamefont {Pawbake}}, \bibinfo {author} {\bibfnamefont {N.}~\bibnamefont {Aubergier}}, \bibinfo {author} {\bibfnamefont {A.}~\bibnamefont {Barra}}, \bibinfo {author} {\bibfnamefont {K.}~\bibnamefont {Mosina}}, \bibinfo {author} {\bibfnamefont {Z.}~\bibnamefont {Sofer}}, \bibinfo {author} {\bibfnamefont {M.}~\bibnamefont {Zhitomirsky}}, \bibinfo {author} {\bibfnamefont {C.}~\bibnamefont {Faugeras}},\ and\ \bibinfo {author} {\bibfnamefont {B.}~\bibnamefont {Piot}},\ }\bibfield  {title} {\bibinfo {title} {{Microscopic parameters of the van der Waals CrSBr antiferromagnet from microwave absorption experiments}},\ }\href {https://doi.org/10.1103/PhysRevB.107.094403} {\bibfield  {journal} {\bibinfo  {journal} {Physical Review B}\ }\textbf {\bibinfo {volume} {107}},\ \bibinfo {pages} {094403} (\bibinfo {year} {2023})}\BibitemShut {NoStop}%
\bibitem [{\citenamefont {Pal}\ \emph {et~al.}(2024{\natexlab{a}})\citenamefont {Pal}, \citenamefont {Abraham}, \citenamefont {Mistonov}, \citenamefont {Mishra}, \citenamefont {Stilkerich}, \citenamefont {Mondal}, \citenamefont {Mandal}, \citenamefont {Pal}, \citenamefont {Geck}, \citenamefont {B{\"{u}}chner}, \citenamefont {Kataev},\ and\ \citenamefont {Alfonsov}}]{Pal2024_ESR}%
  \BibitemOpen
  \bibfield  {author} {\bibinfo {author} {\bibfnamefont {R.}~\bibnamefont {Pal}}, \bibinfo {author} {\bibfnamefont {J.~J.}\ \bibnamefont {Abraham}}, \bibinfo {author} {\bibfnamefont {A.}~\bibnamefont {Mistonov}}, \bibinfo {author} {\bibfnamefont {S.}~\bibnamefont {Mishra}}, \bibinfo {author} {\bibfnamefont {N.}~\bibnamefont {Stilkerich}}, \bibinfo {author} {\bibfnamefont {S.}~\bibnamefont {Mondal}}, \bibinfo {author} {\bibfnamefont {P.}~\bibnamefont {Mandal}}, \bibinfo {author} {\bibfnamefont {A.~N.}\ \bibnamefont {Pal}}, \bibinfo {author} {\bibfnamefont {J.}~\bibnamefont {Geck}}, \bibinfo {author} {\bibfnamefont {B.}~\bibnamefont {B{\"{u}}chner}}, \bibinfo {author} {\bibfnamefont {V.}~\bibnamefont {Kataev}},\ and\ \bibinfo {author} {\bibfnamefont {A.}~\bibnamefont {Alfonsov}},\ }\bibfield  {title} {\bibinfo {title} {{Disentangling the Unusual Magnetic Anisotropy of the Near-Room-Temperature Ferromagnet Fe$_4$GeTe$_2$}},\ }\href {https://onlinelibrary.wiley.com/doi/10.1002/adfm.202402551} {\bibfield
  {journal} {\bibinfo  {journal} {Advanced Functional Materials}\ }\textbf {\bibinfo {volume} {34}},\ \bibinfo {pages} {2402551} (\bibinfo {year} {2024}{\natexlab{a}})}\BibitemShut {NoStop}%
\bibitem [{\citenamefont {Moro}\ \emph {et~al.}(2022)\citenamefont {Moro}, \citenamefont {Ke}, \citenamefont {del {\'A}guila}, \citenamefont {S{\"o}ll}, \citenamefont {Sofer}, \citenamefont {Wu}, \citenamefont {Yue}, \citenamefont {Li}, \citenamefont {Liu},\ and\ \citenamefont {Fanciulli}}]{Moro2022}%
  \BibitemOpen
  \bibfield  {author} {\bibinfo {author} {\bibfnamefont {F.}~\bibnamefont {Moro}}, \bibinfo {author} {\bibfnamefont {S.}~\bibnamefont {Ke}}, \bibinfo {author} {\bibfnamefont {A.~G.}\ \bibnamefont {del {\'A}guila}}, \bibinfo {author} {\bibfnamefont {A.}~\bibnamefont {S{\"o}ll}}, \bibinfo {author} {\bibfnamefont {Z.}~\bibnamefont {Sofer}}, \bibinfo {author} {\bibfnamefont {Q.}~\bibnamefont {Wu}}, \bibinfo {author} {\bibfnamefont {M.}~\bibnamefont {Yue}}, \bibinfo {author} {\bibfnamefont {L.}~\bibnamefont {Li}}, \bibinfo {author} {\bibfnamefont {X.}~\bibnamefont {Liu}},\ and\ \bibinfo {author} {\bibfnamefont {M.}~\bibnamefont {Fanciulli}},\ }\bibfield  {title} {\bibinfo {title} {{Revealing 2D magnetism in a bulk CrSBr single crystal by electron spin resonance}},\ }\href {https://doi.org/10.1002/adfm.202207044} {\bibfield  {journal} {\bibinfo  {journal} {Advanced Functional Materials}\ }\textbf {\bibinfo {volume} {32}},\ \bibinfo {pages} {2207044} (\bibinfo {year} {2022})}\BibitemShut {NoStop}%
\bibitem [{\citenamefont {Lee}\ \emph {et~al.}(2020)\citenamefont {Lee}, \citenamefont {Utermohlen}, \citenamefont {Weber}, \citenamefont {Hwang}, \citenamefont {Zhang}, \citenamefont {Van~Tol}, \citenamefont {Goldberger}, \citenamefont {Trivedi},\ and\ \citenamefont {Hammel}}]{Lee2020fmr}%
  \BibitemOpen
  \bibfield  {author} {\bibinfo {author} {\bibfnamefont {I.}~\bibnamefont {Lee}}, \bibinfo {author} {\bibfnamefont {F.~G.}\ \bibnamefont {Utermohlen}}, \bibinfo {author} {\bibfnamefont {D.}~\bibnamefont {Weber}}, \bibinfo {author} {\bibfnamefont {K.}~\bibnamefont {Hwang}}, \bibinfo {author} {\bibfnamefont {C.}~\bibnamefont {Zhang}}, \bibinfo {author} {\bibfnamefont {J.}~\bibnamefont {Van~Tol}}, \bibinfo {author} {\bibfnamefont {J.~E.}\ \bibnamefont {Goldberger}}, \bibinfo {author} {\bibfnamefont {N.}~\bibnamefont {Trivedi}},\ and\ \bibinfo {author} {\bibfnamefont {P.~C.}\ \bibnamefont {Hammel}},\ }\bibfield  {title} {\bibinfo {title} {{Fundamental spin interactions underlying the magnetic anisotropy in the Kitaev ferromagnet CrI$_3$}},\ }\href {https://doi.org/10.1103/PhysRevLett.124.017201} {\bibfield  {journal} {\bibinfo  {journal} {Physical Review Letters}\ }\textbf {\bibinfo {volume} {124}},\ \bibinfo {pages} {017201} (\bibinfo {year} {2020})}\BibitemShut {NoStop}%
\bibitem [{\citenamefont {Pal}\ \emph {et~al.}(2025{\natexlab{a}})\citenamefont {Pal}, \citenamefont {Deb}, \citenamefont {Kumar}, \citenamefont {B{\"u}chner}, \citenamefont {Alfonsov},\ and\ \citenamefont {Kataev}}]{Pal_MMS_2025}%
  \BibitemOpen
  \bibfield  {author} {\bibinfo {author} {\bibfnamefont {R.}~\bibnamefont {Pal}}, \bibinfo {author} {\bibfnamefont {K.}~\bibnamefont {Deb}}, \bibinfo {author} {\bibfnamefont {N.}~\bibnamefont {Kumar}}, \bibinfo {author} {\bibfnamefont {B.}~\bibnamefont {B{\"u}chner}}, \bibinfo {author} {\bibfnamefont {A.}~\bibnamefont {Alfonsov}},\ and\ \bibinfo {author} {\bibfnamefont {V.}~\bibnamefont {Kataev}},\ }\bibfield  {title} {\bibinfo {title} {{Ferromagnetic Resonance Spectroscopy on the Kagome Magnet MgMn$_6$Sn$_6$}},\ }\href {https://doi.org/10.1007/s00723-025-01806-8} {\bibfield  {journal} {\bibinfo  {journal} {Applied Magnetic Resonance}\ }\textbf {\bibinfo {volume} {56}},\ \bibinfo {pages} {1507} (\bibinfo {year} {2025}{\natexlab{a}})}\BibitemShut {NoStop}%
\bibitem [{\citenamefont {Wang}\ \emph {et~al.}(2023{\natexlab{a}})\citenamefont {Wang}, \citenamefont {Shang}, \citenamefont {Zhang}, \citenamefont {Kang}, \citenamefont {Liu}, \citenamefont {Wang}, \citenamefont {Chen}, \citenamefont {Liu}, \citenamefont {Tang}, \citenamefont {Zeng} \emph {et~al.}}]{Wang2023fmr}%
  \BibitemOpen
  \bibfield  {author} {\bibinfo {author} {\bibfnamefont {X.}~\bibnamefont {Wang}}, \bibinfo {author} {\bibfnamefont {Z.}~\bibnamefont {Shang}}, \bibinfo {author} {\bibfnamefont {C.}~\bibnamefont {Zhang}}, \bibinfo {author} {\bibfnamefont {J.}~\bibnamefont {Kang}}, \bibinfo {author} {\bibfnamefont {T.}~\bibnamefont {Liu}}, \bibinfo {author} {\bibfnamefont {X.}~\bibnamefont {Wang}}, \bibinfo {author} {\bibfnamefont {S.}~\bibnamefont {Chen}}, \bibinfo {author} {\bibfnamefont {H.}~\bibnamefont {Liu}}, \bibinfo {author} {\bibfnamefont {W.}~\bibnamefont {Tang}}, \bibinfo {author} {\bibfnamefont {Y.-J.}\ \bibnamefont {Zeng}}, \emph {et~al.},\ }\bibfield  {title} {\bibinfo {title} {{Electrical and magnetic anisotropies in van der Waals multiferroic CuCrP$_2$S$_6$}},\ }\href {https://doi.org/10.1038/s41467-023-36512-1} {\bibfield  {journal} {\bibinfo  {journal} {Nature Communications}\ }\textbf {\bibinfo {volume} {14}},\ \bibinfo {pages} {840} (\bibinfo {year} {2023}{\natexlab{a}})}\BibitemShut {NoStop}%
\bibitem [{\citenamefont {Beier}\ \emph {et~al.}(2025)\citenamefont {Beier}, \citenamefont {Beier}, \citenamefont {Arneth}, \citenamefont {Br\"ucher}, \citenamefont {Kremer},\ and\ \citenamefont {Klingeler}}]{Beier2025}%
  \BibitemOpen
  \bibfield  {author} {\bibinfo {author} {\bibfnamefont {B.~G.}\ \bibnamefont {Beier}}, \bibinfo {author} {\bibfnamefont {E.~S.}\ \bibnamefont {Beier}}, \bibinfo {author} {\bibfnamefont {J.}~\bibnamefont {Arneth}}, \bibinfo {author} {\bibfnamefont {E.}~\bibnamefont {Br\"ucher}}, \bibinfo {author} {\bibfnamefont {R.~K.}\ \bibnamefont {Kremer}},\ and\ \bibinfo {author} {\bibfnamefont {R.}~\bibnamefont {Klingeler}},\ }\bibfield  {title} {\bibinfo {title} {{Elucidating the origin of long-range ferromagnetic order in Fe$_3$GeTe$_2$ by low-energy magnon excitation studies}},\ }\href {https://doi.org/10.1103/zf74-j18j} {\bibfield  {journal} {\bibinfo  {journal} {Physical Review B}\ }\textbf {\bibinfo {volume} {112}},\ \bibinfo {pages} {214414} (\bibinfo {year} {2025})}\BibitemShut {NoStop}%
\bibitem [{\citenamefont {Alfonsov}\ \emph {et~al.}(2021)\citenamefont {Alfonsov}, \citenamefont {Mehlawat}, \citenamefont {Zeugner}, \citenamefont {Isaeva}, \citenamefont {B{\"{u}}chner},\ and\ \citenamefont {Kataev}}]{Alfonsov2021b}%
  \BibitemOpen
  \bibfield  {author} {\bibinfo {author} {\bibfnamefont {A.}~\bibnamefont {Alfonsov}}, \bibinfo {author} {\bibfnamefont {K.}~\bibnamefont {Mehlawat}}, \bibinfo {author} {\bibfnamefont {A.}~\bibnamefont {Zeugner}}, \bibinfo {author} {\bibfnamefont {A.}~\bibnamefont {Isaeva}}, \bibinfo {author} {\bibfnamefont {B.}~\bibnamefont {B{\"{u}}chner}},\ and\ \bibinfo {author} {\bibfnamefont {V.}~\bibnamefont {Kataev}},\ }\bibfield  {title} {\bibinfo {title} {{Magnetic-field tuning of the spin dynamics in the magnetic topological insulators (MnBi$_2$Te$_4$)(Bi$_2$Te$_3$)$_n$}},\ }\href {http://dx.doi.org/10.1103/PhysRevB.104.195139} {\bibfield  {journal} {\bibinfo  {journal} {Physical Review B}\ }\textbf {\bibinfo {volume} {104}},\ \bibinfo {pages} {195139} (\bibinfo {year} {2021})}\BibitemShut {NoStop}%
\bibitem [{\citenamefont {Alahmed}\ \emph {et~al.}(2021)\citenamefont {Alahmed}, \citenamefont {Nepal}, \citenamefont {Macy}, \citenamefont {Zheng}, \citenamefont {Casas}, \citenamefont {Sapkota}, \citenamefont {Jones}, \citenamefont {Mazza}, \citenamefont {Brahlek}, \citenamefont {Jin}, \citenamefont {Mahjouri-Samani}, \citenamefont {Zhang}, \citenamefont {Mewes}, \citenamefont {Balicas}, \citenamefont {Mewes},\ and\ \citenamefont {Li}}]{Alahmed2021}%
  \BibitemOpen
  \bibfield  {author} {\bibinfo {author} {\bibfnamefont {L.}~\bibnamefont {Alahmed}}, \bibinfo {author} {\bibfnamefont {B.}~\bibnamefont {Nepal}}, \bibinfo {author} {\bibfnamefont {J.}~\bibnamefont {Macy}}, \bibinfo {author} {\bibfnamefont {W.}~\bibnamefont {Zheng}}, \bibinfo {author} {\bibfnamefont {B.}~\bibnamefont {Casas}}, \bibinfo {author} {\bibfnamefont {A.}~\bibnamefont {Sapkota}}, \bibinfo {author} {\bibfnamefont {N.}~\bibnamefont {Jones}}, \bibinfo {author} {\bibfnamefont {A.~R.}\ \bibnamefont {Mazza}}, \bibinfo {author} {\bibfnamefont {M.}~\bibnamefont {Brahlek}}, \bibinfo {author} {\bibfnamefont {W.}~\bibnamefont {Jin}}, \bibinfo {author} {\bibfnamefont {M.}~\bibnamefont {Mahjouri-Samani}}, \bibinfo {author} {\bibfnamefont {S.~S.-L.}\ \bibnamefont {Zhang}}, \bibinfo {author} {\bibfnamefont {C.}~\bibnamefont {Mewes}}, \bibinfo {author} {\bibfnamefont {L.}~\bibnamefont {Balicas}}, \bibinfo {author} {\bibfnamefont {T.}~\bibnamefont {Mewes}},\ and\ \bibinfo {author} {\bibfnamefont {P.}~\bibnamefont
  {Li}},\ }\bibfield  {title} {\bibinfo {title} {{Magnetism and spin dynamics in room-temperature van der Waals magnet Fe$_5$GeTe$_2$}},\ }\href {https://doi.org/10.1088/2053-1583/ac2028} {\bibfield  {journal} {\bibinfo  {journal} {2D Materials}\ }\textbf {\bibinfo {volume} {8}},\ \bibinfo {pages} {045030} (\bibinfo {year} {2021})}\BibitemShut {NoStop}%
\bibitem [{\citenamefont {Zeisner}\ \emph {et~al.}(2020)\citenamefont {Zeisner}, \citenamefont {Mehlawat}, \citenamefont {Alfonsov}, \citenamefont {Roslova}, \citenamefont {Doert}, \citenamefont {Isaeva}, \citenamefont {B{\"{u}}chner},\ and\ \citenamefont {Kataev}}]{Zeisner2020}%
  \BibitemOpen
  \bibfield  {author} {\bibinfo {author} {\bibfnamefont {J.}~\bibnamefont {Zeisner}}, \bibinfo {author} {\bibfnamefont {K.}~\bibnamefont {Mehlawat}}, \bibinfo {author} {\bibfnamefont {A.}~\bibnamefont {Alfonsov}}, \bibinfo {author} {\bibfnamefont {M.}~\bibnamefont {Roslova}}, \bibinfo {author} {\bibfnamefont {T.}~\bibnamefont {Doert}}, \bibinfo {author} {\bibfnamefont {A.}~\bibnamefont {Isaeva}}, \bibinfo {author} {\bibfnamefont {B.}~\bibnamefont {B{\"{u}}chner}},\ and\ \bibinfo {author} {\bibfnamefont {V.}~\bibnamefont {Kataev}},\ }\bibfield  {title} {\bibinfo {title} {{Electron spin resonance and ferromagnetic resonance spectroscopy in the high-field phase of the van der Waals magnet CrCl$_3$}},\ }\href {https://link.aps.org/doi/10.1103/PhysRevMaterials.4.064406} {\bibfield  {journal} {\bibinfo  {journal} {Physical Review Materials}\ }\textbf {\bibinfo {volume} {4}},\ \bibinfo {pages} {064406} (\bibinfo {year} {2020})}\BibitemShut {NoStop}%
\bibitem [{\citenamefont {Huang}\ \emph {et~al.}(2017)\citenamefont {Huang}, \citenamefont {Clark}, \citenamefont {Navarro-Moratalla}, \citenamefont {Klein}, \citenamefont {Cheng}, \citenamefont {Seyler}, \citenamefont {Zhong}, \citenamefont {Schmidgall}, \citenamefont {McGuire}, \citenamefont {Cobden}, \citenamefont {Yao}, \citenamefont {Xiao}, \citenamefont {Jarillo-Herrero},\ and\ \citenamefont {Xu}}]{Huang2017}%
  \BibitemOpen
  \bibfield  {author} {\bibinfo {author} {\bibfnamefont {B.}~\bibnamefont {Huang}}, \bibinfo {author} {\bibfnamefont {G.}~\bibnamefont {Clark}}, \bibinfo {author} {\bibfnamefont {E.}~\bibnamefont {Navarro-Moratalla}}, \bibinfo {author} {\bibfnamefont {D.~R.}\ \bibnamefont {Klein}}, \bibinfo {author} {\bibfnamefont {R.}~\bibnamefont {Cheng}}, \bibinfo {author} {\bibfnamefont {K.~L.}\ \bibnamefont {Seyler}}, \bibinfo {author} {\bibfnamefont {D.}~\bibnamefont {Zhong}}, \bibinfo {author} {\bibfnamefont {E.}~\bibnamefont {Schmidgall}}, \bibinfo {author} {\bibfnamefont {M.~A.}\ \bibnamefont {McGuire}}, \bibinfo {author} {\bibfnamefont {D.~H.}\ \bibnamefont {Cobden}}, \bibinfo {author} {\bibfnamefont {W.}~\bibnamefont {Yao}}, \bibinfo {author} {\bibfnamefont {D.}~\bibnamefont {Xiao}}, \bibinfo {author} {\bibfnamefont {P.}~\bibnamefont {Jarillo-Herrero}},\ and\ \bibinfo {author} {\bibfnamefont {X.}~\bibnamefont {Xu}},\ }\bibfield  {title} {\bibinfo {title} {{Layer-dependent ferromagnetism in a van der Waals crystal
  down to the monolayer limit}},\ }\href {https://doi.org/10.1038/nature22391} {\bibfield  {journal} {\bibinfo  {journal} {Nature}\ }\textbf {\bibinfo {volume} {546}},\ \bibinfo {pages} {270} (\bibinfo {year} {2017})}\BibitemShut {NoStop}%
\bibitem [{\citenamefont {Gong}\ \emph {et~al.}(2017)\citenamefont {Gong}, \citenamefont {Li}, \citenamefont {Li}, \citenamefont {Ji}, \citenamefont {Stern}, \citenamefont {Xia}, \citenamefont {Cao}, \citenamefont {Bao}, \citenamefont {Wang}, \citenamefont {Wang}, \citenamefont {Qiu}, \citenamefont {Cava}, \citenamefont {Louie}, \citenamefont {Xia},\ and\ \citenamefont {Zhang}}]{Gong2017}%
  \BibitemOpen
  \bibfield  {author} {\bibinfo {author} {\bibfnamefont {C.}~\bibnamefont {Gong}}, \bibinfo {author} {\bibfnamefont {L.}~\bibnamefont {Li}}, \bibinfo {author} {\bibfnamefont {Z.}~\bibnamefont {Li}}, \bibinfo {author} {\bibfnamefont {H.}~\bibnamefont {Ji}}, \bibinfo {author} {\bibfnamefont {A.}~\bibnamefont {Stern}}, \bibinfo {author} {\bibfnamefont {Y.}~\bibnamefont {Xia}}, \bibinfo {author} {\bibfnamefont {T.}~\bibnamefont {Cao}}, \bibinfo {author} {\bibfnamefont {W.}~\bibnamefont {Bao}}, \bibinfo {author} {\bibfnamefont {C.}~\bibnamefont {Wang}}, \bibinfo {author} {\bibfnamefont {Y.}~\bibnamefont {Wang}}, \bibinfo {author} {\bibfnamefont {Z.~Q.}\ \bibnamefont {Qiu}}, \bibinfo {author} {\bibfnamefont {R.~J.}\ \bibnamefont {Cava}}, \bibinfo {author} {\bibfnamefont {S.~G.}\ \bibnamefont {Louie}}, \bibinfo {author} {\bibfnamefont {J.}~\bibnamefont {Xia}},\ and\ \bibinfo {author} {\bibfnamefont {X.}~\bibnamefont {Zhang}},\ }\bibfield  {title} {\bibinfo {title} {{Discovery of intrinsic ferromagnetism in
  two-dimensional van der Waals crystals}},\ }\href {http://dx.doi.org/10.1038/nature22060} {\bibfield  {journal} {\bibinfo  {journal} {Nature}\ }\textbf {\bibinfo {volume} {546}},\ \bibinfo {pages} {265} (\bibinfo {year} {2017})}\BibitemShut {NoStop}%
\bibitem [{\citenamefont {Jiang}\ \emph {et~al.}(2021)\citenamefont {Jiang}, \citenamefont {Liu}, \citenamefont {Xing}, \citenamefont {Liu}, \citenamefont {Guo}, \citenamefont {Liu},\ and\ \citenamefont {Zhao}}]{Jiang2021}%
  \BibitemOpen
  \bibfield  {author} {\bibinfo {author} {\bibfnamefont {X.}~\bibnamefont {Jiang}}, \bibinfo {author} {\bibfnamefont {Q.}~\bibnamefont {Liu}}, \bibinfo {author} {\bibfnamefont {J.}~\bibnamefont {Xing}}, \bibinfo {author} {\bibfnamefont {N.}~\bibnamefont {Liu}}, \bibinfo {author} {\bibfnamefont {Y.}~\bibnamefont {Guo}}, \bibinfo {author} {\bibfnamefont {Z.}~\bibnamefont {Liu}},\ and\ \bibinfo {author} {\bibfnamefont {J.}~\bibnamefont {Zhao}},\ }\bibfield  {title} {\bibinfo {title} {{Recent progress on 2D magnets: Fundamental mechanism, structural design and modification}},\ }\href {http://dx.doi.org/10.1063/5.0039979} {\bibfield  {journal} {\bibinfo  {journal} {Applied Physics Reviews}\ }\textbf {\bibinfo {volume} {8}} (\bibinfo {year} {2021})}\BibitemShut {NoStop}%
\bibitem [{\citenamefont {Shen}\ \emph {et~al.}(2021)\citenamefont {Shen}, \citenamefont {Chen}, \citenamefont {Li}, \citenamefont {Xia}, \citenamefont {Zeng}, \citenamefont {Xu}, \citenamefont {Kwon}, \citenamefont {Ji}, \citenamefont {Won}, \citenamefont {Zhang},\ and\ \citenamefont {Wu}}]{Shen2021}%
  \BibitemOpen
  \bibfield  {author} {\bibinfo {author} {\bibfnamefont {X.}~\bibnamefont {Shen}}, \bibinfo {author} {\bibfnamefont {H.}~\bibnamefont {Chen}}, \bibinfo {author} {\bibfnamefont {Y.}~\bibnamefont {Li}}, \bibinfo {author} {\bibfnamefont {H.}~\bibnamefont {Xia}}, \bibinfo {author} {\bibfnamefont {F.}~\bibnamefont {Zeng}}, \bibinfo {author} {\bibfnamefont {J.}~\bibnamefont {Xu}}, \bibinfo {author} {\bibfnamefont {H.~Y.}\ \bibnamefont {Kwon}}, \bibinfo {author} {\bibfnamefont {Y.}~\bibnamefont {Ji}}, \bibinfo {author} {\bibfnamefont {C.}~\bibnamefont {Won}}, \bibinfo {author} {\bibfnamefont {W.}~\bibnamefont {Zhang}},\ and\ \bibinfo {author} {\bibfnamefont {Y.}~\bibnamefont {Wu}},\ }\bibfield  {title} {\bibinfo {title} {{Multi-domain ferromagnetic resonance in magnetic van der Waals crystals CrI$_3$ and CrBr$_3$}},\ }\href {http://dx.doi.org/10.1016/j.jmmm.2021.167772} {\bibfield  {journal} {\bibinfo  {journal} {Journal of Magnetism and Magnetic Materials}\ }\textbf {\bibinfo {volume} {528}} (\bibinfo {year}
  {2021})}\BibitemShut {NoStop}%
\bibitem [{\citenamefont {Williams}\ \emph {et~al.}(2015)\citenamefont {Williams}, \citenamefont {Aczel}, \citenamefont {Lumsden}, \citenamefont {Nagler}, \citenamefont {Stone}, \citenamefont {Yan},\ and\ \citenamefont {Mandrus}}]{Williams2015}%
  \BibitemOpen
  \bibfield  {author} {\bibinfo {author} {\bibfnamefont {T.~J.}\ \bibnamefont {Williams}}, \bibinfo {author} {\bibfnamefont {A.~A.}\ \bibnamefont {Aczel}}, \bibinfo {author} {\bibfnamefont {M.~D.}\ \bibnamefont {Lumsden}}, \bibinfo {author} {\bibfnamefont {S.~E.}\ \bibnamefont {Nagler}}, \bibinfo {author} {\bibfnamefont {M.~B.}\ \bibnamefont {Stone}}, \bibinfo {author} {\bibfnamefont {J.~Q.}\ \bibnamefont {Yan}},\ and\ \bibinfo {author} {\bibfnamefont {D.}~\bibnamefont {Mandrus}},\ }\bibfield  {title} {\bibinfo {title} {{Magnetic correlations in the quasi-two-dimensional semiconducting ferromagnet CrSiTe$_3$}},\ }\href {https://link.aps.org/doi/10.1103/PhysRevB.92.144404} {\bibfield  {journal} {\bibinfo  {journal} {Physical Review B}\ }\textbf {\bibinfo {volume} {92}},\ \bibinfo {pages} {144404} (\bibinfo {year} {2015})}\BibitemShut {NoStop}%
\bibitem [{\citenamefont {Lee}\ \emph {et~al.}(2016)\citenamefont {Lee}, \citenamefont {Lee}, \citenamefont {Ryoo}, \citenamefont {Kang}, \citenamefont {Kim}, \citenamefont {Kim}, \citenamefont {Park}, \citenamefont {Park},\ and\ \citenamefont {Cheong}}]{Lee2016b}%
  \BibitemOpen
  \bibfield  {author} {\bibinfo {author} {\bibfnamefont {J.~U.}\ \bibnamefont {Lee}}, \bibinfo {author} {\bibfnamefont {S.}~\bibnamefont {Lee}}, \bibinfo {author} {\bibfnamefont {J.~H.}\ \bibnamefont {Ryoo}}, \bibinfo {author} {\bibfnamefont {S.}~\bibnamefont {Kang}}, \bibinfo {author} {\bibfnamefont {T.~Y.}\ \bibnamefont {Kim}}, \bibinfo {author} {\bibfnamefont {P.}~\bibnamefont {Kim}}, \bibinfo {author} {\bibfnamefont {C.~H.}\ \bibnamefont {Park}}, \bibinfo {author} {\bibfnamefont {J.~G.}\ \bibnamefont {Park}},\ and\ \bibinfo {author} {\bibfnamefont {H.}~\bibnamefont {Cheong}},\ }\bibfield  {title} {\bibinfo {title} {{Ising-Type Magnetic Ordering in Atomically Thin FePS$_3$}},\ }\href {https://doi.org/10.1021/acs.nanolett.6b03052} {\bibfield  {journal} {\bibinfo  {journal} {Nano Letters}\ }\textbf {\bibinfo {volume} {16}},\ \bibinfo {pages} {7433} (\bibinfo {year} {2016})}\BibitemShut {NoStop}%
\bibitem [{\citenamefont {Wang}\ \emph {et~al.}(2022)\citenamefont {Wang}, \citenamefont {Cao}, \citenamefont {Li}, \citenamefont {Lu}, \citenamefont {Cohen}, \citenamefont {Haldar}, \citenamefont {Kitadai}, \citenamefont {Tan}, \citenamefont {Burch}, \citenamefont {Smirnov}, \citenamefont {Xu}, \citenamefont {Sharifzadeh}, \citenamefont {Liang},\ and\ \citenamefont {Ling}}]{Wang2022c}%
  \BibitemOpen
  \bibfield  {author} {\bibinfo {author} {\bibfnamefont {X.}~\bibnamefont {Wang}}, \bibinfo {author} {\bibfnamefont {J.}~\bibnamefont {Cao}}, \bibinfo {author} {\bibfnamefont {H.}~\bibnamefont {Li}}, \bibinfo {author} {\bibfnamefont {Z.}~\bibnamefont {Lu}}, \bibinfo {author} {\bibfnamefont {A.}~\bibnamefont {Cohen}}, \bibinfo {author} {\bibfnamefont {A.}~\bibnamefont {Haldar}}, \bibinfo {author} {\bibfnamefont {H.}~\bibnamefont {Kitadai}}, \bibinfo {author} {\bibfnamefont {Q.}~\bibnamefont {Tan}}, \bibinfo {author} {\bibfnamefont {K.~S.}\ \bibnamefont {Burch}}, \bibinfo {author} {\bibfnamefont {D.}~\bibnamefont {Smirnov}}, \bibinfo {author} {\bibfnamefont {W.}~\bibnamefont {Xu}}, \bibinfo {author} {\bibfnamefont {S.}~\bibnamefont {Sharifzadeh}}, \bibinfo {author} {\bibfnamefont {L.}~\bibnamefont {Liang}},\ and\ \bibinfo {author} {\bibfnamefont {X.}~\bibnamefont {Ling}},\ }\bibfield  {title} {\bibinfo {title} {{Electronic Raman scattering in the 2D antiferromagnet NiPS$_3$}},\ }\href
  {http://dx.doi.org/10.1126/sciadv.abl7707} {\bibfield  {journal} {\bibinfo  {journal} {Science Advances}\ }\textbf {\bibinfo {volume} {8}},\ \bibinfo {pages} {1} (\bibinfo {year} {2022})}\BibitemShut {NoStop}%
\bibitem [{\citenamefont {Deng}\ \emph {et~al.}(2018)\citenamefont {Deng}, \citenamefont {Yu}, \citenamefont {Song}, \citenamefont {Zhang}, \citenamefont {Wang}, \citenamefont {Sun}, \citenamefont {Yi}, \citenamefont {Wu}, \citenamefont {Wu}, \citenamefont {Zhu}, \citenamefont {Wang}, \citenamefont {Chen},\ and\ \citenamefont {Zhang}}]{Deng2018b}%
  \BibitemOpen
  \bibfield  {author} {\bibinfo {author} {\bibfnamefont {Y.}~\bibnamefont {Deng}}, \bibinfo {author} {\bibfnamefont {Y.}~\bibnamefont {Yu}}, \bibinfo {author} {\bibfnamefont {Y.}~\bibnamefont {Song}}, \bibinfo {author} {\bibfnamefont {J.}~\bibnamefont {Zhang}}, \bibinfo {author} {\bibfnamefont {N.~Z.}\ \bibnamefont {Wang}}, \bibinfo {author} {\bibfnamefont {Z.}~\bibnamefont {Sun}}, \bibinfo {author} {\bibfnamefont {Y.}~\bibnamefont {Yi}}, \bibinfo {author} {\bibfnamefont {Y.~Z.}\ \bibnamefont {Wu}}, \bibinfo {author} {\bibfnamefont {S.}~\bibnamefont {Wu}}, \bibinfo {author} {\bibfnamefont {J.}~\bibnamefont {Zhu}}, \bibinfo {author} {\bibfnamefont {J.}~\bibnamefont {Wang}}, \bibinfo {author} {\bibfnamefont {X.~H.}\ \bibnamefont {Chen}},\ and\ \bibinfo {author} {\bibfnamefont {Y.}~\bibnamefont {Zhang}},\ }\bibfield  {title} {\bibinfo {title} {{Gate-tunable room-temperature ferromagnetism in two-dimensional Fe$_3$GeTe$_2$}},\ }\href {https://www.nature.com/articles/s41586-018-0626-9} {\bibfield  {journal}
  {\bibinfo  {journal} {Nature}\ }\textbf {\bibinfo {volume} {563}},\ \bibinfo {pages} {94} (\bibinfo {year} {2018})}\BibitemShut {NoStop}%
\bibitem [{\citenamefont {Pal}\ \emph {et~al.}(2024{\natexlab{b}})\citenamefont {Pal}, \citenamefont {Pal}, \citenamefont {Mondal}, \citenamefont {Sharma}, \citenamefont {Das}, \citenamefont {Mandal},\ and\ \citenamefont {Pal}}]{Pal2024}%
  \BibitemOpen
  \bibfield  {author} {\bibinfo {author} {\bibfnamefont {R.}~\bibnamefont {Pal}}, \bibinfo {author} {\bibfnamefont {B.}~\bibnamefont {Pal}}, \bibinfo {author} {\bibfnamefont {S.}~\bibnamefont {Mondal}}, \bibinfo {author} {\bibfnamefont {R.~O.}\ \bibnamefont {Sharma}}, \bibinfo {author} {\bibfnamefont {T.}~\bibnamefont {Das}}, \bibinfo {author} {\bibfnamefont {P.}~\bibnamefont {Mandal}},\ and\ \bibinfo {author} {\bibfnamefont {A.~N.}\ \bibnamefont {Pal}},\ }\bibfield  {title} {\bibinfo {title} {{Spin-reorientation driven emergent phases and unconventional magnetotransport in quasi-2D vdW ferromagnet Fe$_4$GeTe$_2$}},\ }\href {https://doi.org/10.1038/s41699-024-00463-y} {\bibfield  {journal} {\bibinfo  {journal} {npj 2D Materials and Applications}\ }\textbf {\bibinfo {volume} {8}},\ \bibinfo {pages} {30} (\bibinfo {year} {2024}{\natexlab{b}})}\BibitemShut {NoStop}%
\bibitem [{\citenamefont {Seo}\ \emph {et~al.}(2020)\citenamefont {Seo}, \citenamefont {Kim}, \citenamefont {An}, \citenamefont {Kim}, \citenamefont {Kim}, \citenamefont {Hwang}, \citenamefont {Kim}, \citenamefont {Jang}, \citenamefont {Kim}, \citenamefont {Eom}, \citenamefont {Seo}, \citenamefont {Stania}, \citenamefont {Muntwiler}, \citenamefont {Lee}, \citenamefont {Watanabe}, \citenamefont {Taniguchi}, \citenamefont {Jo}, \citenamefont {Lee}, \citenamefont {Min}, \citenamefont {Jo}, \citenamefont {Yeom}, \citenamefont {Choi}, \citenamefont {Shim},\ and\ \citenamefont {Kim}}]{Seo2020}%
  \BibitemOpen
  \bibfield  {author} {\bibinfo {author} {\bibfnamefont {J.}~\bibnamefont {Seo}}, \bibinfo {author} {\bibfnamefont {D.~Y.}\ \bibnamefont {Kim}}, \bibinfo {author} {\bibfnamefont {E.~S.}\ \bibnamefont {An}}, \bibinfo {author} {\bibfnamefont {K.}~\bibnamefont {Kim}}, \bibinfo {author} {\bibfnamefont {G.-Y.}\ \bibnamefont {Kim}}, \bibinfo {author} {\bibfnamefont {S.-Y.}\ \bibnamefont {Hwang}}, \bibinfo {author} {\bibfnamefont {D.~W.}\ \bibnamefont {Kim}}, \bibinfo {author} {\bibfnamefont {B.~G.}\ \bibnamefont {Jang}}, \bibinfo {author} {\bibfnamefont {H.}~\bibnamefont {Kim}}, \bibinfo {author} {\bibfnamefont {G.}~\bibnamefont {Eom}}, \bibinfo {author} {\bibfnamefont {S.~Y.}\ \bibnamefont {Seo}}, \bibinfo {author} {\bibfnamefont {R.}~\bibnamefont {Stania}}, \bibinfo {author} {\bibfnamefont {M.}~\bibnamefont {Muntwiler}}, \bibinfo {author} {\bibfnamefont {J.}~\bibnamefont {Lee}}, \bibinfo {author} {\bibfnamefont {K.}~\bibnamefont {Watanabe}}, \bibinfo {author} {\bibfnamefont {T.}~\bibnamefont {Taniguchi}},
  \bibinfo {author} {\bibfnamefont {Y.~J.}\ \bibnamefont {Jo}}, \bibinfo {author} {\bibfnamefont {J.}~\bibnamefont {Lee}}, \bibinfo {author} {\bibfnamefont {B.~I.}\ \bibnamefont {Min}}, \bibinfo {author} {\bibfnamefont {M.~H.}\ \bibnamefont {Jo}}, \bibinfo {author} {\bibfnamefont {H.~W.}\ \bibnamefont {Yeom}}, \bibinfo {author} {\bibfnamefont {S.-Y.}\ \bibnamefont {Choi}}, \bibinfo {author} {\bibfnamefont {J.~H.}\ \bibnamefont {Shim}},\ and\ \bibinfo {author} {\bibfnamefont {J.~S.}\ \bibnamefont {Kim}},\ }\bibfield  {title} {\bibinfo {title} {{Nearly room temperature ferromagnetism in a magnetic metal-rich van der Waals metal}},\ }\href {https://doi.org/10.1126/sciadv.aay8912} {\bibfield  {journal} {\bibinfo  {journal} {Science Advances}\ }\textbf {\bibinfo {volume} {6}},\ \bibinfo {pages} {eaay8912} (\bibinfo {year} {2020})}\BibitemShut {NoStop}%
\bibitem [{\citenamefont {Bera}\ \emph {et~al.}(2023{\natexlab{a}})\citenamefont {Bera}, \citenamefont {Pradhan}, \citenamefont {Khan}, \citenamefont {Pal}, \citenamefont {Pal}, \citenamefont {Kalimuddin}, \citenamefont {Bera}, \citenamefont {Das}, \citenamefont {Pal},\ and\ \citenamefont {Mondal}}]{Bera2023}%
  \BibitemOpen
  \bibfield  {author} {\bibinfo {author} {\bibfnamefont {S.}~\bibnamefont {Bera}}, \bibinfo {author} {\bibfnamefont {S.~K.}\ \bibnamefont {Pradhan}}, \bibinfo {author} {\bibfnamefont {M.~S.}\ \bibnamefont {Khan}}, \bibinfo {author} {\bibfnamefont {R.}~\bibnamefont {Pal}}, \bibinfo {author} {\bibfnamefont {B.}~\bibnamefont {Pal}}, \bibinfo {author} {\bibfnamefont {S.}~\bibnamefont {Kalimuddin}}, \bibinfo {author} {\bibfnamefont {A.}~\bibnamefont {Bera}}, \bibinfo {author} {\bibfnamefont {B.}~\bibnamefont {Das}}, \bibinfo {author} {\bibfnamefont {A.~N.}\ \bibnamefont {Pal}},\ and\ \bibinfo {author} {\bibfnamefont {M.}~\bibnamefont {Mondal}},\ }\bibfield  {title} {\bibinfo {title} {{Unravelling the nature of spin reorientation transition in quasi-2D vdW magnetic material, Fe$_4$GeTe$_2$}},\ }\bibfield  {journal} {\bibinfo  {journal} {Journal of Magnetism and Magnetic Materials}\ }\textbf {\bibinfo {volume} {565}},\ \href {https://doi.org/10.1016/j.jmmm.2022.170257} {10.1016/j.jmmm.2022.170257} (\bibinfo {year}
  {2023}{\natexlab{a}})\BibitemShut {NoStop}%
\bibitem [{\citenamefont {Tan}\ \emph {et~al.}(2021)\citenamefont {Tan}, \citenamefont {Xie}, \citenamefont {Zheng}, \citenamefont {Aloufi}, \citenamefont {Albarakati}, \citenamefont {Algarni}, \citenamefont {Li}, \citenamefont {Partridge}, \citenamefont {Culcer}, \citenamefont {Wang}, \citenamefont {Yi}, \citenamefont {Tian}, \citenamefont {Xiong}, \citenamefont {Zhao},\ and\ \citenamefont {Wang}}]{Tan2021}%
  \BibitemOpen
  \bibfield  {author} {\bibinfo {author} {\bibfnamefont {C.}~\bibnamefont {Tan}}, \bibinfo {author} {\bibfnamefont {W.-Q.}\ \bibnamefont {Xie}}, \bibinfo {author} {\bibfnamefont {G.}~\bibnamefont {Zheng}}, \bibinfo {author} {\bibfnamefont {N.}~\bibnamefont {Aloufi}}, \bibinfo {author} {\bibfnamefont {S.}~\bibnamefont {Albarakati}}, \bibinfo {author} {\bibfnamefont {M.}~\bibnamefont {Algarni}}, \bibinfo {author} {\bibfnamefont {J.}~\bibnamefont {Li}}, \bibinfo {author} {\bibfnamefont {J.}~\bibnamefont {Partridge}}, \bibinfo {author} {\bibfnamefont {D.}~\bibnamefont {Culcer}}, \bibinfo {author} {\bibfnamefont {X.}~\bibnamefont {Wang}}, \bibinfo {author} {\bibfnamefont {J.~B.}\ \bibnamefont {Yi}}, \bibinfo {author} {\bibfnamefont {M.}~\bibnamefont {Tian}}, \bibinfo {author} {\bibfnamefont {Y.}~\bibnamefont {Xiong}}, \bibinfo {author} {\bibfnamefont {Y.-J.}\ \bibnamefont {Zhao}},\ and\ \bibinfo {author} {\bibfnamefont {L.}~\bibnamefont {Wang}},\ }\bibfield  {title} {\bibinfo {title} {{Gate-Controlled Magnetic
  Phase Transition in a van der Waals Magnet Fe$_5$GeTe$_2$}},\ }\href {https://pubs.acs.org/doi/10.1021/acs.nanolett.1c01108} {\bibfield  {journal} {\bibinfo  {journal} {Nano Letters}\ }\textbf {\bibinfo {volume} {21}},\ \bibinfo {pages} {5599} (\bibinfo {year} {2021})}\BibitemShut {NoStop}%
\bibitem [{\citenamefont {Pal}\ \emph {et~al.}(2025{\natexlab{b}})\citenamefont {Pal}, \citenamefont {Hasan}, \citenamefont {Nayak}, \citenamefont {Deka}, \citenamefont {Salehi}, \citenamefont {Pereiro}, \citenamefont {Mondal}, \citenamefont {Misra}, \citenamefont {Singha}, \citenamefont {Mandal} \emph {et~al.}}]{Pal_Raman_2025}%
  \BibitemOpen
  \bibfield  {author} {\bibinfo {author} {\bibfnamefont {R.}~\bibnamefont {Pal}}, \bibinfo {author} {\bibfnamefont {M.~N.}\ \bibnamefont {Hasan}}, \bibinfo {author} {\bibfnamefont {C.}~\bibnamefont {Nayak}}, \bibinfo {author} {\bibfnamefont {M.}~\bibnamefont {Deka}}, \bibinfo {author} {\bibfnamefont {N.}~\bibnamefont {Salehi}}, \bibinfo {author} {\bibfnamefont {M.}~\bibnamefont {Pereiro}}, \bibinfo {author} {\bibfnamefont {S.}~\bibnamefont {Mondal}}, \bibinfo {author} {\bibfnamefont {A.}~\bibnamefont {Misra}}, \bibinfo {author} {\bibfnamefont {A.}~\bibnamefont {Singha}}, \bibinfo {author} {\bibfnamefont {P.}~\bibnamefont {Mandal}}, \emph {et~al.},\ }\bibfield  {title} {\bibinfo {title} {{Spin Reorientation Driven Renormalization of Spin-Phonon Coupling in Fe$_4$GeTe$_2$}},\ }\href {https://doi.org/10.48550/arXiv.2512.18544} {\bibfield  {journal} {\bibinfo  {journal} {arXiv preprint arXiv:2512.18544}\ } (\bibinfo {year} {2025}{\natexlab{b}})}\BibitemShut {NoStop}%
\bibitem [{\citenamefont {May}\ \emph {et~al.}(2019)\citenamefont {May}, \citenamefont {Bridges},\ and\ \citenamefont {McGuire}}]{May2019}%
  \BibitemOpen
  \bibfield  {author} {\bibinfo {author} {\bibfnamefont {A.~F.}\ \bibnamefont {May}}, \bibinfo {author} {\bibfnamefont {C.~A.}\ \bibnamefont {Bridges}},\ and\ \bibinfo {author} {\bibfnamefont {M.~A.}\ \bibnamefont {McGuire}},\ }\bibfield  {title} {\bibinfo {title} {{Physical properties and thermal stability of ${\mathrm{Fe}}_{5\ensuremath{-}x}{\mathrm{GeTe}}_{2}$ single crystals}},\ }\href {https://doi.org/10.1103/PhysRevMaterials.3.104401} {\bibfield  {journal} {\bibinfo  {journal} {Phys. Rev. Mater.}\ }\textbf {\bibinfo {volume} {3}},\ \bibinfo {pages} {104401} (\bibinfo {year} {2019})}\BibitemShut {NoStop}%
\bibitem [{\citenamefont {Fei}\ \emph {et~al.}(2018)\citenamefont {Fei}, \citenamefont {Huang}, \citenamefont {Malinowski}, \citenamefont {Wang}, \citenamefont {Song}, \citenamefont {Sanchez}, \citenamefont {Yao}, \citenamefont {Xiao}, \citenamefont {Zhu}, \citenamefont {May}, \citenamefont {Wu}, \citenamefont {Cobden}, \citenamefont {Chu},\ and\ \citenamefont {Xu}}]{Fei2018}%
  \BibitemOpen
  \bibfield  {author} {\bibinfo {author} {\bibfnamefont {Z.}~\bibnamefont {Fei}}, \bibinfo {author} {\bibfnamefont {B.}~\bibnamefont {Huang}}, \bibinfo {author} {\bibfnamefont {P.}~\bibnamefont {Malinowski}}, \bibinfo {author} {\bibfnamefont {W.}~\bibnamefont {Wang}}, \bibinfo {author} {\bibfnamefont {T.}~\bibnamefont {Song}}, \bibinfo {author} {\bibfnamefont {J.}~\bibnamefont {Sanchez}}, \bibinfo {author} {\bibfnamefont {W.}~\bibnamefont {Yao}}, \bibinfo {author} {\bibfnamefont {D.}~\bibnamefont {Xiao}}, \bibinfo {author} {\bibfnamefont {X.}~\bibnamefont {Zhu}}, \bibinfo {author} {\bibfnamefont {A.~F.}\ \bibnamefont {May}}, \bibinfo {author} {\bibfnamefont {W.}~\bibnamefont {Wu}}, \bibinfo {author} {\bibfnamefont {D.~H.}\ \bibnamefont {Cobden}}, \bibinfo {author} {\bibfnamefont {J.-H.}\ \bibnamefont {Chu}},\ and\ \bibinfo {author} {\bibfnamefont {X.}~\bibnamefont {Xu}},\ }\bibfield  {title} {\bibinfo {title} {{Two-dimensional itinerant ferromagnetism in atomically thin Fe$_3$GeTe$_2$}},\ }\href
  {https://doi.org/10.1038/s41563-018-0149-7} {\bibfield  {journal} {\bibinfo  {journal} {Nature Materials}\ }\textbf {\bibinfo {volume} {17}},\ \bibinfo {pages} {778} (\bibinfo {year} {2018})}\BibitemShut {NoStop}%
\bibitem [{\citenamefont {Pal}\ \emph {et~al.}(2024{\natexlab{c}})\citenamefont {Pal}, \citenamefont {Bera}, \citenamefont {Pal}, \citenamefont {Mondal},\ and\ \citenamefont {Pal}}]{Pal_2024_aip}%
  \BibitemOpen
  \bibfield  {author} {\bibinfo {author} {\bibfnamefont {R.}~\bibnamefont {Pal}}, \bibinfo {author} {\bibfnamefont {S.}~\bibnamefont {Bera}}, \bibinfo {author} {\bibfnamefont {B.}~\bibnamefont {Pal}}, \bibinfo {author} {\bibfnamefont {M.}~\bibnamefont {Mondal}},\ and\ \bibinfo {author} {\bibfnamefont {A.~N.}\ \bibnamefont {Pal}},\ }\bibfield  {title} {\bibinfo {title} {{Intrinsic room temperature ferromagnetism in van der Waals Fe$_5$GeTe$_2$ crystal}},\ }\href {https://doi.org/10.1063/5.0204358} {\bibfield  {journal} {\bibinfo  {journal} {AIP Conference Proceedings}\ }\textbf {\bibinfo {volume} {3067}},\ \bibinfo {pages} {20002} (\bibinfo {year} {2024}{\natexlab{c}})}\BibitemShut {NoStop}%
\bibitem [{\citenamefont {Albarakati}\ \emph {et~al.}(2019)\citenamefont {Albarakati}, \citenamefont {Tan}, \citenamefont {Chen}, \citenamefont {Partridge}, \citenamefont {Zheng}, \citenamefont {Farrar}, \citenamefont {Mayes}, \citenamefont {Field}, \citenamefont {Lee}, \citenamefont {Wang}, \citenamefont {Xiong}, \citenamefont {Tian}, \citenamefont {Xiang}, \citenamefont {Hamilton}, \citenamefont {Tretiakov}, \citenamefont {Culcer}, \citenamefont {Zhao},\ and\ \citenamefont {Wang}}]{Albarakati2019a}%
  \BibitemOpen
  \bibfield  {author} {\bibinfo {author} {\bibfnamefont {S.}~\bibnamefont {Albarakati}}, \bibinfo {author} {\bibfnamefont {C.}~\bibnamefont {Tan}}, \bibinfo {author} {\bibfnamefont {Z.~J.}\ \bibnamefont {Chen}}, \bibinfo {author} {\bibfnamefont {J.~G.}\ \bibnamefont {Partridge}}, \bibinfo {author} {\bibfnamefont {G.}~\bibnamefont {Zheng}}, \bibinfo {author} {\bibfnamefont {L.}~\bibnamefont {Farrar}}, \bibinfo {author} {\bibfnamefont {E.~L.}\ \bibnamefont {Mayes}}, \bibinfo {author} {\bibfnamefont {M.~R.}\ \bibnamefont {Field}}, \bibinfo {author} {\bibfnamefont {C.}~\bibnamefont {Lee}}, \bibinfo {author} {\bibfnamefont {Y.}~\bibnamefont {Wang}}, \bibinfo {author} {\bibfnamefont {Y.}~\bibnamefont {Xiong}}, \bibinfo {author} {\bibfnamefont {M.}~\bibnamefont {Tian}}, \bibinfo {author} {\bibfnamefont {F.}~\bibnamefont {Xiang}}, \bibinfo {author} {\bibfnamefont {A.~R.}\ \bibnamefont {Hamilton}}, \bibinfo {author} {\bibfnamefont {O.~A.}\ \bibnamefont {Tretiakov}}, \bibinfo {author} {\bibfnamefont {D.}~\bibnamefont
  {Culcer}}, \bibinfo {author} {\bibfnamefont {Y.~J.}\ \bibnamefont {Zhao}},\ and\ \bibinfo {author} {\bibfnamefont {L.}~\bibnamefont {Wang}},\ }\bibfield  {title} {\bibinfo {title} {{Antisymmetric magnetoresistance in van der Waals Fe$_3$GeTe$_2$/graphite/Fe$_3$GeTe$_2$ trilayer heterostructures}},\ }\href {http://advances.sciencemag.org/content/5/7/eaaw0409} {\bibfield  {journal} {\bibinfo  {journal} {Science Advances}\ }\textbf {\bibinfo {volume} {5}},\ \bibinfo {pages} {1} (\bibinfo {year} {2019})}\BibitemShut {NoStop}%
\bibitem [{\citenamefont {Zhou}\ \emph {et~al.}(2022)\citenamefont {Zhou}, \citenamefont {Huang}, \citenamefont {Tang}, \citenamefont {Qiu}, \citenamefont {Qin}, \citenamefont {Zhang}, \citenamefont {Li}, \citenamefont {Wu},\ and\ \citenamefont {Yuan}}]{Zhou2022b}%
  \BibitemOpen
  \bibfield  {author} {\bibinfo {author} {\bibfnamefont {L.}~\bibnamefont {Zhou}}, \bibinfo {author} {\bibfnamefont {J.}~\bibnamefont {Huang}}, \bibinfo {author} {\bibfnamefont {M.}~\bibnamefont {Tang}}, \bibinfo {author} {\bibfnamefont {C.}~\bibnamefont {Qiu}}, \bibinfo {author} {\bibfnamefont {F.}~\bibnamefont {Qin}}, \bibinfo {author} {\bibfnamefont {C.}~\bibnamefont {Zhang}}, \bibinfo {author} {\bibfnamefont {Z.}~\bibnamefont {Li}}, \bibinfo {author} {\bibfnamefont {D.}~\bibnamefont {Wu}},\ and\ \bibinfo {author} {\bibfnamefont {H.}~\bibnamefont {Yuan}},\ }\bibfield  {title} {\bibinfo {title} {{Gate-tunable spin valve effect in Fe$_3$GeTe$_2$-based van der Waals heterostructures}},\ }\href {http://dx.doi.org/10.1002/inf2.12371} {\bibfield  {journal} {\bibinfo  {journal} {InfoMat}\ ,\ \bibinfo {pages} {1}} (\bibinfo {year} {2022})}\BibitemShut {NoStop}%
\bibitem [{\citenamefont {Chen}\ \emph {et~al.}(2013)\citenamefont {Chen}, \citenamefont {Yang}, \citenamefont {Wang}, \citenamefont {Imai}, \citenamefont {Ohta}, \citenamefont {Michioka}, \citenamefont {Yoshimura},\ and\ \citenamefont {Fang}}]{Chen2013}%
  \BibitemOpen
  \bibfield  {author} {\bibinfo {author} {\bibfnamefont {B.}~\bibnamefont {Chen}}, \bibinfo {author} {\bibfnamefont {J.}~\bibnamefont {Yang}}, \bibinfo {author} {\bibfnamefont {H.}~\bibnamefont {Wang}}, \bibinfo {author} {\bibfnamefont {M.}~\bibnamefont {Imai}}, \bibinfo {author} {\bibfnamefont {H.}~\bibnamefont {Ohta}}, \bibinfo {author} {\bibfnamefont {C.}~\bibnamefont {Michioka}}, \bibinfo {author} {\bibfnamefont {K.}~\bibnamefont {Yoshimura}},\ and\ \bibinfo {author} {\bibfnamefont {M.}~\bibnamefont {Fang}},\ }\bibfield  {title} {\bibinfo {title} {{Magnetic Properties of Layered Itinerant Electron Ferromagnet Fe$_3$GeTe$_2$}},\ }\href {https://doi.org/10.7566/JPSJ.82.124711} {\bibfield  {journal} {\bibinfo  {journal} {Journal of the Physical Society of Japan}\ }\textbf {\bibinfo {volume} {82}},\ \bibinfo {pages} {124711} (\bibinfo {year} {2013})}\BibitemShut {NoStop}%
\bibitem [{\citenamefont {Wang}\ \emph {et~al.}(2017)\citenamefont {Wang}, \citenamefont {Xian}, \citenamefont {Wang}, \citenamefont {Liu}, \citenamefont {Ling}, \citenamefont {Zhang}, \citenamefont {Cao}, \citenamefont {Qu},\ and\ \citenamefont {Xiong}}]{Wang2017}%
  \BibitemOpen
  \bibfield  {author} {\bibinfo {author} {\bibfnamefont {Y.}~\bibnamefont {Wang}}, \bibinfo {author} {\bibfnamefont {C.}~\bibnamefont {Xian}}, \bibinfo {author} {\bibfnamefont {J.}~\bibnamefont {Wang}}, \bibinfo {author} {\bibfnamefont {B.}~\bibnamefont {Liu}}, \bibinfo {author} {\bibfnamefont {L.}~\bibnamefont {Ling}}, \bibinfo {author} {\bibfnamefont {L.}~\bibnamefont {Zhang}}, \bibinfo {author} {\bibfnamefont {L.}~\bibnamefont {Cao}}, \bibinfo {author} {\bibfnamefont {Z.}~\bibnamefont {Qu}},\ and\ \bibinfo {author} {\bibfnamefont {Y.}~\bibnamefont {Xiong}},\ }\bibfield  {title} {\bibinfo {title} {{Anisotropic anomalous Hall effect in triangular itinerant ferromagnet Fe$_3$GeTe$_2$}},\ }\href {http://dx.doi.org/10.1103/PhysRevB.96.134428} {\bibfield  {journal} {\bibinfo  {journal} {Physical Review B}\ }\textbf {\bibinfo {volume} {96}},\ \bibinfo {pages} {1} (\bibinfo {year} {2017})}\BibitemShut {NoStop}%
\bibitem [{\citenamefont {Zhang}\ \emph {et~al.}(2018)\citenamefont {Zhang}, \citenamefont {Lu}, \citenamefont {Zhu}, \citenamefont {Tan}, \citenamefont {Feng}, \citenamefont {Liu}, \citenamefont {Zhang}, \citenamefont {Chen}, \citenamefont {Liu}, \citenamefont {Luo}, \citenamefont {Xie}, \citenamefont {Luo}, \citenamefont {Zhang},\ and\ \citenamefont {Lai}}]{Zhang2018}%
  \BibitemOpen
  \bibfield  {author} {\bibinfo {author} {\bibfnamefont {Y.}~\bibnamefont {Zhang}}, \bibinfo {author} {\bibfnamefont {H.}~\bibnamefont {Lu}}, \bibinfo {author} {\bibfnamefont {X.}~\bibnamefont {Zhu}}, \bibinfo {author} {\bibfnamefont {S.}~\bibnamefont {Tan}}, \bibinfo {author} {\bibfnamefont {W.}~\bibnamefont {Feng}}, \bibinfo {author} {\bibfnamefont {Q.}~\bibnamefont {Liu}}, \bibinfo {author} {\bibfnamefont {W.}~\bibnamefont {Zhang}}, \bibinfo {author} {\bibfnamefont {Q.}~\bibnamefont {Chen}}, \bibinfo {author} {\bibfnamefont {Y.}~\bibnamefont {Liu}}, \bibinfo {author} {\bibfnamefont {X.}~\bibnamefont {Luo}}, \bibinfo {author} {\bibfnamefont {D.}~\bibnamefont {Xie}}, \bibinfo {author} {\bibfnamefont {L.}~\bibnamefont {Luo}}, \bibinfo {author} {\bibfnamefont {Z.}~\bibnamefont {Zhang}},\ and\ \bibinfo {author} {\bibfnamefont {X.}~\bibnamefont {Lai}},\ }\bibfield  {title} {\bibinfo {title} {{Emergence of Kondo lattice behavior in a van der Waals itinerant ferromagnet, Fe$_3$GeTe$_2$}},\ }\href
  {http://dx.doi.org/10.1126/sciadv.aao6791} {\bibfield  {journal} {\bibinfo  {journal} {Science Advances}\ }\textbf {\bibinfo {volume} {4}},\ \bibinfo {pages} {1} (\bibinfo {year} {2018})}\BibitemShut {NoStop}%
\bibitem [{\citenamefont {Xu}\ \emph {et~al.}(2024)\citenamefont {Xu}, \citenamefont {Wang}, \citenamefont {Jin}, \citenamefont {Liu}, \citenamefont {Liu}, \citenamefont {Song},\ and\ \citenamefont {Tian}}]{Xu2024}%
  \BibitemOpen
  \bibfield  {author} {\bibinfo {author} {\bibfnamefont {Y.}~\bibnamefont {Xu}}, \bibinfo {author} {\bibfnamefont {Y.-C.}\ \bibnamefont {Wang}}, \bibinfo {author} {\bibfnamefont {X.}~\bibnamefont {Jin}}, \bibinfo {author} {\bibfnamefont {H.}~\bibnamefont {Liu}}, \bibinfo {author} {\bibfnamefont {Y.}~\bibnamefont {Liu}}, \bibinfo {author} {\bibfnamefont {H.}~\bibnamefont {Song}},\ and\ \bibinfo {author} {\bibfnamefont {F.}~\bibnamefont {Tian}},\ }\bibfield  {title} {\bibinfo {title} {{Mechanism of magnetic phase transition in correlated magnetic metal: insight into itinerant ferromagnet Fe$_{3-\delta}$GeTe$_2$}},\ }\href {https://doi.org/10.1038/s42005-024-01874-5} {\bibfield  {journal} {\bibinfo  {journal} {Communications Physics}\ }\textbf {\bibinfo {volume} {7}},\ \bibinfo {pages} {381} (\bibinfo {year} {2024})}\BibitemShut {NoStop}%
\bibitem [{\citenamefont {Kim}\ \emph {et~al.}(2022)\citenamefont {Kim}, \citenamefont {Ryee},\ and\ \citenamefont {Han}}]{Kim2022a}%
  \BibitemOpen
  \bibfield  {author} {\bibinfo {author} {\bibfnamefont {T.~J.}\ \bibnamefont {Kim}}, \bibinfo {author} {\bibfnamefont {S.}~\bibnamefont {Ryee}},\ and\ \bibinfo {author} {\bibfnamefont {M.~J.}\ \bibnamefont {Han}},\ }\bibfield  {title} {\bibinfo {title} {{Fe$_3$GeTe$_2$: a site-differentiated Hund metal}},\ }\href {https://doi.org/10.1038/s41524-022-00937-x} {\bibfield  {journal} {\bibinfo  {journal} {npj Computational Materials}\ }\textbf {\bibinfo {volume} {8}},\ \bibinfo {pages} {245} (\bibinfo {year} {2022})}\BibitemShut {NoStop}%
\bibitem [{\citenamefont {Kim}\ \emph {et~al.}(2018)\citenamefont {Kim}, \citenamefont {Seo}, \citenamefont {Lee}, \citenamefont {Ko}, \citenamefont {Kim}, \citenamefont {Jang}, \citenamefont {Ok}, \citenamefont {Lee}, \citenamefont {Jo}, \citenamefont {Kang}, \citenamefont {Shim}, \citenamefont {Kim}, \citenamefont {Yeom}, \citenamefont {Il~Min}, \citenamefont {Yang},\ and\ \citenamefont {Kim}}]{Kim2018b}%
  \BibitemOpen
  \bibfield  {author} {\bibinfo {author} {\bibfnamefont {K.}~\bibnamefont {Kim}}, \bibinfo {author} {\bibfnamefont {J.}~\bibnamefont {Seo}}, \bibinfo {author} {\bibfnamefont {E.}~\bibnamefont {Lee}}, \bibinfo {author} {\bibfnamefont {K.~T.}\ \bibnamefont {Ko}}, \bibinfo {author} {\bibfnamefont {B.~S.}\ \bibnamefont {Kim}}, \bibinfo {author} {\bibfnamefont {B.~G.}\ \bibnamefont {Jang}}, \bibinfo {author} {\bibfnamefont {J.~M.}\ \bibnamefont {Ok}}, \bibinfo {author} {\bibfnamefont {J.}~\bibnamefont {Lee}}, \bibinfo {author} {\bibfnamefont {Y.~J.}\ \bibnamefont {Jo}}, \bibinfo {author} {\bibfnamefont {W.}~\bibnamefont {Kang}}, \bibinfo {author} {\bibfnamefont {J.~H.}\ \bibnamefont {Shim}}, \bibinfo {author} {\bibfnamefont {C.}~\bibnamefont {Kim}}, \bibinfo {author} {\bibfnamefont {H.~W.}\ \bibnamefont {Yeom}}, \bibinfo {author} {\bibfnamefont {B.}~\bibnamefont {Il~Min}}, \bibinfo {author} {\bibfnamefont {B.~J.}\ \bibnamefont {Yang}},\ and\ \bibinfo {author} {\bibfnamefont {J.~S.}\ \bibnamefont {Kim}},\
  }\bibfield  {title} {\bibinfo {title} {{Large anomalous Hall current induced by topological nodal lines in a ferromagnetic van der Waals semimetal}},\ }\href {https://doi.org/10.1038/s41563-018-0132-3} {\bibfield  {journal} {\bibinfo  {journal} {Nature Materials}\ }\textbf {\bibinfo {volume} {17}},\ \bibinfo {pages} {794} (\bibinfo {year} {2018})}\BibitemShut {NoStop}%
\bibitem [{\citenamefont {Ding}\ \emph {et~al.}(2020)\citenamefont {Ding}, \citenamefont {Li}, \citenamefont {Xu}, \citenamefont {Li}, \citenamefont {Hou}, \citenamefont {Liu}, \citenamefont {Xi}, \citenamefont {Xu}, \citenamefont {Yao},\ and\ \citenamefont {Wang}}]{Ding2020a}%
  \BibitemOpen
  \bibfield  {author} {\bibinfo {author} {\bibfnamefont {B.}~\bibnamefont {Ding}}, \bibinfo {author} {\bibfnamefont {Z.}~\bibnamefont {Li}}, \bibinfo {author} {\bibfnamefont {G.}~\bibnamefont {Xu}}, \bibinfo {author} {\bibfnamefont {H.}~\bibnamefont {Li}}, \bibinfo {author} {\bibfnamefont {Z.}~\bibnamefont {Hou}}, \bibinfo {author} {\bibfnamefont {E.}~\bibnamefont {Liu}}, \bibinfo {author} {\bibfnamefont {X.}~\bibnamefont {Xi}}, \bibinfo {author} {\bibfnamefont {F.}~\bibnamefont {Xu}}, \bibinfo {author} {\bibfnamefont {Y.}~\bibnamefont {Yao}},\ and\ \bibinfo {author} {\bibfnamefont {W.}~\bibnamefont {Wang}},\ }\bibfield  {title} {\bibinfo {title} {{Observation of Magnetic Skyrmion Bubbles in a van der Waals Ferromagnet Fe$_3$GeTe$_2$}},\ }\href {http://dx.doi.org/10.1021/acs.nanolett.9b03453} {\bibfield  {journal} {\bibinfo  {journal} {Nano Letters}\ }\textbf {\bibinfo {volume} {20}},\ \bibinfo {pages} {868} (\bibinfo {year} {2020})}\BibitemShut {NoStop}%
\bibitem [{\citenamefont {Tan}\ \emph {et~al.}(2018)\citenamefont {Tan}, \citenamefont {Lee}, \citenamefont {Jung}, \citenamefont {Park}, \citenamefont {Albarakati}, \citenamefont {Partridge}, \citenamefont {Field}, \citenamefont {McCulloch}, \citenamefont {Wang},\ and\ \citenamefont {Lee}}]{Tan2018}%
  \BibitemOpen
  \bibfield  {author} {\bibinfo {author} {\bibfnamefont {C.}~\bibnamefont {Tan}}, \bibinfo {author} {\bibfnamefont {J.}~\bibnamefont {Lee}}, \bibinfo {author} {\bibfnamefont {S.~G.}\ \bibnamefont {Jung}}, \bibinfo {author} {\bibfnamefont {T.}~\bibnamefont {Park}}, \bibinfo {author} {\bibfnamefont {S.}~\bibnamefont {Albarakati}}, \bibinfo {author} {\bibfnamefont {J.}~\bibnamefont {Partridge}}, \bibinfo {author} {\bibfnamefont {M.~R.}\ \bibnamefont {Field}}, \bibinfo {author} {\bibfnamefont {D.~G.}\ \bibnamefont {McCulloch}}, \bibinfo {author} {\bibfnamefont {L.}~\bibnamefont {Wang}},\ and\ \bibinfo {author} {\bibfnamefont {C.}~\bibnamefont {Lee}},\ }\bibfield  {title} {\bibinfo {title} {{Hard magnetic properties in nanoflake van der Waals Fe$_3$GeTe$_2$}},\ }\href {https://www.nature.com/articles/s41467-018-04018-w} {\bibfield  {journal} {\bibinfo  {journal} {Nature Communications}\ }\textbf {\bibinfo {volume} {9}},\ \bibinfo {pages} {1554} (\bibinfo {year} {2018})}\BibitemShut {NoStop}%
\bibitem [{\citenamefont {Hu}\ \emph {et~al.}(2020)\citenamefont {Hu}, \citenamefont {Zhao}, \citenamefont {Shen}, \citenamefont {Krasheninnikov}, \citenamefont {Chen},\ and\ \citenamefont {Sun}}]{Hu2020d}%
  \BibitemOpen
  \bibfield  {author} {\bibinfo {author} {\bibfnamefont {X.}~\bibnamefont {Hu}}, \bibinfo {author} {\bibfnamefont {Y.}~\bibnamefont {Zhao}}, \bibinfo {author} {\bibfnamefont {X.}~\bibnamefont {Shen}}, \bibinfo {author} {\bibfnamefont {A.~V.}\ \bibnamefont {Krasheninnikov}}, \bibinfo {author} {\bibfnamefont {Z.}~\bibnamefont {Chen}},\ and\ \bibinfo {author} {\bibfnamefont {L.}~\bibnamefont {Sun}},\ }\bibfield  {title} {\bibinfo {title} {{Enhanced ferromagnetism and tunable magnetism in Fe$_3$GeTe$_2$ monolayer by strain engineering}},\ }\href {https://pubs.acs.org/doi/10.1021/acsami.0c05530} {\bibfield  {journal} {\bibinfo  {journal} {ACS applied materials \& interfaces}\ }\textbf {\bibinfo {volume} {12}},\ \bibinfo {pages} {26367} (\bibinfo {year} {2020})}\BibitemShut {NoStop}%
\bibitem [{\citenamefont {Weerahennedige}\ \emph {et~al.}(2024)\citenamefont {Weerahennedige}, \citenamefont {Irziqat}, \citenamefont {Vithanage}, \citenamefont {Weerarathne}, \citenamefont {Ronau}, \citenamefont {Sumanasekera},\ and\ \citenamefont {Jasinski}}]{ThicknessF3GT}%
  \BibitemOpen
  \bibfield  {author} {\bibinfo {author} {\bibfnamefont {H.}~\bibnamefont {Weerahennedige}}, \bibinfo {author} {\bibfnamefont {M.}~\bibnamefont {Irziqat}}, \bibinfo {author} {\bibfnamefont {D.}~\bibnamefont {Vithanage}}, \bibinfo {author} {\bibfnamefont {H.}~\bibnamefont {Weerarathne}}, \bibinfo {author} {\bibfnamefont {Z.}~\bibnamefont {Ronau}}, \bibinfo {author} {\bibfnamefont {G.}~\bibnamefont {Sumanasekera}},\ and\ \bibinfo {author} {\bibfnamefont {J.~B.}\ \bibnamefont {Jasinski}},\ }\bibfield  {title} {\bibinfo {title} {{The effects of thickness, polarization, and strain on vibrational modes of 2D Fe$_3$GeTe$_2$}},\ }\href {https://doi.org/10.1016/j.surfin.2024.104797} {\bibfield  {journal} {\bibinfo  {journal} {Surfaces and Interfaces}\ }\textbf {\bibinfo {volume} {51}},\ \bibinfo {pages} {104797} (\bibinfo {year} {2024})}\BibitemShut {NoStop}%
\bibitem [{\citenamefont {Liu}\ \emph {et~al.}(2017)\citenamefont {Liu}, \citenamefont {Ivanovski},\ and\ \citenamefont {Petrovic}}]{Liu2017c}%
  \BibitemOpen
  \bibfield  {author} {\bibinfo {author} {\bibfnamefont {Y.}~\bibnamefont {Liu}}, \bibinfo {author} {\bibfnamefont {V.~N.}\ \bibnamefont {Ivanovski}},\ and\ \bibinfo {author} {\bibfnamefont {C.}~\bibnamefont {Petrovic}},\ }\bibfield  {title} {\bibinfo {title} {{Critical behavior of the van der Waals bonded ferromagnet Fe$_{3-x}$GeTe$_2$}},\ }\href {http://dx.doi.org/10.1103/PhysRevB.96.144429} {\bibfield  {journal} {\bibinfo  {journal} {Physical Review B}\ }\textbf {\bibinfo {volume} {96}} (\bibinfo {year} {2017})}\BibitemShut {NoStop}%
\bibitem [{\citenamefont {May}\ \emph {et~al.}(2016)\citenamefont {May}, \citenamefont {Calder}, \citenamefont {Cantoni}, \citenamefont {Cao},\ and\ \citenamefont {McGuire}}]{May2016}%
  \BibitemOpen
  \bibfield  {author} {\bibinfo {author} {\bibfnamefont {A.~F.}\ \bibnamefont {May}}, \bibinfo {author} {\bibfnamefont {S.}~\bibnamefont {Calder}}, \bibinfo {author} {\bibfnamefont {C.}~\bibnamefont {Cantoni}}, \bibinfo {author} {\bibfnamefont {H.}~\bibnamefont {Cao}},\ and\ \bibinfo {author} {\bibfnamefont {M.~A.}\ \bibnamefont {McGuire}},\ }\bibfield  {title} {\bibinfo {title} {{Magnetic structure and phase stability of the van der Waals bonded ferromagnet Fe$_{3-x}$GeTe$_2$}},\ }\href {http://dx.doi.org/10.1103/PhysRevB.93.014411} {\bibfield  {journal} {\bibinfo  {journal} {Physical Review B}\ }\textbf {\bibinfo {volume} {93}} (\bibinfo {year} {2016})}\BibitemShut {NoStop}%
\bibitem [{\citenamefont {Mayoh}\ \emph {et~al.}(2021)\citenamefont {Mayoh}, \citenamefont {Wood}, \citenamefont {Holt}, \citenamefont {Beckett}, \citenamefont {Dekker}, \citenamefont {Lees},\ and\ \citenamefont {Balakrishnan}}]{Mayoh2021}%
  \BibitemOpen
  \bibfield  {author} {\bibinfo {author} {\bibfnamefont {D.~A.}\ \bibnamefont {Mayoh}}, \bibinfo {author} {\bibfnamefont {G.~D.~A.}\ \bibnamefont {Wood}}, \bibinfo {author} {\bibfnamefont {S.~J.~R.}\ \bibnamefont {Holt}}, \bibinfo {author} {\bibfnamefont {G.}~\bibnamefont {Beckett}}, \bibinfo {author} {\bibfnamefont {E.~J.~L.}\ \bibnamefont {Dekker}}, \bibinfo {author} {\bibfnamefont {M.~R.}\ \bibnamefont {Lees}},\ and\ \bibinfo {author} {\bibfnamefont {G.}~\bibnamefont {Balakrishnan}},\ }\bibfield  {title} {\bibinfo {title} {{Effects of Fe Deficiency and Co Substitution in Polycrystalline and Single Crystals of Fe$_{3}$GeTe$_2$}},\ }\href {https://doi.org/10.1021/acs.cgd.1c00684} {\bibfield  {journal} {\bibinfo  {journal} {Crystal Growth \& Design}\ }\textbf {\bibinfo {volume} {21}},\ \bibinfo {pages} {6786} (\bibinfo {year} {2021})},\ \Eprint {https://arxiv.org/abs/https://doi.org/10.1021/acs.cgd.1c00684} {https://doi.org/10.1021/acs.cgd.1c00684} \BibitemShut {NoStop}%
\bibitem [{\citenamefont {Backes}\ \emph {et~al.}(2024)\citenamefont {Backes}, \citenamefont {Fujita}, \citenamefont {Veiga}, \citenamefont {Mayoh}, \citenamefont {Wood}, \citenamefont {Dhesi}, \citenamefont {Balakrishnan}, \citenamefont {van~der Laan},\ and\ \citenamefont {Hesjedal}}]{Backes2024}%
  \BibitemOpen
  \bibfield  {author} {\bibinfo {author} {\bibfnamefont {D.}~\bibnamefont {Backes}}, \bibinfo {author} {\bibfnamefont {R.}~\bibnamefont {Fujita}}, \bibinfo {author} {\bibfnamefont {L.~S.~I.}\ \bibnamefont {Veiga}}, \bibinfo {author} {\bibfnamefont {D.~A.}\ \bibnamefont {Mayoh}}, \bibinfo {author} {\bibfnamefont {G.~D.~A.}\ \bibnamefont {Wood}}, \bibinfo {author} {\bibfnamefont {S.~S.}\ \bibnamefont {Dhesi}}, \bibinfo {author} {\bibfnamefont {G.}~\bibnamefont {Balakrishnan}}, \bibinfo {author} {\bibfnamefont {G.}~\bibnamefont {van~der Laan}},\ and\ \bibinfo {author} {\bibfnamefont {T.}~\bibnamefont {Hesjedal}},\ }\bibfield  {title} {\bibinfo {title} {{Valence-state mixing and reduced magnetic moment in Fe$_{3-\delta}$GeTe$_2$ single crystals with varying Fe content probed by x-ray spectroscopy}},\ }\href {https://doi.org/10.1088/1361-6528/ad5e87} {\bibfield  {journal} {\bibinfo  {journal} {Nanotechnology}\ }\textbf {\bibinfo {volume} {35}},\ \bibinfo {pages} {395709} (\bibinfo {year} {2024})}\BibitemShut
  {NoStop}%
\bibitem [{\citenamefont {Bera}\ \emph {et~al.}(2023{\natexlab{b}})\citenamefont {Bera}, \citenamefont {Pradhan}, \citenamefont {Pal}, \citenamefont {Pal}, \citenamefont {Bera}, \citenamefont {Kalimuddin}, \citenamefont {Das}, \citenamefont {Roy}, \citenamefont {Afzal}, \citenamefont {Pal},\ and\ \citenamefont {Mondal}}]{Bera2023a}%
  \BibitemOpen
  \bibfield  {author} {\bibinfo {author} {\bibfnamefont {S.}~\bibnamefont {Bera}}, \bibinfo {author} {\bibfnamefont {S.~K.}\ \bibnamefont {Pradhan}}, \bibinfo {author} {\bibfnamefont {R.}~\bibnamefont {Pal}}, \bibinfo {author} {\bibfnamefont {B.}~\bibnamefont {Pal}}, \bibinfo {author} {\bibfnamefont {A.}~\bibnamefont {Bera}}, \bibinfo {author} {\bibfnamefont {S.}~\bibnamefont {Kalimuddin}}, \bibinfo {author} {\bibfnamefont {M.}~\bibnamefont {Das}}, \bibinfo {author} {\bibfnamefont {D.~S.}\ \bibnamefont {Roy}}, \bibinfo {author} {\bibfnamefont {H.}~\bibnamefont {Afzal}}, \bibinfo {author} {\bibfnamefont {A.~N.}\ \bibnamefont {Pal}},\ and\ \bibinfo {author} {\bibfnamefont {M.}~\bibnamefont {Mondal}},\ }\bibfield  {title} {\bibinfo {title} {{Enhanced coercivity and emergent spin-cluster-glass state in 2D ferromagnetic material, Fe$_3$GeTe$_2$}},\ }\href {https://doi.org/https://doi.org/10.1016/j.jmmm.2023.171052} {\bibfield  {journal} {\bibinfo  {journal} {Journal of Magnetism and Magnetic Materials}\ }\textbf
  {\bibinfo {volume} {583}},\ \bibinfo {pages} {171052} (\bibinfo {year} {2023}{\natexlab{b}})}\BibitemShut {NoStop}%
\bibitem [{\citenamefont {Yi}\ \emph {et~al.}(2017)\citenamefont {Yi}, \citenamefont {Zhuang}, \citenamefont {Zou}, \citenamefont {Wu}, \citenamefont {Cao}, \citenamefont {Tang}, \citenamefont {Calder}, \citenamefont {Kent}, \citenamefont {Mandrus},\ and\ \citenamefont {Gai}}]{Yi2017b}%
  \BibitemOpen
  \bibfield  {author} {\bibinfo {author} {\bibfnamefont {J.}~\bibnamefont {Yi}}, \bibinfo {author} {\bibfnamefont {H.}~\bibnamefont {Zhuang}}, \bibinfo {author} {\bibfnamefont {Q.}~\bibnamefont {Zou}}, \bibinfo {author} {\bibfnamefont {Z.}~\bibnamefont {Wu}}, \bibinfo {author} {\bibfnamefont {G.}~\bibnamefont {Cao}}, \bibinfo {author} {\bibfnamefont {S.}~\bibnamefont {Tang}}, \bibinfo {author} {\bibfnamefont {S.~A.}\ \bibnamefont {Calder}}, \bibinfo {author} {\bibfnamefont {P.~R.}\ \bibnamefont {Kent}}, \bibinfo {author} {\bibfnamefont {D.}~\bibnamefont {Mandrus}},\ and\ \bibinfo {author} {\bibfnamefont {Z.}~\bibnamefont {Gai}},\ }\bibfield  {title} {\bibinfo {title} {{Competing antiferromagnetism in a quasi-2D itinerant ferromagnet: Fe$_3$GeTe$_2$}},\ }\href {https://iopscience.iop.org/article/10.1088/2053-1583/4/1/011005} {\bibfield  {journal} {\bibinfo  {journal} {2D Materials}\ }\textbf {\bibinfo {volume} {4}},\ \bibinfo {pages} {011005} (\bibinfo {year} {2017})}\BibitemShut {NoStop}%
\bibitem [{\citenamefont {Farle}(1998)}]{Farle1998}%
  \BibitemOpen
  \bibfield  {author} {\bibinfo {author} {\bibfnamefont {M.}~\bibnamefont {Farle}},\ }\bibfield  {title} {\bibinfo {title} {{Ferromagnetic resonance of ultrathin metallic layers}},\ }\href {https://iopscience.iop.org/article/10.1088/0034-4885/61/7/001} {\bibfield  {journal} {\bibinfo  {journal} {Reports on Progress in Physics}\ }\textbf {\bibinfo {volume} {61}},\ \bibinfo {pages} {755} (\bibinfo {year} {1998})}\BibitemShut {NoStop}%
\bibitem [{\citenamefont {Abragam}\ and\ \citenamefont {Bleaney}(2012)}]{Abragam2012}%
  \BibitemOpen
  \bibfield  {author} {\bibinfo {author} {\bibfnamefont {A.}~\bibnamefont {Abragam}}\ and\ \bibinfo {author} {\bibfnamefont {B.}~\bibnamefont {Bleaney}},\ }\href@noop {} {\emph {\bibinfo {title} {{Electron paramagnetic resonance of transition ions}}}}\ (\bibinfo  {publisher} {Oxford University Press, Oxford},\ \bibinfo {year} {2012})\BibitemShut {NoStop}%
\bibitem [{\citenamefont {Turov}(1965)}]{Turov}%
  \BibitemOpen
  \bibfield  {author} {\bibinfo {author} {\bibfnamefont {E.~A.}\ \bibnamefont {Turov}},\ }\href@noop {} {\emph {\bibinfo {title} {Physical Properties of Magnetically Ordered Crystals}}},\ edited by\ \bibinfo {editor} {\bibfnamefont {A.}~\bibnamefont {Tybulewicz}}\ and\ \bibinfo {editor} {\bibfnamefont {S.}~\bibnamefont {Chomet}}\ (\bibinfo  {publisher} {Academic Press},\ \bibinfo {address} {New York},\ \bibinfo {year} {1965})\BibitemShut {NoStop}%
\bibitem [{\citenamefont {Holstein}\ and\ \citenamefont {Primakoff}(1940)}]{Holstein1940}%
  \BibitemOpen
  \bibfield  {author} {\bibinfo {author} {\bibfnamefont {T.}~\bibnamefont {Holstein}}\ and\ \bibinfo {author} {\bibfnamefont {H.}~\bibnamefont {Primakoff}},\ }\bibfield  {title} {\bibinfo {title} {{Field Dependence of the Intrinsic Domain Magnetization of a Ferromagnet}},\ }\href {https://doi.org/10.1103/PhysRev.58.1098} {\bibfield  {journal} {\bibinfo  {journal} {Phys. Rev.}\ }\textbf {\bibinfo {volume} {58}},\ \bibinfo {pages} {1098} (\bibinfo {year} {1940})}\BibitemShut {NoStop}%
\bibitem [{\citenamefont {Yang}\ \emph {et~al.}(2025)\citenamefont {Yang}, \citenamefont {Meng}, \citenamefont {Liu}, \citenamefont {Li}, \citenamefont {Zhang}, \citenamefont {Fan}, \citenamefont {Ma}, \citenamefont {Ge}, \citenamefont {Pi}, \citenamefont {Qu},\ and\ \citenamefont {Zhang}}]{Yang2025}%
  \BibitemOpen
  \bibfield  {author} {\bibinfo {author} {\bibfnamefont {Y.}~\bibnamefont {Yang}}, \bibinfo {author} {\bibfnamefont {F.}~\bibnamefont {Meng}}, \bibinfo {author} {\bibfnamefont {W.}~\bibnamefont {Liu}}, \bibinfo {author} {\bibfnamefont {J.}~\bibnamefont {Li}}, \bibinfo {author} {\bibfnamefont {J.}~\bibnamefont {Zhang}}, \bibinfo {author} {\bibfnamefont {J.}~\bibnamefont {Fan}}, \bibinfo {author} {\bibfnamefont {C.}~\bibnamefont {Ma}}, \bibinfo {author} {\bibfnamefont {M.}~\bibnamefont {Ge}}, \bibinfo {author} {\bibfnamefont {L.}~\bibnamefont {Pi}}, \bibinfo {author} {\bibfnamefont {Z.}~\bibnamefont {Qu}},\ and\ \bibinfo {author} {\bibfnamefont {L.}~\bibnamefont {Zhang}},\ }\bibfield  {title} {\bibinfo {title} {{Electron spin resonance study of the van der Waals ferromagnet Fe$_5$GeTe$_2$}},\ }\href {https://doi.org/10.1016/j.jmmm.2025.173251} {\bibfield  {journal} {\bibinfo  {journal} {Journal of Magnetism and Magnetic Materials}\ }\textbf {\bibinfo {volume} {629}},\ \bibinfo {pages} {173251} (\bibinfo {year}
  {2025})}\BibitemShut {NoStop}%
\bibitem [{\citenamefont {Jonak}\ \emph {et~al.}(2022)\citenamefont {Jonak}, \citenamefont {Walendy}, \citenamefont {Arneth}, \citenamefont {Abdel-Hafiez},\ and\ \citenamefont {Klingeler}}]{Jonak2022}%
  \BibitemOpen
  \bibfield  {author} {\bibinfo {author} {\bibfnamefont {M.}~\bibnamefont {Jonak}}, \bibinfo {author} {\bibfnamefont {E.}~\bibnamefont {Walendy}}, \bibinfo {author} {\bibfnamefont {J.}~\bibnamefont {Arneth}}, \bibinfo {author} {\bibfnamefont {M.}~\bibnamefont {Abdel-Hafiez}},\ and\ \bibinfo {author} {\bibfnamefont {R.}~\bibnamefont {Klingeler}},\ }\bibfield  {title} {\bibinfo {title} {{Low-energy magnon excitations and emerging anisotropic nature of short-range order in CrI$_3$}},\ }\href {https://doi.org/10.1103/PhysRevB.106.214412} {\bibfield  {journal} {\bibinfo  {journal} {Physical Review B}\ }\textbf {\bibinfo {volume} {106}},\ \bibinfo {pages} {214412} (\bibinfo {year} {2022})}\BibitemShut {NoStop}%
\bibitem [{\citenamefont {Dillon~Jr}\ and\ \citenamefont {Olson}(1965)}]{Dillon1965}%
  \BibitemOpen
  \bibfield  {author} {\bibinfo {author} {\bibfnamefont {J.}~\bibnamefont {Dillon~Jr}}\ and\ \bibinfo {author} {\bibfnamefont {C.}~\bibnamefont {Olson}},\ }\bibfield  {title} {\bibinfo {title} {{Magnetization, resonance, and optical properties of the ferromagnet CrI$_3$}},\ }\href {https://doi.org/https://doi.org/10.1063/1.1714194} {\bibfield  {journal} {\bibinfo  {journal} {Journal of Applied Physics}\ }\textbf {\bibinfo {volume} {36}},\ \bibinfo {pages} {1259} (\bibinfo {year} {1965})}\BibitemShut {NoStop}%
\bibitem [{\citenamefont {Li}\ \emph {et~al.}(2022)\citenamefont {Li}, \citenamefont {Xu}, \citenamefont {Li}, \citenamefont {Liao}, \citenamefont {Xi}, \citenamefont {Yu},\ and\ \citenamefont {Wang}}]{Li2022anomalous}%
  \BibitemOpen
  \bibfield  {author} {\bibinfo {author} {\bibfnamefont {Z.}~\bibnamefont {Li}}, \bibinfo {author} {\bibfnamefont {D.-H.}\ \bibnamefont {Xu}}, \bibinfo {author} {\bibfnamefont {X.}~\bibnamefont {Li}}, \bibinfo {author} {\bibfnamefont {H.-J.}\ \bibnamefont {Liao}}, \bibinfo {author} {\bibfnamefont {X.}~\bibnamefont {Xi}}, \bibinfo {author} {\bibfnamefont {Y.-C.}\ \bibnamefont {Yu}},\ and\ \bibinfo {author} {\bibfnamefont {W.}~\bibnamefont {Wang}},\ }\bibfield  {title} {\bibinfo {title} {Anomalous spin dynamics in a two-dimensional magnet induced by anisotropic critical fluctuations},\ }\href {https://link.aps.org/doi/10.1103/PhysRevB.106.054427} {\bibfield  {journal} {\bibinfo  {journal} {Physical Review B}\ }\textbf {\bibinfo {volume} {106}},\ \bibinfo {pages} {054427} (\bibinfo {year} {2022})}\BibitemShut {NoStop}%
\bibitem [{\citenamefont {Khan}\ \emph {et~al.}(2019)\citenamefont {Khan}, \citenamefont {Zollitsch}, \citenamefont {Arroo}, \citenamefont {Cheng}, \citenamefont {Verzhbitskiy}, \citenamefont {Sud}, \citenamefont {Feng}, \citenamefont {Eda},\ and\ \citenamefont {Kurebayashi}}]{khan2019}%
  \BibitemOpen
  \bibfield  {author} {\bibinfo {author} {\bibfnamefont {S.}~\bibnamefont {Khan}}, \bibinfo {author} {\bibfnamefont {C.}~\bibnamefont {Zollitsch}}, \bibinfo {author} {\bibfnamefont {D.}~\bibnamefont {Arroo}}, \bibinfo {author} {\bibfnamefont {H.}~\bibnamefont {Cheng}}, \bibinfo {author} {\bibfnamefont {I.}~\bibnamefont {Verzhbitskiy}}, \bibinfo {author} {\bibfnamefont {A.}~\bibnamefont {Sud}}, \bibinfo {author} {\bibfnamefont {Y.}~\bibnamefont {Feng}}, \bibinfo {author} {\bibfnamefont {G.}~\bibnamefont {Eda}},\ and\ \bibinfo {author} {\bibfnamefont {H.}~\bibnamefont {Kurebayashi}},\ }\bibfield  {title} {\bibinfo {title} {{Spin dynamics study in layered van der Waals single-crystal Cr$_2$Ge$_2$Te$_6$}},\ }\href {https://doi.org/https://doi.org/10.1103/PhysRevB.100.134437} {\bibfield  {journal} {\bibinfo  {journal} {Physical Review B}\ }\textbf {\bibinfo {volume} {100}},\ \bibinfo {pages} {134437} (\bibinfo {year} {2019})}\BibitemShut {NoStop}%
\bibitem [{\citenamefont {Zeisner}\ \emph {et~al.}(2019)\citenamefont {Zeisner}, \citenamefont {Alfonsov}, \citenamefont {Selter}, \citenamefont {Aswartham}, \citenamefont {Ghimire}, \citenamefont {Richter}, \citenamefont {Van Den~Brink}, \citenamefont {B{\"{u}}chner},\ and\ \citenamefont {Kataev}}]{Zeisner2018}%
  \BibitemOpen
  \bibfield  {author} {\bibinfo {author} {\bibfnamefont {J.}~\bibnamefont {Zeisner}}, \bibinfo {author} {\bibfnamefont {A.}~\bibnamefont {Alfonsov}}, \bibinfo {author} {\bibfnamefont {S.}~\bibnamefont {Selter}}, \bibinfo {author} {\bibfnamefont {S.}~\bibnamefont {Aswartham}}, \bibinfo {author} {\bibfnamefont {M.~P.}\ \bibnamefont {Ghimire}}, \bibinfo {author} {\bibfnamefont {M.}~\bibnamefont {Richter}}, \bibinfo {author} {\bibfnamefont {J.}~\bibnamefont {Van Den~Brink}}, \bibinfo {author} {\bibfnamefont {B.}~\bibnamefont {B{\"{u}}chner}},\ and\ \bibinfo {author} {\bibfnamefont {V.}~\bibnamefont {Kataev}},\ }\bibfield  {title} {\bibinfo {title} {{Magnetic anisotropy and spin-polarized two-dimensional electron gas in the van der Waals ferromagnet Cr$_2$Ge$_2$Te$_6$}},\ }\href {https://link.aps.org/doi/10.1103/PhysRevB.99.165109} {\bibfield  {journal} {\bibinfo  {journal} {Physical Review B}\ }\textbf {\bibinfo {volume} {99}},\ \bibinfo {pages} {165109} (\bibinfo {year} {2019})}\BibitemShut {NoStop}%
\bibitem [{\citenamefont {Wang}\ \emph {et~al.}(2023{\natexlab{b}})\citenamefont {Wang}, \citenamefont {Fu}, \citenamefont {Ji}, \citenamefont {Yin}, \citenamefont {Chen}, \citenamefont {Liu},\ and\ \citenamefont {Li}}]{wang2023esr}%
  \BibitemOpen
  \bibfield  {author} {\bibinfo {author} {\bibfnamefont {Y.}~\bibnamefont {Wang}}, \bibinfo {author} {\bibfnamefont {Q.}~\bibnamefont {Fu}}, \bibinfo {author} {\bibfnamefont {S.}~\bibnamefont {Ji}}, \bibinfo {author} {\bibfnamefont {X.}~\bibnamefont {Yin}}, \bibinfo {author} {\bibfnamefont {Y.}~\bibnamefont {Chen}}, \bibinfo {author} {\bibfnamefont {R.}~\bibnamefont {Liu}},\ and\ \bibinfo {author} {\bibfnamefont {X.}~\bibnamefont {Li}},\ }\bibfield  {title} {\bibinfo {title} {{Temperature-dependent magnetic order of two-dimensional ferromagnetic Cr$_2$Ge$_2$Te$_6$ single crystal}},\ }\href {https://doi.org/https://doi.org/10.1016/j.jallcom.2022.167401} {\bibfield  {journal} {\bibinfo  {journal} {Journal of Alloys and Compounds}\ }\textbf {\bibinfo {volume} {931}},\ \bibinfo {pages} {167401} (\bibinfo {year} {2023}{\natexlab{b}})}\BibitemShut {NoStop}%
\bibitem [{\citenamefont {Osborn}(1945)}]{Osborn1945}%
  \BibitemOpen
  \bibfield  {author} {\bibinfo {author} {\bibfnamefont {J.~A.}\ \bibnamefont {Osborn}},\ }\bibfield  {title} {\bibinfo {title} {Demagnetizing factors of the general ellipsoid},\ }\href {https://doi.org/10.1103/PhysRev.67.351} {\bibfield  {journal} {\bibinfo  {journal} {Phys. Rev.}\ }\textbf {\bibinfo {volume} {67}},\ \bibinfo {pages} {351} (\bibinfo {year} {1945})}\BibitemShut {NoStop}%
\bibitem [{\citenamefont {Cronemeyer}(1991)}]{Cronemeyer1991}%
  \BibitemOpen
  \bibfield  {author} {\bibinfo {author} {\bibfnamefont {D.~C.}\ \bibnamefont {Cronemeyer}},\ }\bibfield  {title} {\bibinfo {title} {Demagnetization factors for general ellipsoids},\ }\href {https://doi.org/10.1063/1.349315} {\bibfield  {journal} {\bibinfo  {journal} {J. Appl. Phys.}\ }\textbf {\bibinfo {volume} {70}},\ \bibinfo {pages} {2911} (\bibinfo {year} {1991})}\BibitemShut {NoStop}%
\bibitem [{\citenamefont {Cullity}\ and\ \citenamefont {Graham}(2011)}]{Cullity}%
  \BibitemOpen
  \bibfield  {author} {\bibinfo {author} {\bibfnamefont {B.~D.}\ \bibnamefont {Cullity}}\ and\ \bibinfo {author} {\bibfnamefont {C.~D.}\ \bibnamefont {Graham}},\ }\href@noop {} {\emph {\bibinfo {title} {Introduction to magnetic materials}}}\ (\bibinfo  {publisher} {John Wiley \& Sons},\ \bibinfo {year} {2011})\BibitemShut {NoStop}%
\bibitem [{\citenamefont {Zhu}\ \emph {et~al.}(2016)\citenamefont {Zhu}, \citenamefont {Janoschek}, \citenamefont {Chaves}, \citenamefont {Cezar}, \citenamefont {Durakiewicz}, \citenamefont {Ronning}, \citenamefont {Sassa}, \citenamefont {Mansson}, \citenamefont {Scott}, \citenamefont {Wakeham}, \citenamefont {Bauer},\ and\ \citenamefont {Thompson}}]{Zhu2016}%
  \BibitemOpen
  \bibfield  {author} {\bibinfo {author} {\bibfnamefont {J.-X.}\ \bibnamefont {Zhu}}, \bibinfo {author} {\bibfnamefont {M.}~\bibnamefont {Janoschek}}, \bibinfo {author} {\bibfnamefont {D.~S.}\ \bibnamefont {Chaves}}, \bibinfo {author} {\bibfnamefont {J.~C.}\ \bibnamefont {Cezar}}, \bibinfo {author} {\bibfnamefont {T.}~\bibnamefont {Durakiewicz}}, \bibinfo {author} {\bibfnamefont {F.}~\bibnamefont {Ronning}}, \bibinfo {author} {\bibfnamefont {Y.}~\bibnamefont {Sassa}}, \bibinfo {author} {\bibfnamefont {M.}~\bibnamefont {Mansson}}, \bibinfo {author} {\bibfnamefont {B.~L.}\ \bibnamefont {Scott}}, \bibinfo {author} {\bibfnamefont {N.}~\bibnamefont {Wakeham}}, \bibinfo {author} {\bibfnamefont {E.~D.}\ \bibnamefont {Bauer}},\ and\ \bibinfo {author} {\bibfnamefont {J.~D.}\ \bibnamefont {Thompson}},\ }\bibfield  {title} {\bibinfo {title} {{Electronic correlation and magnetism in the ferromagnetic metal Fe$_3$GeTe$_2$}},\ }\href {https://link.aps.org/doi/10.1103/PhysRevB.93.144404} {\bibfield  {journal} {\bibinfo
  {journal} {Physical Review B}\ }\textbf {\bibinfo {volume} {93}},\ \bibinfo {pages} {144404} (\bibinfo {year} {2016})}\BibitemShut {NoStop}%
\bibitem [{\citenamefont {Liu}\ \emph {et~al.}(2022)\citenamefont {Liu}, \citenamefont {Xing}, \citenamefont {Jiang}, \citenamefont {Guo}, \citenamefont {Jiang}, \citenamefont {Qi},\ and\ \citenamefont {Zhao}}]{Liu2022}%
  \BibitemOpen
  \bibfield  {author} {\bibinfo {author} {\bibfnamefont {Q.}~\bibnamefont {Liu}}, \bibinfo {author} {\bibfnamefont {J.}~\bibnamefont {Xing}}, \bibinfo {author} {\bibfnamefont {Z.}~\bibnamefont {Jiang}}, \bibinfo {author} {\bibfnamefont {Y.}~\bibnamefont {Guo}}, \bibinfo {author} {\bibfnamefont {X.}~\bibnamefont {Jiang}}, \bibinfo {author} {\bibfnamefont {Y.}~\bibnamefont {Qi}},\ and\ \bibinfo {author} {\bibfnamefont {J.}~\bibnamefont {Zhao}},\ }\bibfield  {title} {\bibinfo {title} {{Layer-dependent magnetic phase diagram in Fe$_n$GeTe$_2$ (3 $\leq$ n $\leq$ 7) ultrathin films}},\ }\href {https://doi.org/10.1038/s42005-022-00921-3} {\bibfield  {journal} {\bibinfo  {journal} {Communications Physics}\ }\textbf {\bibinfo {volume} {5}},\ \bibinfo {pages} {140} (\bibinfo {year} {2022})}\BibitemShut {NoStop}%
\bibitem [{\citenamefont {Calder}\ \emph {et~al.}(2019)\citenamefont {Calder}, \citenamefont {Kolesnikov},\ and\ \citenamefont {May}}]{Calder2019}%
  \BibitemOpen
  \bibfield  {author} {\bibinfo {author} {\bibfnamefont {S.}~\bibnamefont {Calder}}, \bibinfo {author} {\bibfnamefont {A.~I.}\ \bibnamefont {Kolesnikov}},\ and\ \bibinfo {author} {\bibfnamefont {A.~F.}\ \bibnamefont {May}},\ }\bibfield  {title} {\bibinfo {title} {{Magnetic excitations in the quasi-two-dimensional ferromagnet Fe$_{3-x}$GeTe$_2$ measured with inelastic neutron scattering}},\ }\href {https://doi.org/10.1103/PhysRevB.99.094423} {\bibfield  {journal} {\bibinfo  {journal} {Physical Review B}\ }\textbf {\bibinfo {volume} {99}},\ \bibinfo {pages} {094423} (\bibinfo {year} {2019})}\BibitemShut {NoStop}%
\bibitem [{\citenamefont {Trainer}\ \emph {et~al.}(2022)\citenamefont {Trainer}, \citenamefont {Armitage}, \citenamefont {Lane}, \citenamefont {Rhodes}, \citenamefont {Chan}, \citenamefont {Benedi{\v{c}}i{\v{c}}}, \citenamefont {Rodriguez-Rivera}, \citenamefont {Fabelo}, \citenamefont {Stock},\ and\ \citenamefont {Wahl}}]{Trainer2022}%
  \BibitemOpen
  \bibfield  {author} {\bibinfo {author} {\bibfnamefont {C.}~\bibnamefont {Trainer}}, \bibinfo {author} {\bibfnamefont {O.~R.}\ \bibnamefont {Armitage}}, \bibinfo {author} {\bibfnamefont {H.}~\bibnamefont {Lane}}, \bibinfo {author} {\bibfnamefont {L.~C.}\ \bibnamefont {Rhodes}}, \bibinfo {author} {\bibfnamefont {E.}~\bibnamefont {Chan}}, \bibinfo {author} {\bibfnamefont {I.}~\bibnamefont {Benedi{\v{c}}i{\v{c}}}}, \bibinfo {author} {\bibfnamefont {J.~A.}\ \bibnamefont {Rodriguez-Rivera}}, \bibinfo {author} {\bibfnamefont {O.}~\bibnamefont {Fabelo}}, \bibinfo {author} {\bibfnamefont {C.}~\bibnamefont {Stock}},\ and\ \bibinfo {author} {\bibfnamefont {P.}~\bibnamefont {Wahl}},\ }\bibfield  {title} {\bibinfo {title} {{Relating spin-polarized STM imaging and inelastic neutron scattering in the van der Waals ferromagnet Fe$_{3}$GeTe$_2$}},\ }\href {https://doi.org/https://doi.org/10.1103/PhysRevB.106.L081405} {\bibfield  {journal} {\bibinfo  {journal} {Physical Review B}\ }\textbf {\bibinfo {volume} {106}},\
  \bibinfo {pages} {L081405} (\bibinfo {year} {2022})}\BibitemShut {NoStop}%
\bibitem [{\citenamefont {Bao}\ \emph {et~al.}(2023)\citenamefont {Bao}, \citenamefont {Wang}, \citenamefont {Yano}, \citenamefont {Shangguan}, \citenamefont {Huang}, \citenamefont {Liao}, \citenamefont {Wang}, \citenamefont {Gao}, \citenamefont {Zhang}, \citenamefont {Cheng} \emph {et~al.}}]{Bao2023}%
  \BibitemOpen
  \bibfield  {author} {\bibinfo {author} {\bibfnamefont {S.}~\bibnamefont {Bao}}, \bibinfo {author} {\bibfnamefont {J.}~\bibnamefont {Wang}}, \bibinfo {author} {\bibfnamefont {S.-i.}\ \bibnamefont {Yano}}, \bibinfo {author} {\bibfnamefont {Y.}~\bibnamefont {Shangguan}}, \bibinfo {author} {\bibfnamefont {Z.}~\bibnamefont {Huang}}, \bibinfo {author} {\bibfnamefont {J.}~\bibnamefont {Liao}}, \bibinfo {author} {\bibfnamefont {W.}~\bibnamefont {Wang}}, \bibinfo {author} {\bibfnamefont {Y.}~\bibnamefont {Gao}}, \bibinfo {author} {\bibfnamefont {B.}~\bibnamefont {Zhang}}, \bibinfo {author} {\bibfnamefont {S.}~\bibnamefont {Cheng}}, \emph {et~al.},\ }\bibfield  {title} {\bibinfo {title} {{Observation of Magnon Damping Minimum Induced by Kondo Coupling in a van der Waals Ferromagnet Fe$_{3-x}$GeTe$_2$}},\ }\href {https://doi.org/10.48550/arXiv.2312.15961} {\bibfield  {journal} {\bibinfo  {journal} {arXiv preprint arXiv:2312.15961}\ } (\bibinfo {year} {2023})}\BibitemShut {NoStop}%
\bibitem [{\citenamefont {Bai}\ \emph {et~al.}(2022)\citenamefont {Bai}, \citenamefont {Lechermann}, \citenamefont {Liu}, \citenamefont {Cheng}, \citenamefont {Kolesnikov}, \citenamefont {Ye}, \citenamefont {Williams}, \citenamefont {Chi}, \citenamefont {Hong}, \citenamefont {Granroth} \emph {et~al.}}]{Bai2022}%
  \BibitemOpen
  \bibfield  {author} {\bibinfo {author} {\bibfnamefont {X.}~\bibnamefont {Bai}}, \bibinfo {author} {\bibfnamefont {F.}~\bibnamefont {Lechermann}}, \bibinfo {author} {\bibfnamefont {Y.}~\bibnamefont {Liu}}, \bibinfo {author} {\bibfnamefont {Y.}~\bibnamefont {Cheng}}, \bibinfo {author} {\bibfnamefont {A.~I.}\ \bibnamefont {Kolesnikov}}, \bibinfo {author} {\bibfnamefont {F.}~\bibnamefont {Ye}}, \bibinfo {author} {\bibfnamefont {T.~J.}\ \bibnamefont {Williams}}, \bibinfo {author} {\bibfnamefont {S.}~\bibnamefont {Chi}}, \bibinfo {author} {\bibfnamefont {T.}~\bibnamefont {Hong}}, \bibinfo {author} {\bibfnamefont {G.~E.}\ \bibnamefont {Granroth}}, \emph {et~al.},\ }\bibfield  {title} {\bibinfo {title} {{Antiferromagnetic fluctuations and orbital-selective Mott transition in the van der Waals ferromagnet Fe$_{3-x}$GeTe$_2$}},\ }\href {https://doi.org/10.1103/PhysRevB.106.L180409} {\bibfield  {journal} {\bibinfo  {journal} {Physical Review B}\ }\textbf {\bibinfo {volume} {106}},\ \bibinfo {pages} {L180409} (\bibinfo
  {year} {2022})}\BibitemShut {NoStop}%
\bibitem [{\citenamefont {Bao}\ \emph {et~al.}(2022)\citenamefont {Bao}, \citenamefont {Wang}, \citenamefont {Shangguan}, \citenamefont {Cai}, \citenamefont {Dong}, \citenamefont {Huang}, \citenamefont {Si}, \citenamefont {Ma}, \citenamefont {Kajimoto}, \citenamefont {Ikeuchi} \emph {et~al.}}]{Bao2022neutron}%
  \BibitemOpen
  \bibfield  {author} {\bibinfo {author} {\bibfnamefont {S.}~\bibnamefont {Bao}}, \bibinfo {author} {\bibfnamefont {W.}~\bibnamefont {Wang}}, \bibinfo {author} {\bibfnamefont {Y.}~\bibnamefont {Shangguan}}, \bibinfo {author} {\bibfnamefont {Z.}~\bibnamefont {Cai}}, \bibinfo {author} {\bibfnamefont {Z.-Y.}\ \bibnamefont {Dong}}, \bibinfo {author} {\bibfnamefont {Z.}~\bibnamefont {Huang}}, \bibinfo {author} {\bibfnamefont {W.}~\bibnamefont {Si}}, \bibinfo {author} {\bibfnamefont {Z.}~\bibnamefont {Ma}}, \bibinfo {author} {\bibfnamefont {R.}~\bibnamefont {Kajimoto}}, \bibinfo {author} {\bibfnamefont {K.}~\bibnamefont {Ikeuchi}}, \emph {et~al.},\ }\bibfield  {title} {\bibinfo {title} {{Neutron spectroscopy evidence on the dual nature of magnetic excitations in a van der Waals metallic ferromagnet Fe$_{2.72}$GeTe$_2$}},\ }\href {https://doi.org/10.1103/PhysRevX.12.011022} {\bibfield  {journal} {\bibinfo  {journal} {Physical Review X}\ }\textbf {\bibinfo {volume} {12}},\ \bibinfo {pages} {011022} (\bibinfo {year}
  {2022})}\BibitemShut {NoStop}%
\bibitem [{\citenamefont {Corasaniti}\ \emph {et~al.}(2020)\citenamefont {Corasaniti}, \citenamefont {Yang}, \citenamefont {Sen}, \citenamefont {Willa}, \citenamefont {Merz}, \citenamefont {Haghighirad}, \citenamefont {Le~Tacon},\ and\ \citenamefont {Degiorgi}}]{cora2020}%
  \BibitemOpen
  \bibfield  {author} {\bibinfo {author} {\bibfnamefont {M.}~\bibnamefont {Corasaniti}}, \bibinfo {author} {\bibfnamefont {R.}~\bibnamefont {Yang}}, \bibinfo {author} {\bibfnamefont {K.}~\bibnamefont {Sen}}, \bibinfo {author} {\bibfnamefont {K.}~\bibnamefont {Willa}}, \bibinfo {author} {\bibfnamefont {M.}~\bibnamefont {Merz}}, \bibinfo {author} {\bibfnamefont {A.~A.}\ \bibnamefont {Haghighirad}}, \bibinfo {author} {\bibfnamefont {M.}~\bibnamefont {Le~Tacon}},\ and\ \bibinfo {author} {\bibfnamefont {L.}~\bibnamefont {Degiorgi}},\ }\bibfield  {title} {\bibinfo {title} {{Electronic correlations in the van der Waals ferromagnet Fe$_3$GeTe$_2$ revealed by its charge dynamics}},\ }\href {https://doi.org/https://doi.org/10.1103/PhysRevB.102.161109} {\bibfield  {journal} {\bibinfo  {journal} {Physical Review B}\ }\textbf {\bibinfo {volume} {102}},\ \bibinfo {pages} {161109} (\bibinfo {year} {2020})}\BibitemShut {NoStop}%
\bibitem [{\citenamefont {Liu}\ \emph {et~al.}(2024)\citenamefont {Liu}, \citenamefont {Zhou}, \citenamefont {Li}, \citenamefont {Zhao},\ and\ \citenamefont {Tang}}]{Liu2024}%
  \BibitemOpen
  \bibfield  {author} {\bibinfo {author} {\bibfnamefont {J.}~\bibnamefont {Liu}}, \bibinfo {author} {\bibfnamefont {L.}~\bibnamefont {Zhou}}, \bibinfo {author} {\bibfnamefont {S.}~\bibnamefont {Li}}, \bibinfo {author} {\bibfnamefont {E.}~\bibnamefont {Zhao}},\ and\ \bibinfo {author} {\bibfnamefont {N.}~\bibnamefont {Tang}},\ }\bibfield  {title} {\bibinfo {title} {{Anomalous temperature dependence of uniaxial magnetocrystalline anisotropy in the van der Waals ferromagnet Fe$_3$GeTe$_2$}},\ }\href {https://doi.org/10.1103/PhysRevB.109.L060404} {\bibfield  {journal} {\bibinfo  {journal} {Physical Review B}\ }\textbf {\bibinfo {volume} {109}},\ \bibinfo {pages} {L060404} (\bibinfo {year} {2024})}\BibitemShut {NoStop}%
\bibitem [{\citenamefont {Shu}\ \emph {et~al.}(2024)\citenamefont {Shu}, \citenamefont {Zhang},\ and\ \citenamefont {Kong}}]{Shu2024}%
  \BibitemOpen
  \bibfield  {author} {\bibinfo {author} {\bibfnamefont {Z.}~\bibnamefont {Shu}}, \bibinfo {author} {\bibfnamefont {S.}~\bibnamefont {Zhang}},\ and\ \bibinfo {author} {\bibfnamefont {T.}~\bibnamefont {Kong}},\ }\bibfield  {title} {\bibinfo {title} {{Spin stiffness and spin excitation gap of van der Waals ferromagnetic Fe$_{3+\delta}$GeTe$_2$}},\ }\href {https://doi.org/10.1088/1361-648X/ad581f} {\bibfield  {journal} {\bibinfo  {journal} {Journal of Physics: Condensed Matter}\ }\textbf {\bibinfo {volume} {36}},\ \bibinfo {pages} {385801} (\bibinfo {year} {2024})}\BibitemShut {NoStop}%
\bibitem [{\citenamefont {Johnson}\ \emph {et~al.}(1996)\citenamefont {Johnson}, \citenamefont {Bloemen}, \citenamefont {den Broeder},\ and\ \citenamefont {de~Vries}}]{Johnson1996}%
  \BibitemOpen
  \bibfield  {author} {\bibinfo {author} {\bibfnamefont {M.~T.}\ \bibnamefont {Johnson}}, \bibinfo {author} {\bibfnamefont {P.~J.~H.}\ \bibnamefont {Bloemen}}, \bibinfo {author} {\bibfnamefont {F.~J.~A.}\ \bibnamefont {den Broeder}},\ and\ \bibinfo {author} {\bibfnamefont {J.~J.}\ \bibnamefont {de~Vries}},\ }\bibfield  {title} {\bibinfo {title} {Magnetic anisotropy in metallic multilayers},\ }\href {https://doi.org/10.1088/0034-4885/59/11/002} {\bibfield  {journal} {\bibinfo  {journal} {Reports on Progress in Physics}\ }\textbf {\bibinfo {volume} {59}},\ \bibinfo {pages} {1409} (\bibinfo {year} {1996})}\BibitemShut {NoStop}%
\bibitem [{\citenamefont {Williams}(1937)}]{Williams1937}%
  \BibitemOpen
  \bibfield  {author} {\bibinfo {author} {\bibfnamefont {H.}~\bibnamefont {Williams}},\ }\bibfield  {title} {\bibinfo {title} {Magnetic properties of single crystals of silicon iron},\ }\href {https://doi.org/https://doi.org/10.1103/PhysRev.52.747} {\bibfield  {journal} {\bibinfo  {journal} {Physical Review}\ }\textbf {\bibinfo {volume} {52}},\ \bibinfo {pages} {747} (\bibinfo {year} {1937})}\BibitemShut {NoStop}%
\bibitem [{\citenamefont {Huang}\ \emph {et~al.}(2025)\citenamefont {Huang}, \citenamefont {Zuo}, \citenamefont {Zhang}, \citenamefont {Xing}, \citenamefont {Yao}, \citenamefont {Zhang}, \citenamefont {Ma}, \citenamefont {Xu}, \citenamefont {Jiao}, \citenamefont {Zhou}, \citenamefont {Sankar}, \citenamefont {Qian},\ and\ \citenamefont {Xu}}]{Huang2025}%
  \BibitemOpen
  \bibfield  {author} {\bibinfo {author} {\bibfnamefont {Y.}~\bibnamefont {Huang}}, \bibinfo {author} {\bibfnamefont {N.}~\bibnamefont {Zuo}}, \bibinfo {author} {\bibfnamefont {Z.}~\bibnamefont {Zhang}}, \bibinfo {author} {\bibfnamefont {X.}~\bibnamefont {Xing}}, \bibinfo {author} {\bibfnamefont {X.}~\bibnamefont {Yao}}, \bibinfo {author} {\bibfnamefont {A.}~\bibnamefont {Zhang}}, \bibinfo {author} {\bibfnamefont {H.}~\bibnamefont {Ma}}, \bibinfo {author} {\bibfnamefont {C.}~\bibnamefont {Xu}}, \bibinfo {author} {\bibfnamefont {W.}~\bibnamefont {Jiao}}, \bibinfo {author} {\bibfnamefont {W.}~\bibnamefont {Zhou}}, \bibinfo {author} {\bibfnamefont {R.}~\bibnamefont {Sankar}}, \bibinfo {author} {\bibfnamefont {D.}~\bibnamefont {Qian}},\ and\ \bibinfo {author} {\bibfnamefont {X.}~\bibnamefont {Xu}},\ }\bibfield  {title} {\bibinfo {title} {{In-Plane Magnetic Anisotropy and Large Topological Hall Effect in Self-Intercalated Ferromagnet Cr$_{1.61}$Te$_2$}},\ }\href {https://doi.org/10.1002/adfm.202510351} {\bibfield
  {journal} {\bibinfo  {journal} {Advanced Functional Materials}\ }\textbf {\bibinfo {volume} {n/a}},\ \bibinfo {pages} {e10351} (\bibinfo {year} {2025})}\BibitemShut {NoStop}%
\bibitem [{\citenamefont {Tran}\ \emph {et~al.}(2022)\citenamefont {Tran}, \citenamefont {Momida}, \citenamefont {Matsushita}, \citenamefont {Sato}, \citenamefont {Makino}, \citenamefont {Shirai},\ and\ \citenamefont {Oguchi}}]{Tran2022}%
  \BibitemOpen
  \bibfield  {author} {\bibinfo {author} {\bibfnamefont {H.~B.}\ \bibnamefont {Tran}}, \bibinfo {author} {\bibfnamefont {H.}~\bibnamefont {Momida}}, \bibinfo {author} {\bibfnamefont {Y.-i.}\ \bibnamefont {Matsushita}}, \bibinfo {author} {\bibfnamefont {K.}~\bibnamefont {Sato}}, \bibinfo {author} {\bibfnamefont {Y.}~\bibnamefont {Makino}}, \bibinfo {author} {\bibfnamefont {K.}~\bibnamefont {Shirai}},\ and\ \bibinfo {author} {\bibfnamefont {T.}~\bibnamefont {Oguchi}},\ }\bibfield  {title} {\bibinfo {title} {{Effect of magnetocrystalline anisotropy on magnetocaloric properties of an AlFe$_2$B$_2$ compound}},\ }\href {https://link.aps.org/doi/10.1103/PhysRevB.105.134402} {\bibfield  {journal} {\bibinfo  {journal} {Physical Review B}\ }\textbf {\bibinfo {volume} {105}},\ \bibinfo {pages} {134402} (\bibinfo {year} {2022})}\BibitemShut {NoStop}%
\end{thebibliography}

%

\end{document}